\documentclass[11pt,a4paper]{article}
\usepackage{color,amsmath,amsfonts,graphics,epsfig,amssymb,caption}
\begin{document}
\title{Unsolved problems in particle physics}
\author{Sergey Troitsky\\
\textit{Institute for Nuclear Research of the Russian Academy of Sciences}}
\date{}
\maketitle

\begin{abstract}
I consider selected (most important according to my own choice)
unsolved problems in particle theory, both those related to extensions of
the Standard Model (neutrino oscillations, which probably do not fit the
usual three-generation scheme; indications in favour of new physics from
astrophysical observations; electroweak symmetry breaking and hierarchy of
parameters) and those which appear in the Standard Model (description of
strong interactions at low and intermediate energies).
\end{abstract}
\thispagestyle{empty}
\setcounter{page}{1}
\tableofcontents

\captionsetup{width=\textwidth,format=hang,font={small},labelfont=bf}

\section{Introduction: status and parameters of the Standard Model}
\label{sec:intro}

One may compare the current state of quantum field theory and its
applications to particle physics with the situation 20-30 years ago and
discover, amusingly, that all principal statements of this field of physics
are practically unchanged, which is in contrast with rapid progress in
condensed-matter physics. Indeed, most of the experiments, held during the
last two decades, supported the correctness of predictions which had been
made earlier, derived from the models developed earlier. This success of
the particle theory resulted in considerable stagnation in its
development. However, one may expect that in the next few years, the
particle physics will again become an intensively developing area.
Firstly, there is a certain amount of collected experimental results
(first of all related to cosmology and astrophysics, but also obtained in
laboratories) which suggest that the Standard Model (SM) is incomplete.
Secondly, the theory was developing under the guidance of the principle of
naturalness, that is the requirement to explain quantitatively any
hierarchy in model parameters (in the case of SM, it is possible only
within a larger fundamental theory yet to be constructed). Finally, one of
the most important arguments for the coming excitement in particle physics
is the expectation of new results from the Large Hadron Collider. As it
will become clear soon, this accelerator will be able to study the
\textit{full} range of energies where the physics responsible for the
electroweak symmetry breaking should appear, so we are expecting
interesting discoveries in the next few years in any case: either it will
be the Higgs boson, or some other new particles, or (in the most
interesting case) no new particle will be found which would suggest a
serious reconsideration of the Standard Model.

The Large Hadron Collider (LHC, see e.g. \cite{LHC}) is an accelerator
which allows to collide protons with the center-of-mass energy up to
14~TeV (currently working at 7~TeV) and heavy nuclei. In a 30-km length
tunnel, at the border of Switzerland and France, there are four main
experimental installations (general-purpose detectors ATLAS and CMS; LHCb
which is oriented to the study of $B$ mesons and ALICE, specialized in
heavy-ion physics) as well as a few smaller experiments. The first results
of the work of the collider have brought a lot of new information on
particle interactions which we will mention when necessary.

The purpose of the present review is to discuss briefly the current state
of particle physics and possible prospects for its development. For such a
wide subject, the selection of topics is necessarily subjective and
estimates of importance of particular problems and of potential of
particular approaches reflect the author's personal opinion,
while the bibliography cannot be made exhaustive.

The contemporary situation in the particle physics may be described as
follows. Most of the modern experimental data are well described by the
Standard Model of particle physics which was created in 1970s. At the same
time, there are a considerable amount of indications that SM is not
complete and is not more than a good approximation to the correct
description of particles and interactions. We are not speaking now about
minor deviations of certain measured observables from theoretically
calculated ones -- these deviations may be related to insufficient
precision of either the measurement or the calculations, to
unaccounted systematic errors or insufficient sets of experimental data
(statistical fluctuations); it happens that these deviations disappear
after a few years of more detailed study. Contrary, we will emphasise more
serious qualitative problems of SM, the latter being considered as an
instrument of quantitative description of elementary particles. These
problems include the following:

(1)~experimental indications to the incompleteness of SM, namely the
well-established experimental observations of neutrino oscillations (which
are impossible, see Sec.~\ref{sec:neutrino:mass}, in SM) and incapability
of SM to describe the results of astrophysical observations, in particular
of those related to the structure and evolution of the Universe;

(2)~not fully natural and not calculable in the theory values of the SM
parameters, in particular, the fermion mass hierarchy, the hierarchy of
symmetry-breaking scales and the absence of a light (with mass $\lesssim
100$~GeV) Higgs boson;

(3)~purely theoretical difficulties in description of hadrons by means of
the available methods of quantum field theory.

We will discuss these unsolved problems of SM and related prospects for
the development of the particle theory.

For future reference, it is useful to recall briefly the structure of SM
(see e.g.\ \cite{SMprimer, ChengLi} and the appendix to \cite{RuGorby1}).
The model includes a certain set of particles and their interactions.

\begin{figure}
\centering \includegraphics[width=0.65\columnwidth]{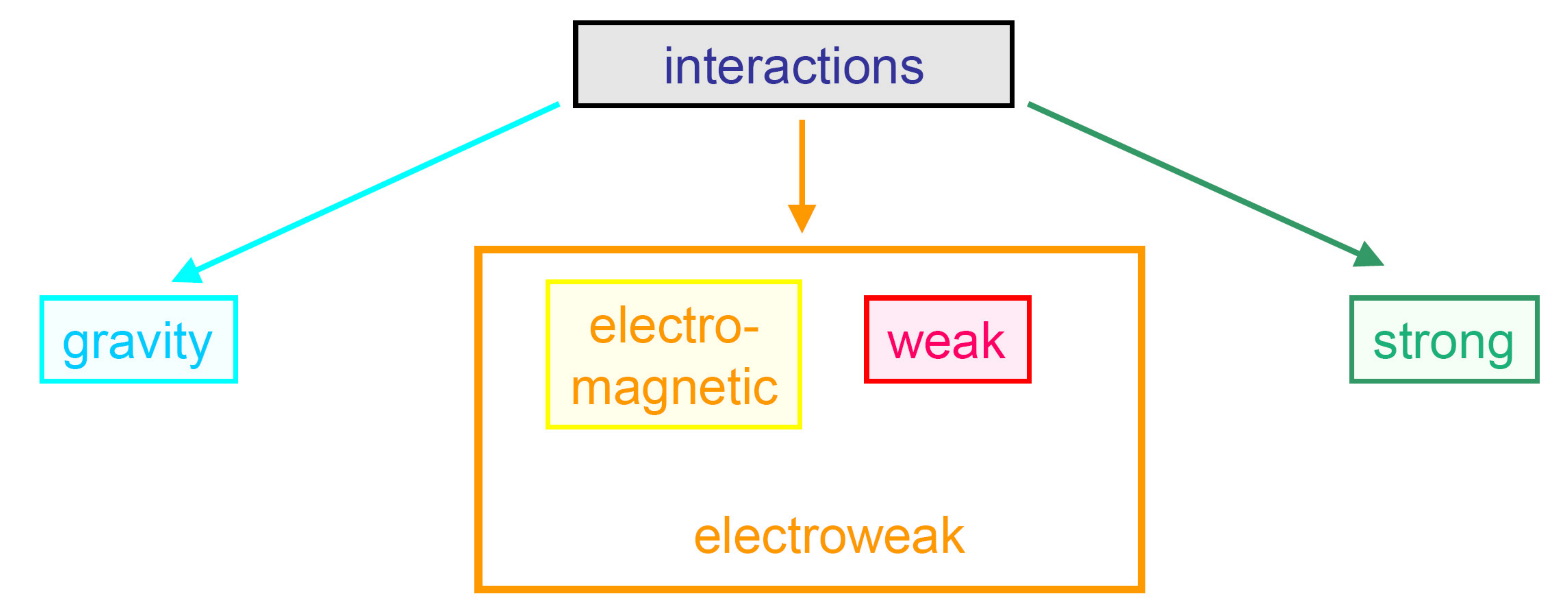}
\caption{
\label{fig:SMinteractions}
Particle interactions.}
\end{figure}

Out of four known interactions (see
Fig.~\ref{fig:SMinteractions}),
three are described by SM -- the electromagnetic, weak and strong ones.
The first two of them have a common \textit{electroweak} gauge interaction
behind them. The symmetry of this interaction,
$SU(2)_{\rm L} \times U(1)_{\rm Y}$,
manifests itself at energies higher than
$\sim 200$~GeV. At lower energies, this symmetry is broken down to
$U(1)_{\rm EM} \ne U(1)_{\rm Y}$ (the electroweak symmetry breaking);
in SM, this breaking is related to the vacuum expectation value of a
scalar field, the Higgs boson. Parameters of the electroweak breaking are
known up to a high precision; experimental data are in a perfect agreement
with the theory. The Higgs boson has not been observed yet; its mass,
being a free parameter of the theory, is bound by direct experimental
searches (see Table~\ref{tab:SMparameters} and more details
in Sec.~\ref{sec:EW}).

The \textit{strong} interaction in SM is described by the quantum
chromodynamics (QCD), a theory with the gauge group
$SU(3)_{\rm C}$. The effective coupling constant of this theory grows when
the energy is decreased. As a result, particles which feel this
interaction cannot exist as free states and appear only in the form of
bound states called hadrons. Most of modern methods of quantum field
theory work for small values of coupling constants, that is, for QCD, at
high energies.

The fourth known interaction, the \textit{gravitational} one, is not
described by SM, but its effect on the microscopic physics is negligible.

\begin{figure}
\centering \includegraphics[width=0.85\columnwidth]{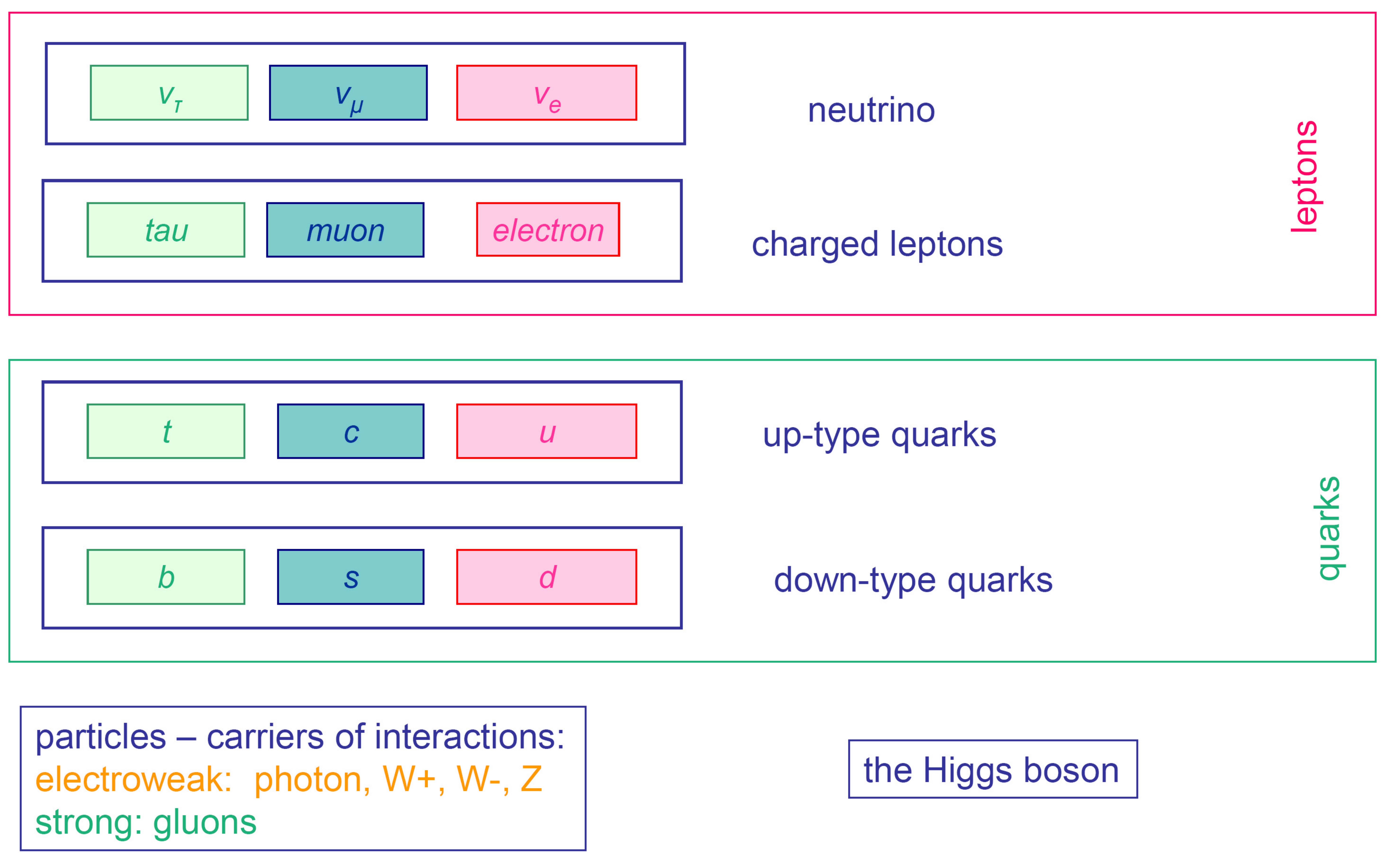}
\caption{
\label{fig:SMparticles}
Particles described by the Standard Model.}
\end{figure}

The particle content of SM is summarized in Fig.~\ref{fig:SMparticles}.
Quarks and leptons, the so-called SM matter fields, are described by
fermionic fields. Quarks take part in strong interactions and compose
observable bound states, hadrons. Both quarks and leptons participate in
the electroweak interaction. The matter fields constitute three
generations; particles from different generations interact identically but
have different masses. The full electroweak symmetry forbids fermion
masses, so nonzero masses of quarks and leptons are directly related to
the electroweak breaking -- in SM, they appear due to the Yukava
interaction with the Higgs field and are proportional to the vacuum
expectation value of the latter.   For the case of neutrino, these Yukawa
interactions are forbidden as well, so neutrinos are strictly massless in
SM. The gauge bosons, which are carriers of interactions,  are massless
for unbroken gauge groups
$U(1)_{\rm EM}$ (electromagnetism -- photons) and
$SU(3)_{\rm C}$ (QCD -- gluons); masses of
$W^{\pm}$ and $Z$ bosons are determined by the mechanism of the
electroweak symmetry breaking. All SM particles, except for the Higgs
boson, are found experimentally.

From the quantum-field-theory point of view, quarks and leptons may be
described as states with definite mass. At the same time, gauge bosons
interact with superpositions of these states; in another formulation, when
the base is chosen to consist of the states interacting with the gauge
bosons, the SM symmetries allow not only for the mass terms,
$m_{ii}\bar\psi_{i}\psi_{i}$, for each $i$th fermion
$\psi_{i}$,
but also for a nondiagonal \textit{mass matrix}
$m_{ij}\bar\psi_{i}\psi_{j}$.
Up to unphysical parameters, in SM, this matrix is trivial in the leptonic
sector, while in the quark sector it is related to the
Cabibbo-Kobayasi-Maskava (CKM) matrix. The latter may be expressed through
three independent real parameters (quark mixing angles) and one complex
phase (for more details, see
\cite{ChengLi, KobayashiNobel}).

The Standard Model has therefore 19 independent parameters, values of 18
of which are determined experimentally. They include three gauge coupling
constants,
$\alpha _{s}$, $\alpha _{2}$ and $\alpha_{1}$ for gauge groups $SU(3)_{C}$,
$SU(2)_{W}$ and $U(1)_{Y}$, respectively
(the latter two are often expressed through the electromagnetic coupling
constant $\alpha$ and the mixing angle
$\theta_{W}$), the QCD $\Theta$-parameter, nine charged-fermion masses
$m_{u,d,s,c,b,t,e,\mu ,\tau }$,
three quark mixing angles $\theta_{12,13,23}$, one complex phase
$\delta$ of the CKM matrix and two parameters of the Higgs sector, which
are conveniently expressed through the known Higgs-boson vacuum expectation
value $v$ and  its unknown mass $M_{H}$. Experimental values of these
parameters, recalculated from the 2010 data
\cite{PDG2010}
(bounds on the mass of the Higgs boson based on LEP, Tevatron and LHC data
are given as of December, 2011), may be found in
Table~\ref{tab:SMparameters}.
\begin{table}
\begin{center}
\begin{tabular}{|ccc|}
\hline
$\alpha_s(M_{Z})$ &$=$& $0.114 \pm 0.0007$\\
$1/\alpha(M_{z}) $ &$=$& $ 127.916 \pm 0.015$\\
$\sin^2 \theta_W (M_{Z}) $ &$=$& $ 0.23108 \pm 0.00005 $\\
$\Theta$ & $\lesssim$ & $10^{-10}$\\
$m_{u}(2~\mbox{GeV}) $ &$=$& $ 2.5^{+0.8}_{-1.0}$~MeV\\
$m_{d}(2~\mbox{GeV}) $ &$=$& $ 5.0^{+1.0}_{-1.5}$~MeV\\
$m_{s}(2~\mbox{GeV}) $ &$=$& $ 105^{+25}_{-35}$~MeV\\
$m_{c}(m_{c})$ &$=$& $ 1.266^{+0.031}_{-0.036}$~GeV\\
$m_{b}(m_{b})$ &$=$& $ 4.198 \pm 0.023$~GeV\\
$m_{t}(m_{t}) $ &$=$& $ 173.1 \pm 1.35$~GeV\\
$m_{e} $ &$=$& $510.998910\pm 0.000013 $~keV\\
$m_{\mu} $ &$=$& $105.658367\pm 0.000004 $~MeV\\
$m_{\tau}$ &$=$& $1.77682 \pm 0.00016$~GeV\\
$ \theta_{12} $ &$=$& $13.02^{\circ}\pm 0.05^{\circ}$\\
$ \theta_{23} $ &$=$& $2.35^{\circ}\pm 0.06^{\circ}$\\
$ \theta_{13} $ &$=$& $0.199^{\circ}\pm 0.011^{\circ}$\\
$\delta $ &$=$& $1.20 \pm 0.08$\\
$v (m_{\mu })$ & $=$ & $246.221\pm 0.002 $~GeV\\
$m_{H}$ &  & 115~GeV \dots 127~GeV\\
\hline
\end{tabular}
\end{center}
\caption{
\label{tab:SMparameters}
Parameters of the Standard Model. For parameters with significant energy
dependence, the energy scales, to which the numerical values correspond,
are given in parentheses.}
\end{table}

It is worth reminding that the observable world is mostly mad eof atoms,
so, out of the full manifold of elementary particles, only few are met ``in
the everyday life''. These are $u$ and $d$ quarks in the form of protons
(udd) and neutrons (uud), electrons and, out of interaction carriers, the
photon. The reasons for that are different for different particles. In
particular, neutrino does not interact with the electromagnetic field and
is therefore very hard to detect; heavy particles are unstable and decay
to lighter ones; strongly interacting quarks and gluons are confined in
hadrons. The full manyfold of SM particles reveal themselves either in
complicated dedicated experiments, or indirectly by their effect seen in
astrophysical observations.

Thus, before to proceed with the description of unsolved problems, let us
recall that all experimental results concerning physics of charged
leptons, photons, $W$ and $Z$ bosons at all available energies and quarks
and gluons at high energies are in excellent agreement with SM for a given
set of its parameters.

\section{The observed deviation from the Standard Model: neutrino
oscillations.}
\label{sec:neutrino}
Let us discuss the unique, well-established in laboratory experiments,
evidence in favour of incomleteness of SM, the phenomenon of neutrino
oscillations, that is mutual conversion of neutrinos of different
generations to each other. A more detailed modern description of the
problem may be found in the book
\cite{Giunti-book}, in the Appendix to the textbook~\cite{RuGorby1}, in
reviews \cite{Bilenky-UFN, Akhmedov-UFN, Kudenko-UFN} etc.

\subsection{Theoretical description.}
\label{sec:neutrino:general}
In analogy with the case of charged leptons, let us consider three
generations of neutrino: electron neutrino  ($\nu_{e}$), muon neutrino
($\nu_{\mu}$) and tau neutrino
($\nu_{\tau}$). The corresponding fermion fields interact with the gauge
bosons $W$ and $Z$ through weak charged and neutral currents. These
interactions are responsible for both creation and experimental detection
of neutrinos.

Similarly to the quark case, one may suppose that neutrinos have a nonzero
mass matrix (though it cannot be incorporated in SM, the low-energy
effective theory, electrodynamics, does not forbid it) which may be
nondiagonal. It is convenient to describe this system in terms of linear
combinations
$\nu_{1,2,3}$ of the original fields
$\nu_{e,\mu,\tau}$
with the diagonal mass matrix,
\[
\nu_{i}=\sum\limits_{\alpha=e,\mu,\tau} U_{i\alpha }\nu_{\alpha },
\]
where $U_{i\alpha }$, $i=1,2,3$; $\alpha=e,\mu,\tau$, are the elements of
the leptonic mixing matrix.

To demonstrate the phenomenon of neutrino oscillations, let us restrict
ourselves to the case of two flavours, $\nu _{e}$ and $\nu _{\mu }$. Let
their linear combinations,
\begin{equation}
\nu _{1}=\cos\theta_{12}\, \nu_{e} +\sin\theta_{12} \,\nu_{\mu},
\label{Eq:2neutrino}
\end{equation}
\[
\nu _{2}=-\sin\theta_{12}\, \nu_{e} +\cos\theta_{12} \,\nu_{\mu},
\]
be the eigenvectors of the mass matrix with eigenvalues
$m_{1}^{2}$, $m_{2}^{2}$, respectively.
The inverse transformation expresses
$(\nu _{e},\nu _{\mu })$ through $(\nu _{1},\nu _{2})$:
\[
\nu _{e}=\cos\theta_{12}\, \nu_{1} -\sin\theta_{12}\, \nu_{2},
\]
\[
\nu _{\mu}=\sin\theta_{12}\, \nu_{1} +\cos\theta_{12}\, \nu_{2}.
\]
Suppose that at the moment
$t=0$, in a certain weak-interaction event, an electron neutrino
$\nu _{e}$ was created, that is the superposition of
$\nu _{1}$ and $\nu _{2}$ with known coefficients:
\[
\nu _{1}(0)=\cos\theta_{12}\, \nu_{e}(0),
\]
\[
\nu _{2}(0)=-\sin\theta_{12}\, \nu_{e}(0).
\]
The evolution of mass eigenstates for a plane monochromatic wave moving
in the direction $z$ mas be described as
\[
\nu_{i}(z,t)={\rm e}^{-i\omega t+i\sqrt{\omega^{2}-m_{i}^{2}}z}
\nu_{i}(0), ~~~i=1,2,
\]
where $\omega$ is the energy and $\sqrt{\omega^{2}-m_{i}^{2}}$ is the
momentum. While propagating, the wave packets corresponding to
$\nu _{1}$ and $\nu _{2}$ disperse in different ways, so that the relation
$(\cos\theta_{12}, -\sin\theta_{12})$ between their coefficient no longer
holds, which means that an admixture of the orthogonal state,
$\nu_{\mu}$, appears. In the (commonly considered) ultrarelativistic limit,
$\omega \gg m_{i}$ and $\sqrt{\omega^{2}-m_{i}^{2}}\simeq
\omega-\frac{m_{i}^{2}}{2\omega}$. The probability to detect $\nu_{\mu}$
at a point $(t,z)$ for each emitted  $\nu_{e}$ is then
\begin{equation}
P(\nu_{\mu}; z, t)= \left| \nu_{\mu} (z,t) \right|^{2} = \sin^{2} 2\theta_{12}
\, \sin^{2} \left( \frac{m_{2}^{2}-m_{1}^{2}}{4\omega}z \right).
\label{Eq:PneutrinoOsc}
\end{equation}
One may see that this probability is an oscillating function of the
distance $z$, hence the term ``neutrino oscillations''. As expected, no
oscillations happen either in the case of equal (even nonzero) masses
(similar dispersions of $\nu_{1}$ and $\nu_{2}$) or for a diagonal mass
matrix ($\theta_{12}=0$, $\nu_{1}=\nu_{e}$  etc.). A similar description
of oscillations of three neutrino flavours determines, in analogy with
Eq.~(\ref{Eq:2neutrino}), three mixing angles $\theta_{12}$,
$\theta_{13}$, $\theta_{23}$.

When individual neutrinos propagate to large distances, the oscillation
formalism described above stops to work, because the particles of
different mass require different time to propagate from the source, hence
loss of coherence; nevertheless the transformations of neutrinos are
possible and their probability is calculable.

\subsection{Experimental results: standard three-flavour oscillations.}
\label{sec:neutrino:usual}
Let us turn now to the history
(see e.g.\ \cite{Giunti-book}) and the modern state
(see e.g.\ \cite{1107.3846}) of the question of neutrino oscillations. In
1957, Pontecorvo \cite{g880, g881} suggested the possibility of
oscillations in the ``neutrino--antineutrino'' system, similar to $K$
meson oscillations already known at that time. This first mentioning of
the possibility of neutrino oscillations was aimed at the explanation of
preliminary Davis' results about observation of the reaction
$\bar\nu +^{37}{\rm Cl} \to ^{37}{\rm Ar}+e^{-}$ with reactor neutrinos.
On one hand, this experimental result has not been confirmed; on the
other one, it has become clear that the Pontecorvo model was not able to
describe it even if it were true. The first mention of mutual
transformations of
$\nu_{e}$ and $\nu_{\mu}$ is due to Maki, Nakagawa and Sakata~\cite{maki},
while the first succesful description of oscillations in the system of
two-flavour neutrinos was given by Pontecorvo \cite{g883} and by Gribov
and Pontecorvo \cite{g567}. The theory of neutrino oscillations in its
present form has been developed in 1975-76 by Bilenky and Pontecorvo
\cite{g236, g239}, Eliser and Swift \cite{g404}, Fritch and Minkowski
\cite{g466}, Mikheyev, Smirnov \cite{g801, g802} and Wolfenstein
\cite{g1065}.

The first experimental evidence in favour of neutrino oscillations have
been obtained more than a half century ago, though for considerable period
of time their interpretation remained an open question. We are speaking
about the so-called ``solar neutrino problem'': the observed flux of
neutrinos form the Sun was considerably lower than it was predicted by a
model of solar nuclear reactions. This solar neutrino deficit was first
found in the Homestake experiment (USA; first as early as 1968
\cite{Homestake}) and subsequently confirmed by Kamiokande (Japan)
\cite{Kamiokande}, SAGE (Russia, Baksan neutrino observatory of INR, RAS)
\cite{SAGE}, GALLEX/GNO (Italy, the Gran-Sasso laboratory) \cite{GALLEX}
and Super-K (Japan) \cite{hep-ex/0404034} experiments, which made use of
various experimental techniques and were sensitive to neutrinos form
different nuclear reactions. Since only electron neutrinos are produced in
the Sun, and only these were detected in the experiments, the deficit
might be explained by transformation of a part of electron neutrinos to
muon ones.

The natural source of \textit{muon} neutrinos is provided by cosmic rays,
that is charged particles (protons and nuclei) of extraterrestrial origin
which interact with atoms in the Earth's atmosphere and produce secondary
particles. A significant part of the latters are charged $\pi$ mesons.
Neutrinos from decays of these $\pi$ mesons, as well as from decays of
secondary muons, are called atmospheric neutrinos. The first indications
to oscillations of the atmospheric neutrinos have been obtained in the end
of 1980s in
Kamiokande \cite{Kamiokande-atm} and IMB \cite{IMB} experiments, with
subsequent
confirmation in Soudan-2
\cite{hep-ex/9611007}, MACRO \cite{hep-ex/9807005} and Super-K
\cite{hep-ex/9812014}. Their result is the anisotropy in the flux of muon
neutrinos: from above, that is from the atmosphere, the flux is higher
than from below (through the Earth). Without oscillations, the flux were
isotropic since it is determined by an isotropic flux of primary cosmic
rays while the interaction of neutrino with the terrestrial matter is
negligible. This anisotropy is not seen for electron neutrinos, hence it
is natural to suppose that
$\nu_{\mu}$ oscillate mainly to $\nu_{\tau}$
(the latters were not detected in these experiments).

In the first decade of our century, a significant experimental progress in
the questions we discuss has been achieved, so that now we have a reliable
experimental proof of neutrino transformations with measured parameters.

\begin{figure}
\centering
\includegraphics[width=0.65\columnwidth]{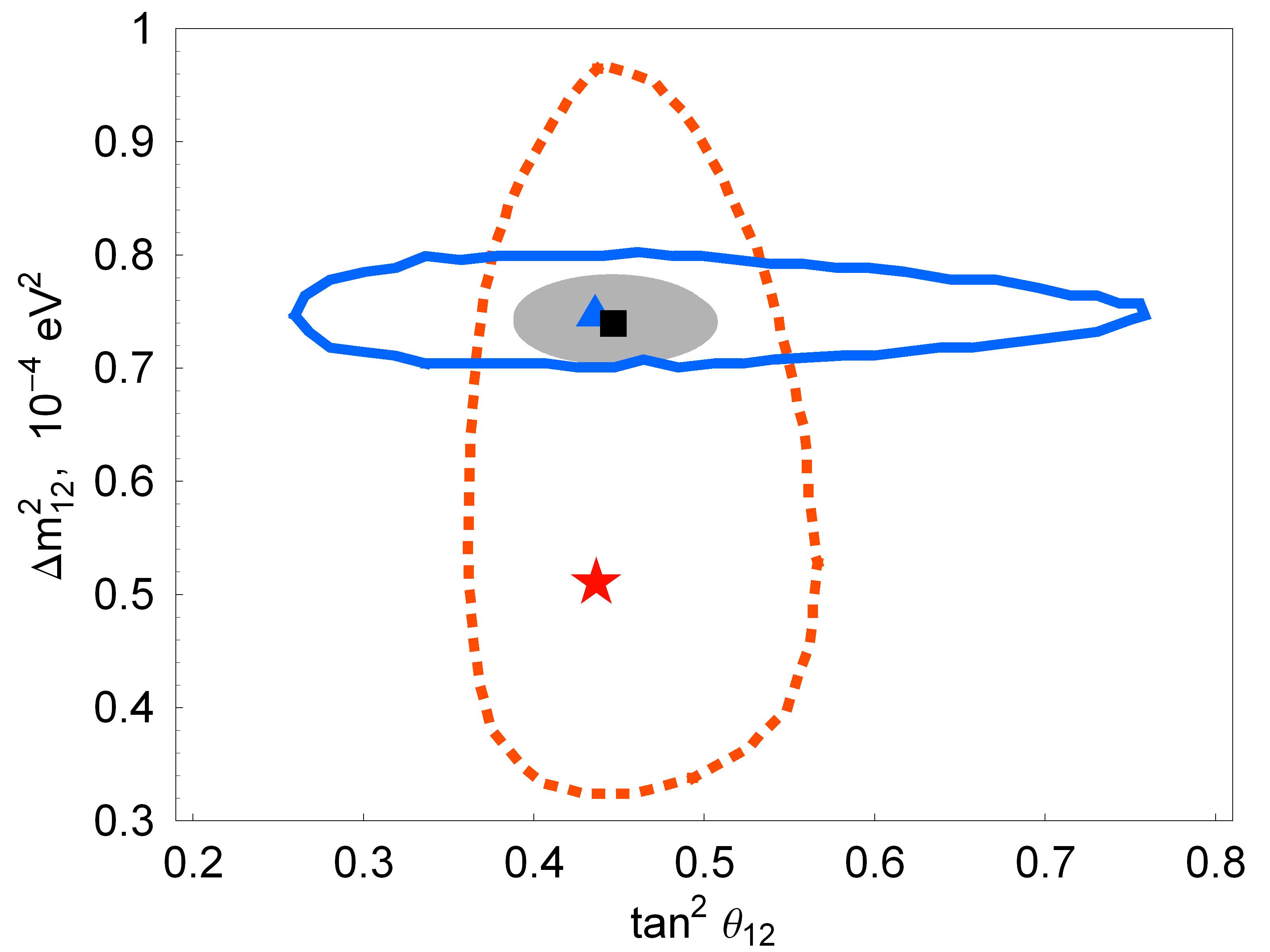}
\caption{
\label{fig:solar-nu}
Limits (95\% confidence level) on the $\nu_{e}-\nu_{\mu }$
oscillation parameters as result from the analysis taking into account
three neutrino flavours
\cite{1109.0763}.
The dotted line corresponds to the combination of solar experiments, the
full line represents the KamLAND constraints, the gray ellipsis gives the
constraints from the combination of all data. The star, the triangle and
the square correspond to the most probable oscillation parameters obtained
in these three analyses, respectively.}
\end{figure}

\textbf{$\nu_{e}-\nu_{\mu }$ oscillations.}
In addition to more or less model-dependent results about the solar
neutrino deficit ($\nu_{e}$ \textit{disappearence}), the SNO experiment
has detected, in 2001
\cite{nucl-ex/0204008}, \textit{appearence} of neutrino of other flavours
from the Sun in a full agreement with the flux expected in the
oscillational picture. It has therefore closed the ``solar neutrino
problem'' and supported, at the same time, the standard solar model.
The KamLAND experiment \cite{0801.4589}
registered disappearence of electron antineutrino born in reactors of
atomic power plants (in contrast with the case of the Sun, the initial
flux of the particles may be directly determined in this case). Parameters
of oscillations, measured in these very different experiments, are in
excellent agreement, see Fig.~\ref{fig:solar-nu}.
The SNO results, together with even more precise results of the BOREXINO
experiment (Italy)
\cite{1104.1816}, confirm the expected energy dependence of the
number of disappeared solar neutrinos in agreement with predictions of
Mikheyev, Smirnov
\cite{g801, g802} and Wolfenstain
\cite{g1065},
who developed a theory of neutrino oscillations in plasma: due to the
fact that electrons are present in plasma, unlike muons and tau leptons,
the interaction with medium goes differently for different types of
neutrino. As a result, the oscillation formalism is modified and the
resonance enhancement of oscillations becomes possible.

\begin{figure}
\centering
\includegraphics[width=0.65\columnwidth]{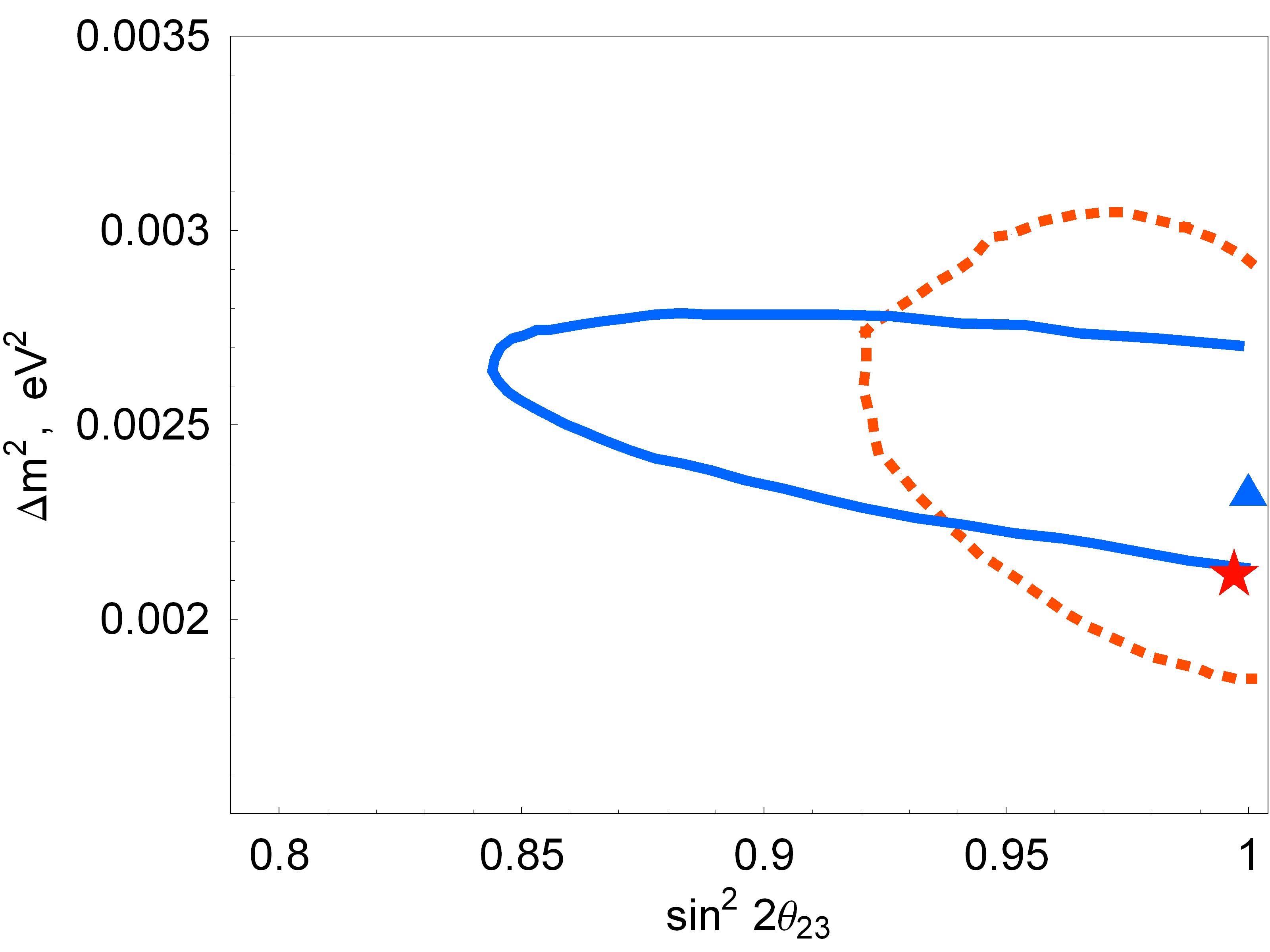}
\caption{
\label{fig:atm-nu}
Limits (90\% confidence level) on the $\nu_{\mu}-\nu_{\tau
}$ oscillation parameters.
The dotted line represents the results of the SuperK analysis with account
of three neutrino flavours \cite{SuperKnu2010}; the full line represents
constraints by MINOS \cite{1103.0340}. The star and the triangle denote
the most probable oscillation parameters for these two analyses,
correspondingly.}
\end{figure}

\textbf{$\nu_{\mu}-\nu_{\tau }$ oscillations.}
In addition to the Super-K experiment, which have measured
\cite{hep-ex/0501064, SuperKnu2010} deviations from isotropy in
atmospheric
$\nu_{\mu}$ and
$\bar\nu_{\mu }$ to a great accuracy, the disappearence of $\nu_{\mu}$ has
been measured directly in neutrino beams created by particle accelerators
(experiments K2K
\cite{hep-ex/0606032} and MINOS
\cite{hep-ex/0607088}), see Fig.~\ref{fig:atm-nu}.
Finally, in 2010, the OPERA detector which is located in the Gran Sasso
laboratory (Italy) has detected
\cite{1006.1623}
the first (and currently unique) case of \textit{appearence} of
$\nu_{\tau}$
in the
$\nu _{\mu }$ beam from the SPS accelerator (CERN, Switzerland).

\textbf{The mixing angle $\theta_{13}$.}
For a long time, the solar
($\nu_{e}-\nu_{\mu}$) and atmospheric
($\nu_{\mu}-\nu_{\tau}$) oscillation data have been described
independently (see discussions in
\cite{RuGorby1}, Appendix~C)
while relatively low precision of experiments allowed for zero value of the
mixing angle
$\theta_{13}$.
The situation has been recently changed and, analyzed commonly, the
data of various experiments point to nonzero $\theta_{13}$
\cite{PRL101-141801}. In summer 2011, two accelerator experiments, T2K
(Japan) \cite{T2K} and MINOS \cite{MINOStheta13}, which both search for
appearence of $\nu_{e}$ in  $\nu_{\mu}$ beams, published their results
which are incompatible with $\theta_{13}=0$. A quantitative analysis of
all data on the solar and atmospheric neutrinos, jointly with
the accelerator and reactor experiments which study the same part of the
parameter space, points \cite{1106.6028} towards a nonzero value of
$\theta_{13}$ at the confidence level better than 99\%. In
Table~\ref{tab:nu},
\begin{table}
\begin{center}
\begin{tabular}{|ccc|}
\hline
$\Delta m_{12}^{2}$  & $=$ & $\left(7.58^{+0.22}_{-0.26}
\right)\times 10^{-5} ~\mbox{eV}^{2}$ \\
$\Delta m_{23}^{2}$  & $=$ & $\left(2.31^{+0.12}_{-0.09}
\right)\times 10^{-3} ~\mbox{eV}^{2}$ \\
$\sin^{2}\theta_{12}$ & $=$ & $0.312^{+0.017}_{-0.016}$\\
$\sin^{2}\theta_{13}$ & $=$ & $0.025\pm 0.007$\\
$\sin^{2}\theta_{23}$ & $=$ & $0.42^{+0.08}_{-0.03}$\\
\hline
\end{tabular}
\end{center}
\caption{
\label{tab:nu}
Parameters of oscillations of three flavours of neutrino obtained with
account of all relevant experimental data as of summer 2011
\cite{1106.6028}.}
\end{table}
the results of this analysis are quoted.

\subsection{Experimental results: non-standard oscillations.}
\label{sec:neutrino:unusual}
The combination of all experiments described above is in a good
quantitative agreement with the picture of oscillations of three types of
neutrino with certain parameters. However, there exist results which do
not fit this picture and may suggest that the fourth (or, maybe, even
the fifth) neutrino exists. As we have seen above, one of the principal
oscillation parameters is the mass-square difference,
$\Delta
m_{ij}^{2}=m_{j}^{2}-m_{i}^{2}$.
The results on atmospheric and solar neutrinos, jointly with the
accelerator and reactor experiments, are explained by two unequal $\Delta
m_{ij}^{2}$, see Table~\ref{tab:nu},
\[
\Delta m_{12}^{2}
\ll
\Delta m_{23}^{2}\sim 2\times 10^{-3}~\mbox{eV}^{2}.
\]
In the case of three neutrinos, these two values compose the set of
linearly independent $\Delta m_{ij}^{2}$ and
\[
|\Delta m_{13}^{2} |=
|\Delta m_{12}^{2}-\Delta m_{23}^{2}|\sim \Delta m_{23}^{2}.
\]
Therefore, the observation of any neutrino oscillations with
$\Delta m_{ij}^{2} \gg \Delta m_{23}^{2}$ implies either the
existence of a new neutrino flavour ($i,j>3$) or some other deviation from
the standard picture. On the other hand, there is a very restrictive bound
on the number of relatively light
($m_{i}<M_{Z}/2$) particles with the quantum numbers of neutrino. This
bound comes from precise measurements of the $Z$-boson width and implies
that there are only three such neutrinos. This means that the fourth
neutrino, if exists, does not interact with the $Z$ boson; in other words,
it is ``sterile''. We will turn now to a certain experimental evidence in
favour of $\Delta m_{ij}^{2}\sim 1 ~\mbox{eV}^{2}$. We note that the
oscillations related to this $\Delta m_{ij}^{2}$ should reveal themselves
at relatively short distances and may be detected in so-called
short-baseline experiments.

\begin{figure}
\centering
\includegraphics[width=0.75\columnwidth]{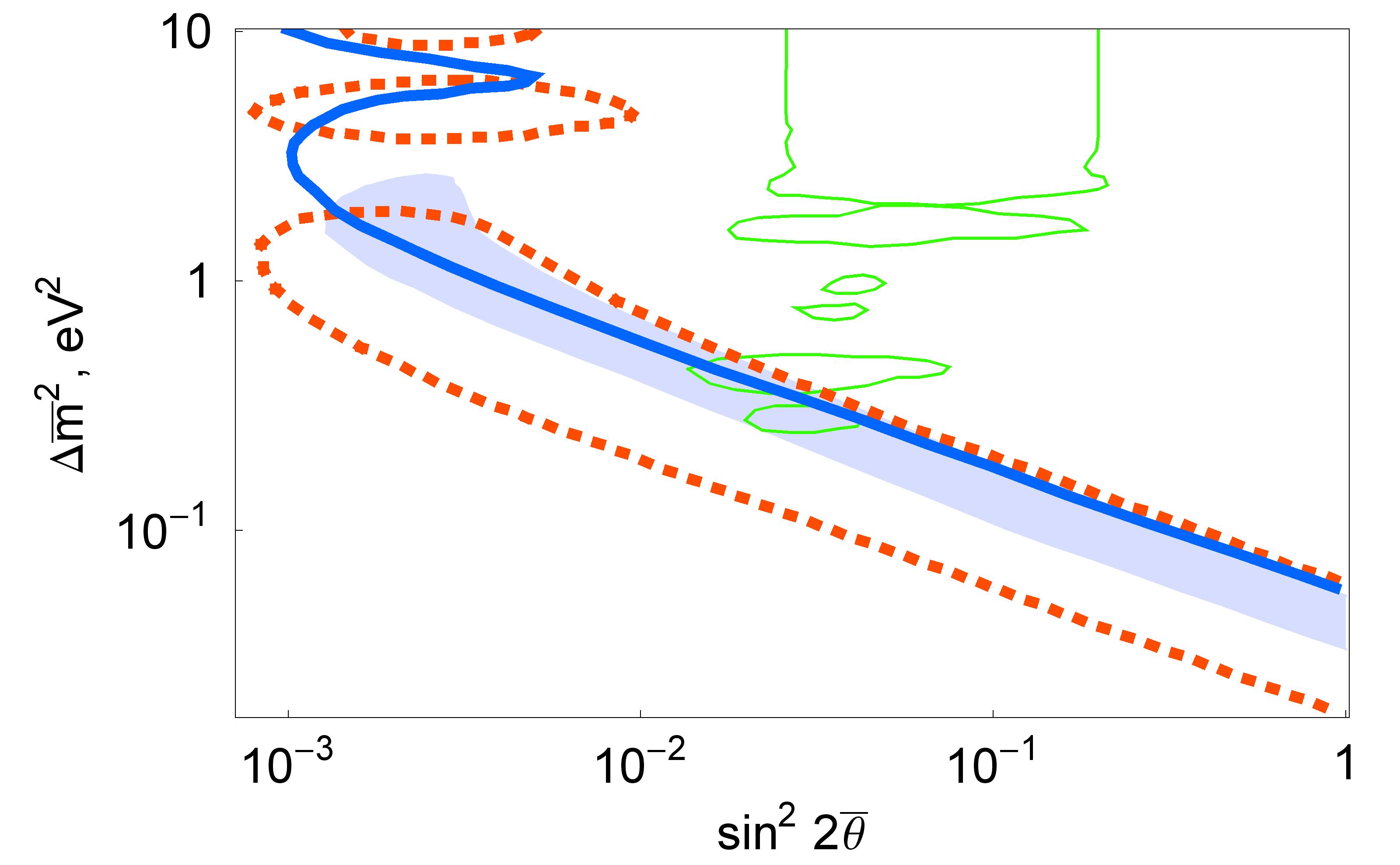}
\caption{
\label{fig:LSND}
Limits (90\% confidence level) on the parameters of
$\bar\nu_{\mu}-\bar\nu_{e}$ oscillations. The shaded region is compatible
with the LSND signal
\cite{LSND}; the region inside the dotted curve -- with the MiniBooNE
signal
\cite{MiniBooNEantiNew}.
Thin full lines bound the region of parameters compatible with a joint
re-analysis of reactor data
\cite{ReactorAnomaly}, see text. The KARMEN2 experiment excludes
\cite{KARMEN} the region above and to the right from the thick full line.
}
\end{figure}

\textbf{$\bar\nu_{\mu}-\bar\nu_{e }$ oscillations.}
The LSND experiment
\cite{LSND} studied muon decay at rest, $\mu^{+}\to
e^{+}\nu_{e}\bar\nu_{\mu}$, and measured the $\bar \nu_{e}$
flux at the distance about 30~m from the place where muons were held.
The excess of this flux over the background rate has been detected and
interpreted as appearence of
$\bar\nu_{e}$ as a result of $\bar\nu_{\mu}$ oscillations,
for a range of possible parameters. A similar experiment, KARMEN
\cite{KARMEN}
excluded a significant part of this parameter space, however, in 2010, the
MiniBooNE experiment \cite{MiniBooNEanti}
has also detected an anomaly which is compatible with the LSND results
and, within statistical uncertainties, does not contradict KARMEN for a
certain range of parameters
(Fig~\ref{fig:LSND}).

Another group of short-baseline experiments which study possible
$\bar\nu_{e} - \bar\nu_{\mu}$
oscillations search for disappearence of
$\bar\nu_{e}$
in the antineutrino flux from nuclear reactors. These experiments continued
for decades; recently, their results have been reanalized jointly
\cite{ReactorAnomaly}
with a more precise theoretical calculation of the expected fluxes. It has
been shown that there is a statistically significant deficit of
$\bar\nu_{e}$ in the detectors which is compatible with
$\Delta
m^{2}\sim 1\mbox{eV}^{2}$, --
the so-called reactor neutrino anomaly. The corresponding limits on the
parameters are also shown in Fig.~\ref{fig:LSND} for convenience. However,
one should keep in mind that while LSND, KARMEN and MiniBooNE detected
$\bar\nu_{e}$  in the $\bar\nu_{\mu}$ flux,
therefore constraining $\bar\nu_{e} - \bar\nu_{\mu}$ oscillations, the
reactor experiments only fix the disappearance of
$\bar\nu_{e}$. While the lack of this disappearence excludes
$\bar\nu_{e} - \bar\nu_{\mu}$ oscillations, the presence of it may be
explained as a transformation of
$\bar\nu_{e}$ into antineutrino of any other type.

One can see that there are several independent indications in
favour of $\Delta m^{2}\sim 1\mbox{eV}^{2}$, which, as we discussed above,
require either introduction of more-than-three neutrino flavours or (see
below) some other new physics.

\textbf{Other anomalies.}
Recent intense exploration of the field of neutrino oscillations revealed
also a range of other anomalies which are currently being discussed and
re-checked intensively.

\begin{figure}
\centering
\includegraphics[width=0.55\columnwidth]{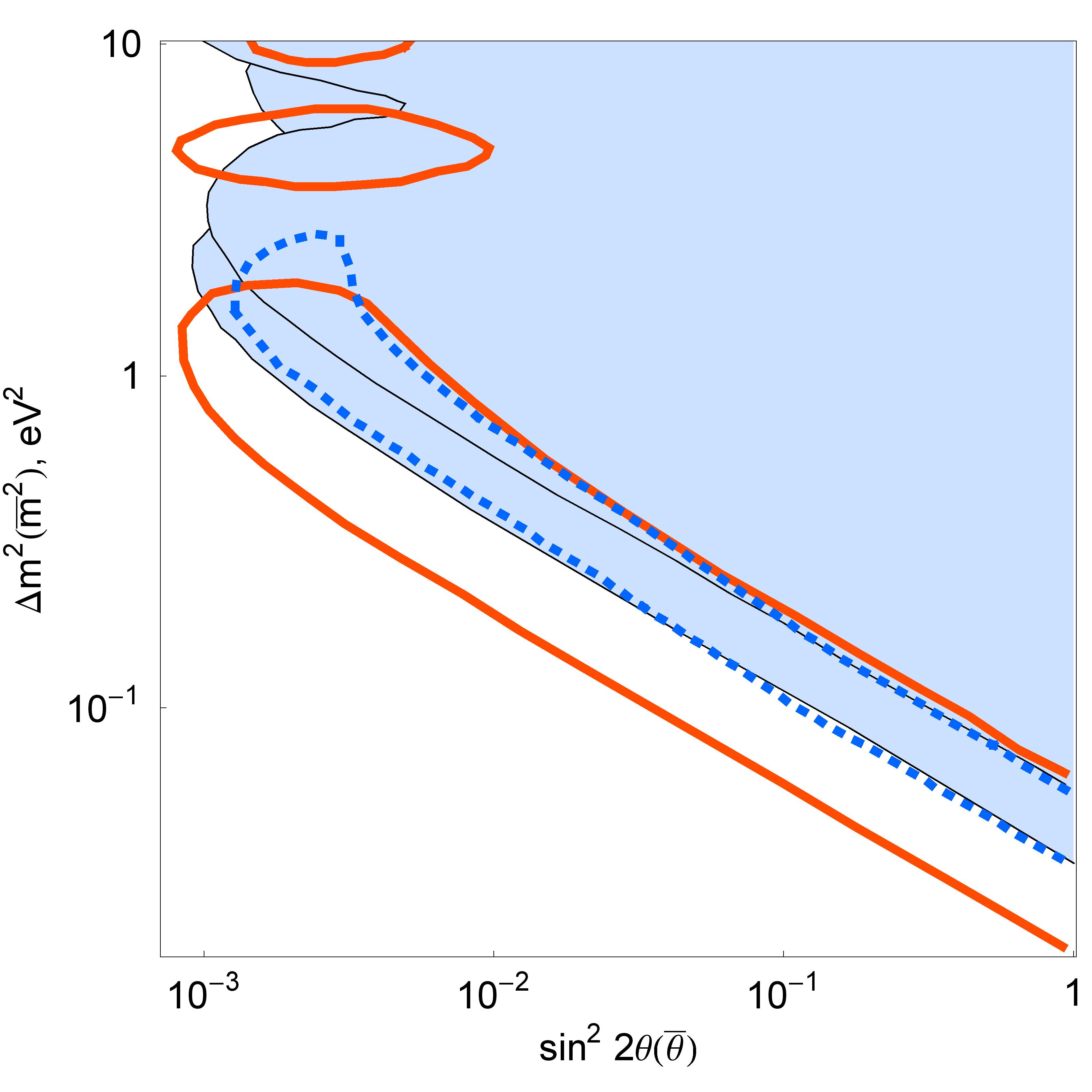}
\caption{
\label{fig:MiniBooNEanti}
Limits (90\% confidence level) on the  $\nu_{\mu}-\nu_{e}$
and $\bar\nu_{\mu}-\bar\nu_{e}$ oscillation parameters.
The shaded area corresponds to the part of the parameter space which is
excluded for neutrino oscillations by
MiniBooNE~\cite{MiniBooNEnu} and KARMEN~\cite{KARMEN},
while thick contours limit the region which corresponds to the signal in
antineutrino oscillations
(full lines, MiniBooNE~\cite{MiniBooNEantiNew}; dotted line,
LSND~\cite{LSND}).}
\end{figure}

\textit{Possible difference between neutrino and antineutrino
oscillations.}
The MiniBooNE experiment  studied separately neutrino and
antineutrino beams. The
$\bar\nu_{e}$ appearence has been detected \cite{MiniBooNEanti,
MiniBooNEantiNew} while that of $\nu_{e}$ has not
 \cite{MiniBooNEnu} (see Fig.~\ref{fig:MiniBooNEanti}).
In assumption of equal oscillation parameters for
$\nu$ and $\bar\nu$, the MiniBooNE result contradicts to LSND, but without
this assumption, contrary, the LSND claim is supported. It is worth noting
that the MINOS experiment also performed separate measurements with
neutrino and antineutrino beams (studying a range of much smaller
$\Delta m^{2}$);
first their results for the two cases were incompatible at the 98\%
confidence level, however, subsequent analysis of a larger amount of data
did not confirm this difference
\cite{MINOSantiNu}.
The latter result agrees with Super-K: though this experiment cannot
distinguish neutrino from antineutrino in each particular case, it may
limit
\cite{SuperKantiNu}
antineutrino oscillation parameters statistically, on the basis of a known
contribution of
$\bar\nu_{\mu}$ to the atmospheric neutrino flux.

\textit{Calibration of gallium detectors.} The GALLEX
\cite{GALLEXcalibration} and SAGE \cite{SAGEcalibration}
experiments, constructed to detect solar neutrinos with the help of the
gallium detectors, calibrated their instruments with the help of
artificial sources of radioactivity. They detected a deficit of electron
neutrino compatible with oscillations with
$\Delta m^{2} \gtrsim 0.1~\mbox{eV}^{2}$ (see also \cite{GiuntiGallium}).
This mass-square difference, which by itself does not agree with the
standard three-neutrino oscillation picture, agrees with the antineutrino
results of LSND, MiniBooNE and reactor experiments, however the
corresponding mixing angle differs from the predictions of the latters
\cite{1008.4750}.

\textit{Other puzzles.}
When speaking about unexplained results of neutrino experiments, one may
mention also the unexpected excess of events with energies $\lesssim
400$~MeV detected by MiniBooNE for neutrinos
\cite{MiniBooNEexcess} and antineutrinos \cite{MiniBooNEantiNew};
possible seasonal variations of the neutrino flux in the Troitsk-$\nu$mass
\cite{Lobashev-season} and MiniBooNE \cite{MiniBooNEseason} experiments;
the result of the OPERA experiment \cite{OPERAsuperluminal} which measured
the velocity of muon neutrinos which happened to be large than the
speed of light. All these very interesting anomalies currently
await their confirmation in independent experiments.

\textit{Possible theoretical explanations.}
The experimental results listed above are rather hard to explain.
On one hand, a series of experiments suggest neutrino transformations
compatible with
$\Delta m^{2} \gtrsim 0.1~\mbox{eV}^{2}$
which cannot be described in the frameworks of a standard three-generation
scheme. On the other hand, the addition of the fourth neutrino cannot
help to explain the difference between the neutrino and antineutrino
oscillations
\cite{hep-ph/0201134, hep-ph/0207157}.
Alternatively, one can consider (a)~two generations of sterile neutrinos
(see e.g.\
\cite{1007.4171} and references therein); (b)~breaking
\cite{hep-ph/0010178} of the CPT invariance\footnote{
Invariance with respect to simultaneous charge conjugation (C) and
reflection of both space (P) and time (T) coordinates is (see e.g.\
\cite{BSh}) a fundamental symmetry which inevitably exists in any
(3+1)-dimensional Lorentz-invariant local quantum field theory. However,
there exist phenomenologically acceptable models with the CPT violation
(either with higher number of space dimensions, or with the Lorentz
invariance violation, or with a nonlocal interaction). In the context of
the neutrino oscillations, they are discussed e.g.\ in
\cite{hep-ph/0010178, Tsukerman, 1108.1799}.}, or (c)~nonstandard
interaction of neutrino with matter which may distinguish particles and
antiparticles
\cite{1002.4452, 1009.0014}. A critical analysis of some of these
suggestions may be found e.g.\ in
\cite{1007.4171, 0805.2234, 1012.3478}.
These scenarios experience considerale difficulties with simultaneous
explanation of the full set of the experimental data, though they cannot
be totally excluded; it might happen that a certain combination of these
possibilities is realized in Nature.

A confirmation of the result about superluminal neutrino motion would
require a serious reconsideration of basic ideas of particle physics. A
successful theory which explains quantitatively the OPERA result should
also agree with very restrictive bounds on the Lorentz-invariance
violation in the sector of charged particles, with the absence of
dispersion of the neutrino signal from the supernova 1987A and with
absence of intense neutrino decays which are characteristic for many
models with deviations from the relativistic invariance.

\subsection{The neutrino mass.}
\label{sec:neutrino:mass}
Conversions of neutrino of one type to another are experimentally proven
and the set of numerous independent and very different experiments are in
a good agreement with the oscillation picture. The oscillatory behaviour
of the neutrino conversions is proven by a comparison of the results
obtained for different energies (cf.\ the argument of the sine squared in
Eq.~(\ref{Eq:PneutrinoOsc})). The last step is to measure the neutrino
flux at different distances along a single path (the distance dependence
in Eq.~(\ref{Eq:PneutrinoOsc})) which is planned for the nearest future. Up
to this last detail, the neutrino oscillations are experimentally
confirmed. Since the oscillations are possible only for different masses
of neutrinos of different types, they prove also that (at least some of)
the neutrino masses are nonzero. At the same time, direct experimental
searches for neutrino masses have not been succesful yet; the most
restrictive bounds, put by the Troitsk-$\nu$mass (INR RAS) and Mainz
experiments, where the tritium beta decay was studied, are
$m_{\nu_{e}}\lesssim 2$~eV \cite{TroitskNuMass, MainzNuMass}. For other
neutrino types, the experimental bounds on the mass are much weaker. An
indirect bound on the sum of the neutrino masses may be obtained from the
studies of anisotropies of the cosmic microwave background and of
hierarchy of structures in the Universe \cite{NuMassCMB}; it reads
$\sum_{i} m_{\nu_{i}}\lesssim 0.35$~eV.

At the same time, in SM the lepton numbers are conserved separately for
each generations, that is changes of the neutrino flavour are forbidden.
By making use of the SM fields, it is impossible to construct a gauge
invariant renormalizable interaction resulting in the neutrino mass, even
after the electroweak symmetry breaking. Therefore, neutrino oscillations
represent an experimental proof of the incompleteness of SM.

How can one modify SM to have massive neutrinos? First note that at
energies below the electroweak breaking scale, the neutrino field is gauge
invariant, it is uncharged and colorless. For such fermion fields one
may write two kinds of mass terms, namely the Dirac term
$m_{\rm D}\bar\nu_{\rm R} \nu_{\rm L} $  (all charged SM fermions have
similar masses) and the Majorana one,
$m_{\rm M}
\nu_{\rm L} C \nu_{\rm L}$, where $C$
is the charge conjugation matrix and
$\nu_{\rm L}$, $\nu_{\rm R}$ denote the left-handed and right-handed
neutrino spinors, respectively.

In SM, only left-handed neutrinos are present, therefore to have Dirac
masses, one must introduce new fields
$\nu_{{\rm R},i}$.
At first sight, the Majorana mass does not require new fields; however,
like the Dirac one, it cannot be obtained from a renormalizable
interaction. Going beyond the renormalizability means that SM is a
low-energy limit of a more complete theory (like the non-renormalizable
Fermi theory is a low-energy limit of SM), so it is again inevitable to
introduce new fields. In any case, neutrinos are several orders of
magnitude lighter than the charged fermions and a succesful theory of
neutrino masses should be able to explain this fact (see also
Sec.~\ref{sec:mass-hierarchy}).

\section{Astrophysical and cosmological indications in favour of new
physics.}
\label{sec:astro}
While laboratory experiments in particle physics give only limited
indications to the incompleteness of SM (neutrino oscillations being the
main one), most scientists are confident that a more complete theory should
be invented. The main reason for this confidence comes from astrophysics
and cosmology. In recent decades, intense development of the observational
astronomy in various energy bands has forced cosmology (that is the branch
of science studying the Universe as a whole) to become an accurate
quantitative discipline based on precise observational data (see e.g.\
textbooks \cite{RuGorby1, RuGorby2}). Today, cosmology has its own
``standard model'' which is in a good agreement with most observational
data. The basis of the model is a concept of the expanding Universe which,
long ago, was very dense and so hot that the energy of thermal motion of
elementary particles did not allow them to compose bound states. As a
result, it were the particle interactions which determined all processes
and, in the end, influenced the Universe development and the state of the
world as we observe it today. The expanding Universe cooled down and
particles were unified into bound states -- first atomic nuclei from
nucleons, then atoms from nuclei and electrons. Unstable particles decayed
and the Universe arrived to its present appearence. As we will see below,
presently the Universe expands with acceleration and is comprised mainly
from unknown particles.

Even a dedicated book would be insufficient to describe all aspects of
interrelations between cosmology and particle physics (the readers of
\textit{Physics Uspekhi} might be interested in reviews
\cite{Ru-UFN1, Ru-UFN2}).
Here, we will briefly consider three principal observational indications
in favour of physics beyond SM, namely, the baryon asymmetry of the
Universe, the dark matter and the accelerated expansion of the Universe
(both the related notion of the dark energy and physical reasons for
inflation).

\subsection{Baryon asymmetry.}
\label{sec:baryon-asymmetry}
Quark-antiquark pairs had to be created intensively in the hot early
Universe. The Universe then expanded and cooled down, quarks and
antiquarks annihilated and the survived ones composed baryons (protons and
neutrons). Notably, there are very few antibaryons in the present
Universe, which means that at the early stages, there were more quarks than
antiquarks. One can determine, by which amount: the number of
quark-antiquark pairs was of the same order as the number of photons while
the baryon-photon ratio may be determined from the analysis of the cosmic
microwave background anisotropy and from studies of primordial
nucleosynthesis. The ratio of the excess of quark number
$n_{q}$
over the antiquark number, $n_{\bar q}$,
is of order
\[
\frac{n_{q}-n_{\bar q}}{n_{q}+n_{\bar q}} \sim 10^{-10},
\]
that is a single ``unpaired'' quark was present for each ten billion
quark-antiquark pairs. It is hard to imagine that this tiny excess of
matter over antimatter was present in the Universe from the very
beginning; moreover, a number of quantitative cosmological models predict
exact baryon symmetry of the very early Universe. Looks like the asymmetry
appeared in course of the evolution of the Universe. For this to happen,
the following Sakharov conditions
\cite{Sakharov} should be fulfilled:
\begin{enumerate}
 \item
baryon number nonconservation;
 \item
$CP$ violation;
 \item
breaking of thermodynamical equilibrium.
\end{enumerate}
Though the classical SM lagrangian conserves the baryon number,
nonperturbative quantum corrections may break it, that is the condition
1 may be fulfilled in SM. The source of $CP$ violation (condition 2) is
also present in SM, it is the phase in the quark mixing matrix. Finally,
in the course of the evolution of the Universe, the state with the zero
vacuum expectation value of the Higgs field (high temperature) has been
replaced by the present state. It can be shown (see e.g.\ \cite{RuShaUFN}
and references therein), that the thermodynamic equilibrium was strongly
broken at that moment, if it were a first-order phase transition.
Therefore, in principle, all three conditions might be met in SM. However,
it has been shown that the first-order electroweak phase transition in SM
is possible only for the Higgs boson mass $M_{H}\lesssim 50$~GeV which was
excluded from direct searches long ago. Also, the amount of $CP$ violation
in the CKM matrix is insufficient. We conclude that the observed baryon
asymmetry of the Universe is an indication to the incompleteness of SM. A
particular mechanism of generation of the baryon asymmetry is yet unknown
(it should also explain the smallness of the asymmetry amount, $\sim
10^{-10}$).

\subsection{Dark matter.}
\label{sec:dark-matter}
Study of dynamics of astrophysical objects (galaxies, galaxy clusters) and
of the Universe as a whole allows one to determine the distribution of mass
which may be subsequently compared to the distribution of the visible
matter. Various independent observational data point to the estimate that
the contribution of the visible matter (mostly baryons) to the energy
density of the Universe is five times smaller than the contribution of
invisible matter. We will first briefly discuss modern observational
evidence for existence of the dark matter and then proceed to the
discussion of implications of these observations to the particle physics.

\begin{figure}
\centering
\includegraphics[width=0.9\columnwidth]{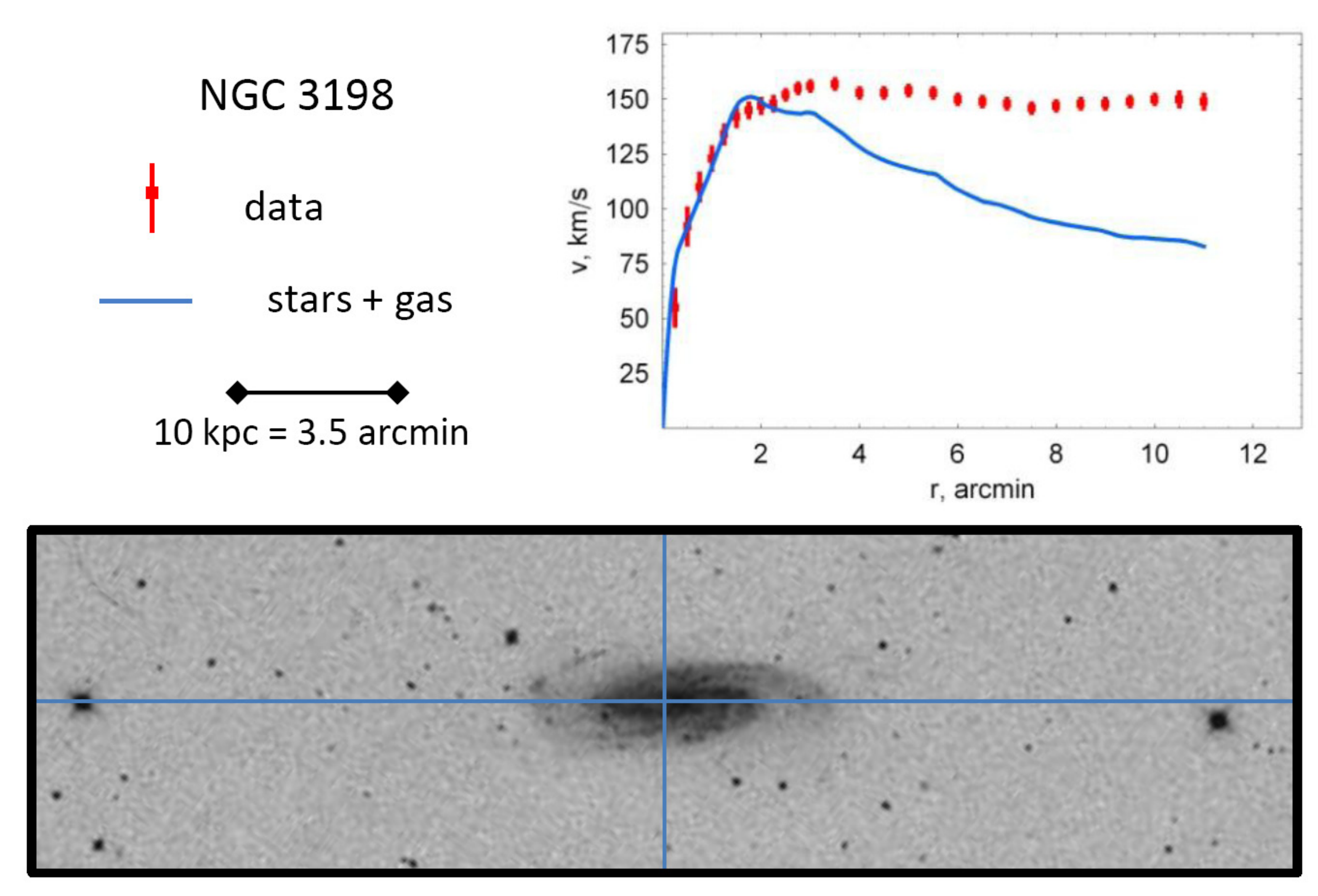}
\caption{
\label{fig:rotation-curve}
Some of the first indications to the existence of dark matter have been
obtained from the analysis of the rotation curves of galaxies.
Observational data on the rotation velocity as a function of the distance
to the axis, given here for the galaxy NGC~3198 (dots), are not described
by the curve which represents the expected velocity calculated from the
distribution of luminous matter (lower line; data and calculation from
\cite{Begeman}). At distances  $\gtrsim 10$~kpc, the luminous matter is
practically absent (as one may see at the lower photograph taken from the
digitalized Palomar sky atlas \cite{DSS}), but the rotation velocities of
gas clouds seen in the radio band are almost constant. This indicates that
at periphery of the galaxy, there is a significant concentration of mass
(the so-called halo).}
\end{figure}

\textit{1. Rotation curves of galaxies.}
Serious attention has been attracted to the question of the invisible
matter after the analysis of rotation curves of galaxies (see
e.g.\ \cite{Rubin}) (Fig.~\ref{fig:rotation-curve}).
For nearby galaxies it is possible to measure, by making use of the
Doppler effect, the velocities of stars and gas clouds at different
distances from the galaxy center, that is from the rotation axis. The
Newton law of gravitation allows to estimate the distribution of mass as a
function of the distance from the center;  it was found that in the outer
parts of galaxies, where luminous matter is practically absent, there is a
significant mass density, so that the visible part of a galaxy is embedded
into a much larger invisible massive halo. These measurements have been
performed for many galaxies, in particular for our Milky Way.

\textit{2. Dynamics of galaxy clusters.}
In a similar way (though based on completely different observations), it is
possible  to determine the mass distribution in galaxy clusters. This
provided for the historically first argument in favour of dark matter
\cite{Zwicky}. Modern observations have demonstrated that the main part of
the baryonic matter sits not in the star systems, galaxies, but in hot gas
clouds in the intergalactic space. This gas emits X rays, so the
observations allow to reconstruct the distribution of the electron density
and temperature. From the latter, by making use of the conditions of
hydrostatic equilibrium, the mass distribution may be determined.
Comparison with the distribution of the luminous matter (that is, mostly
of the gas) points again to the existence of some hidden mass. A similar,
though less precise, conclusion may be obtained from the analysis of
velocities of galaxies inside a cluster.

\begin{figure}
\centering
\includegraphics[width=0.55\columnwidth]{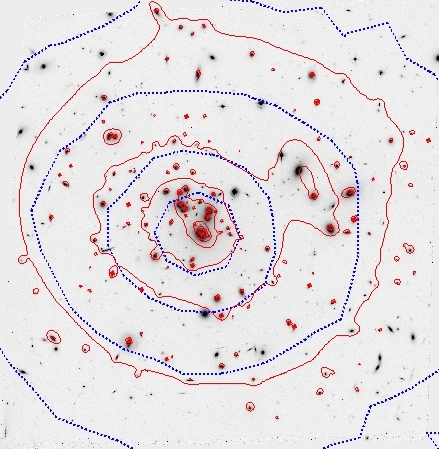}
\caption{
\label{fig:lensing}
The galaxy cluster Abell~1689.
The background image of the cluster in the optical band was obtained by
the Hubble Space Telescope (image from the archive
\cite{Hubble-archive}). The contours describe the model of mass
distribution (full curves, Ref.~\cite{Lensing}) based on the gravitational
lensing and the distribution of luminous gas observed in X rays (dotted
curves based on data from the Chandra X-ray telescope archive,
Ref.~\cite{Chandra-archive}). Currently this mass model is one of the most
precise ones.}
\end{figure}

\textit{3. Gravitational lensing.}
It may happen that a massive object (e.g.\ a galaxy cluster) is located
between a distant source (e.g.\ a galaxy) and the observer. According to
the general relativity, the light from the source is deflected by the
massive object, so the latter may serve as a gravitational lens, which
produces several distorted images of the source. A joint analysis of images
of several sources allows one to model the mass distribution in the lens
and to compare it with the distribution of visible matter (see e.g.\
\cite{Lensing}). The baryon distribution is reconstructed from X-ray
observations of the luminous gas which contains about 90\% of the cluster
mass (Fig.~\ref{fig:lensing}).
The full mass of the cluster calculated in this way far exceeds the mass
of the baryons obtained from observations.

\begin{figure}
\centering
\includegraphics[width=0.55\columnwidth]{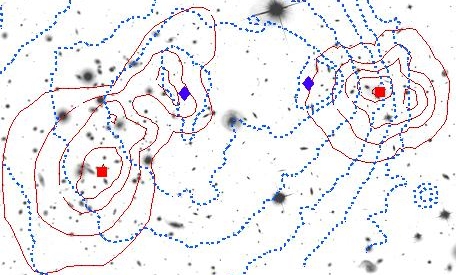}
\caption{
\label{fig:bullet-cluster}
As in Fig.~\ref{fig:lensing} but for colliding clusters
1E~0657--558
(the mass distribution model from \cite{Bullett1}, optical and X-ray
images from archives
\cite{Hubble-archive, Chandra-archive} correspondingly).
Squares denote the positions of maxima of the mass distributions; diamonds
denote the positions of maxima of the gas emission.}
\end{figure}

\textit{4. Colliding clusters of galaxies.}
One of the most beautiful observational proofs of the existence of dark
matter is
\cite{Bullett}
the observation of colliding clusters of galaxies
(Fig.~\ref{fig:bullet-cluster}).
Contrary to the case of a usual cluster, Fig.~\ref{fig:lensing},
one does not need to calculate the mass in this case: comparison of the
mass distribution and the gas distribution demonstrates that the main part
of the mass of the clusters and that of luminous matter are located in
different places. The reason for this dramatic difference, not seen in
normal, noninteracting clusters, is related to the fact that the dark
matter, constituting the dominant part of the mass, behave as a nearly
collisionless gas. During the collision of clusters, the dark matter of
one cluster, together with rare -- and therefore also collisionless --
galaxies kept by its gravitational potential, went through another one,
while the gas clouds collided, stopped and were left behind.

These results, both by themselves and in combination with other results of
quantitative cosmology (first of all those obtained from the analysis of
the cosmic microwave background and the large-scale structure of the
Universe, see e.g.\ \cite{RuGorby1}), point unequivocally  to the
existence of nonluminous matter. One should point out that the terms
``dark'', or ``nonluminous'', mean that the matter does not interact with
the electromagnetic radiation and not only happens to be in a non-emitting
state. Indeed, it should not also absorb electromagnetic waves, since
otherwise the induced radiation would inevitably appear. The usual matter,
that is baryons and electrons, may be put in this state only if packed in
compact, very dense objects (neutron stars, brown dwarfs etc.) which
should be located in the halo of our Galaxy, as well as in other galaxies
and in the intergalactic space within clusters. One may estimate the amount
of these objects which is required to explain the observational results
concerning the nonluminous matter. This amount appears to be so large that
these compact objects should often pass between the observer and some
distant sources, which should result in temporal distortion of the source
image because of the gravitational lensing (the so-called microlensing
effect). These events have been indeed observed, but at a very low rate
which allows for a firm exclusion of this explanation for dark matter
\cite{MACHOlensing}.

We are forced to say that, probably, the dark matter consists of new, yet
unknown, particles, so that its explanation requires to extend SM. The
dark-matter particles should be (almost) stable in order not to decay
during the lifetime of the Universe ($\sim$14 billion years). These
particles should also interact with the ordinary matter only very weakly
to avoid direct experimental detection (direct searches for the dark
matter, which should exist everywhere, in particular in laboratories, last
already for decades). A number of theoretical models of the dark-matter
origin predict the mass of the new particle between
$\sim 1$~GeV and $\sim 1$~TeV and the cross section of interaction with
ordinary particles of order of a typical weak-interaction cross section.
Particles with these properties are called WIMPs
(weakly interacting
massive particles);
they are absent in SM but exist in some of its extensions. One of the most
popular candidates for the WIMP is the lightest superpartner (LSP) in
supersymmetric extensions of SM with conserved $R$ parity (see
Sec.~\ref{sec:gauge-hierarchy} below). The LSP cannot decay because the
conservation of $R$ parity requires that among decay products, at least
one supersymmetric particle should be present, while all other
supersymmetric particles are heavier by definition (in the same way, the
electric-charge conservation provides for the electron stability and the
baryon-number conservation provides for the stability of the proton). In a
wide class of models the LSP is an electrically neutral particle
(neutralino)  which is considered  a good candidate for a dark-matter
particle. Note that there is a plethora of other scenarios in which the
dark-matter particles have very different masses, from
$10^{-5}$~eV (axion) to $10^{22}$~eV (superheavy dark matter). Also, in
 principle, the dark matter may consist of large composite particles
(solitons).

\subsection{Accelerated expansion of the Universe}
\label{sec:dark_energy}
In this section, we briefly discuss several technically interrelated
problems which concern one of the least understandable, from the
particle-physics point of view, part of the modern cosmology. They include:
\begin{enumerate}
 \item
observation of the accelerated expansion of the Universe (``dark energy'');
\item
weakness of the effect of the accelerated expansion as compared to typical
scales of particle physics (the cosmological-constant problem);
\item
indications to intense accelerated expansion of the Universe at one of the
early stages of its evolution (inflation).
\end{enumerate}

Let us start with the observational evidence in favour of (recent and
present) accelerated expansion of the Universe.

\textit{1. The Hubble diagram.}
The first practical instrument of quantitative cosmology, the Hubble
diagram plots distances to remote objects as a function of the
cosmological redshift of their spectral lines. It was the way to discover
the expansion of the Universe and to measure its rate, the Hubble
constant. When methods to measure distances to the objects located really
far away became available for astronomers, they found (see e.g.\
\cite{SN1, SN2})  deviations from a simple Hubble law which indicate
that the expansion rate of the Universe changes with time, namely the
expansion accelerates. The method of distance determination we are
speaking about\footnote{The Nobel prise, 2011.} is based on the study of
type Ia supernovae and deserves a brief discussion (see also
\cite{PerlmPhysToday}).

\begin{figure}
\centering
\includegraphics[width=0.55\columnwidth]{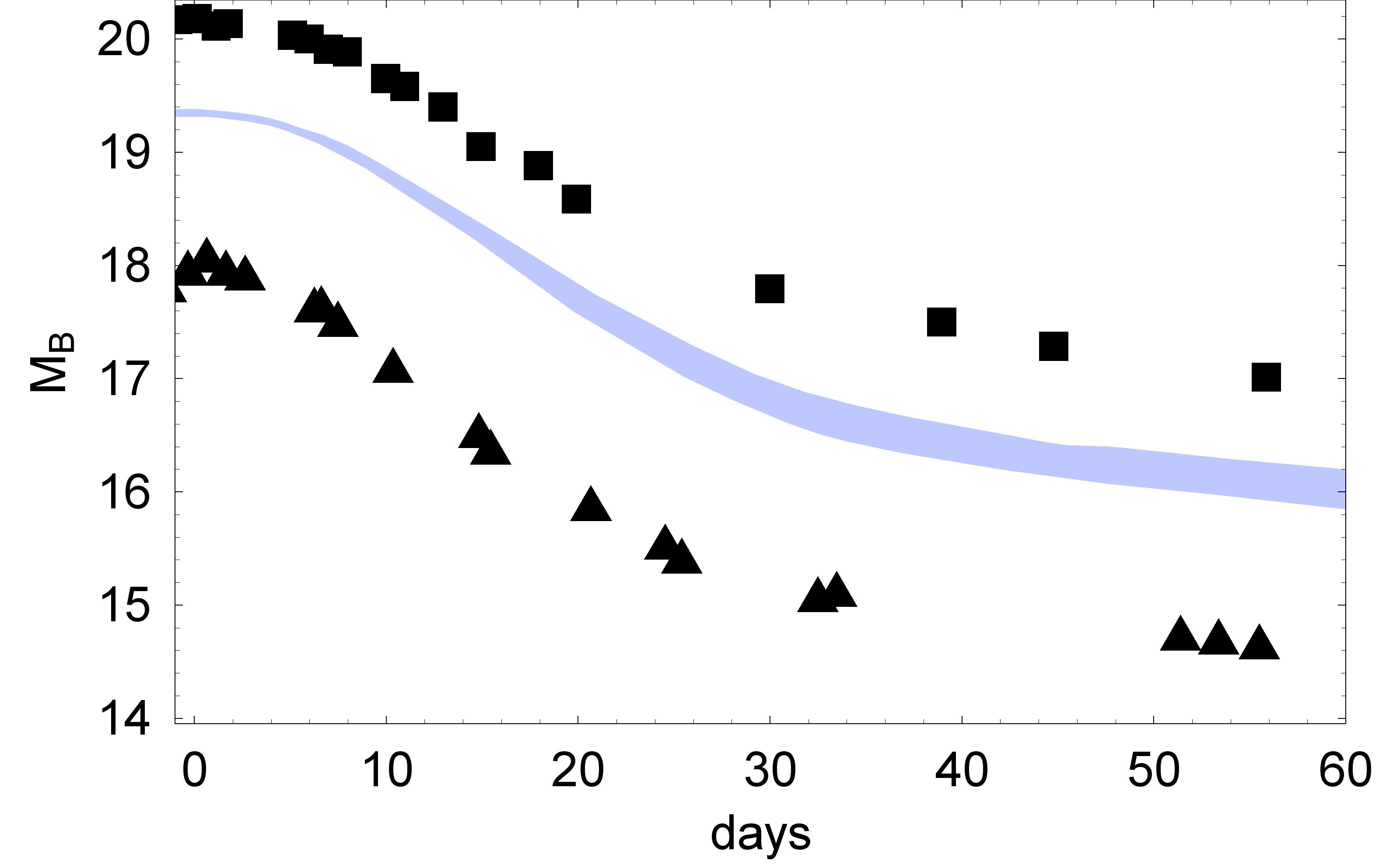}\\
\includegraphics[width=0.55\columnwidth]{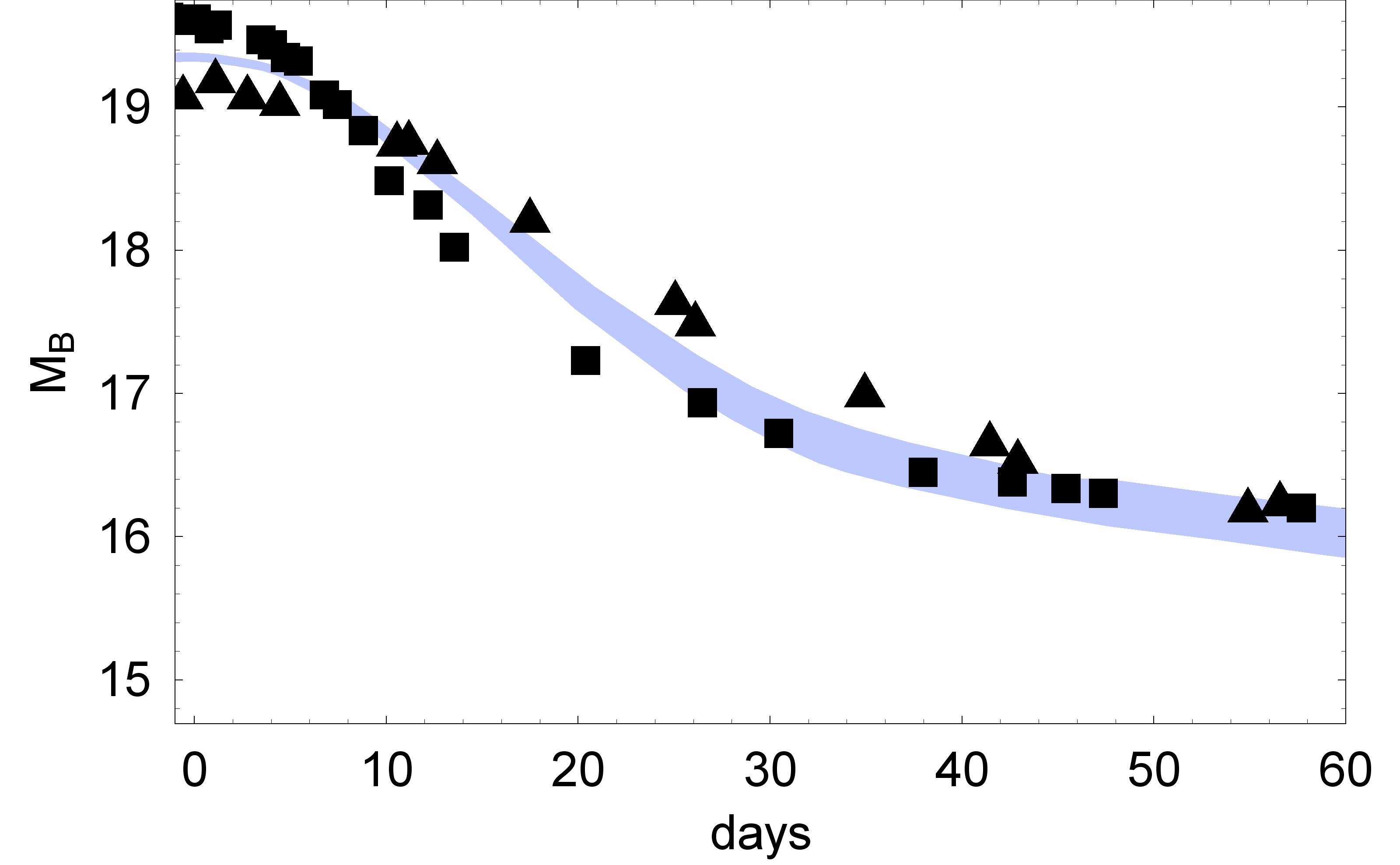}\\
\caption{
\label{fig:Perlmutter}
Temporal dependence of the absolute magnitude of type Ia supernovae.
\textit{Above:}
lightcurves of 68\%
of supernovae are contained within the shaded band, however there are very
rare outsiders (for example, lightcurves of an unusually bright supernova
SN1991T (squares) and an unusually weak supernova SN1986G (triangles) are
presented); light curves and the band are taken from
\cite{astro-ph/9604143}. \textit{Below:}
the same curves but scaled simultaneously in the horizonthal (time) and
vertical (luminosity) axes according to the rules described in
\cite{astro-ph/9608192}.
Introduction of this correction shifts all ``exclusive'' curves to the
band. Therefore, to know the absolute value of the luminosity, it is
sufficient to measure the shape of the light curve.}
\end{figure}

\begin{figure}
\centering
\includegraphics[width=0.55\columnwidth]{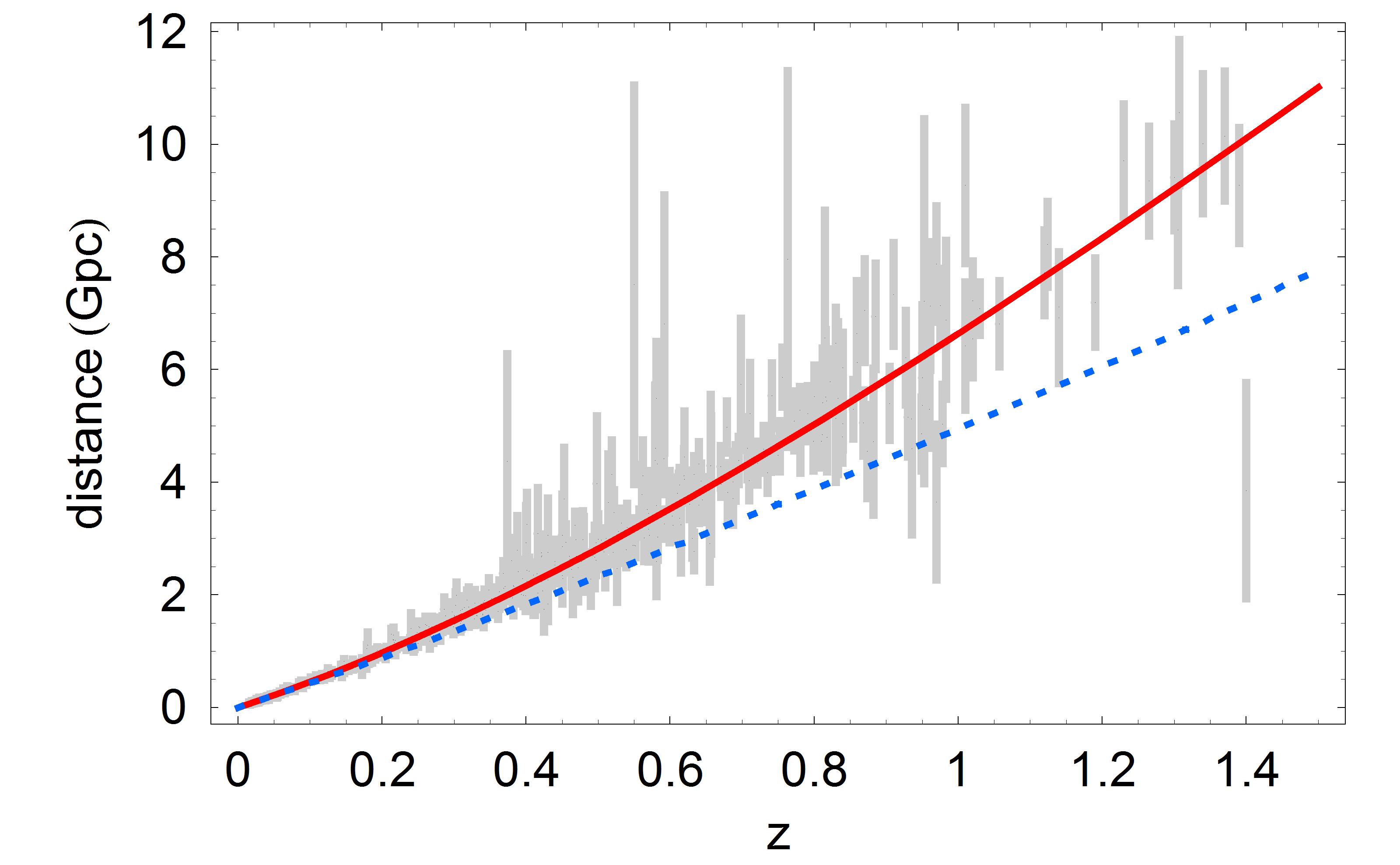}
\caption{
\label{fig:SN}
The Hubble diagram, presenting the dependence of the distance from
the redshift $z$ of spectral lines of distant galaxies, obtained from
observations of type Ia supernovae. Gray lines correspond to data on
individual supernovae with experimental error bars \cite{UNION2}. The
uniform expansion of the Universe corresponds to the lower (dotted) line,
the accelerated expansion -- to the upper (full) line.}
\end{figure}

A probable mechanism of the type Ia supernova explosion is the following. A
white dwarf  (a star at the latest stage of its evolution in which nuclear
reactions have stopped) is rotating in a dense double system with a normal
star. The matter from the normal star flows to the white dwarf and
increases its mass. When the mass exceeds the so-called Chandrasekhar limit
(the limit of stability of a white dwarf whose value depends, in practice,
on the chemical composition of the star only), intense thermonuclear
reactions start and the white dwarf explodes. It is interesting and useful
to note that, therefore, in all cases the exploding stars have roughly the
same mass and constitution (up to details of the chemical composition). As
a consequence, all type Ia supernova explosions resemble each other not
only qualitatively, but quantitatively as well: the energy release is
roughly the same; the time dependence of the luminosity is similar. Even
more amusing is the fact that even for rare outsiders (which differ from
the most of supernovae either by the chemical composition or by some random
circumstances), all curves representing the luminosity as a function of
time are homothetic (Fig.~\ref{fig:Perlmutter}),
that is map one to another at a simultaneous scaling of both time and
luminosity. It means that, upon measurement of a lightcurve of any type Ia
supernova, one may determine  its absolute
luminosity with a good precision. Then, comparison with the observed
magnitude allows to determine the distance to the object. In this way it
is possible to construct the Hubble diagram (Fig.~\ref{fig:SN})
which demonstrates statistically significant deviations from the law which
corresponds to the uniform (or decelerated) expansion of the Universe.

\textit{2. Gravitational lensing.}
The method of the gravitational lensing, discussed above, allows not only
for reconstruction of the mass distribution in the lensing cluster of
galaxies, but also for determination of geometrical distances between
sources, the lens and the observer. If redshifts of the source and the
lens are known, one may compare them with the derived distances and find
\cite{lensingNew} deviations from the Hubble law with a high precision.

\textit{3. Flatness of the Universe and the energy balance.}
A number of measurements point to the spatial flatness of the Universe,
that is to the fact that its three-dimensional curvature is zero. The main
argument here is based on the analysis of the cosmic microwave background
anisotropy
\cite{CMBflatness}.
In the past, the Universe was denser and hotter than now. Various
particles (photons in particular) were in thermodynamical equilibrium, so
that the distribution of photons over energies was Planckian,
corresponding to the temperature of the surrounding plasma. The Universe
cooled down while expanding and at some moment, electrons and protons
started to join into hydrogen atoms. Compared to plasma, the gas of
neutral atoms is practically transparent for radiation; since then,
photons born in the primordial plasma propagate almost freely. We see them
as the cosmic microwave background (CMB) now. At the moment when the
Universe became transparent, the size of the causally connected region
(that is the region which a light signal had time to cross since the Big
Bang), called a horizon, was only $\sim 300$~kpc. This quantity may be
related to a typical scale of the CMB angular anisotropy; the present
Universe is much older and we see at the same moment many regions which
had not been causally connected in the early Universe.  This angular scale
has been directly measured from the CMB anisotropy. The theoretical
relation between this scale and the size of the horizon at the moment when
the Universe became transparent is very sensitive to the value of the
spatial curvature; the analysis of the data from WMAP satellite points to
a flat Universe with a very high accuracy.

Other methods exist to test the flatness of the Universe. One of the most
beautiful among them is the geometric Alcock-Paczinski criterion. If it is
known that an object has a purely spherical shape, one may try to measure
its dimensions along the line of sight and in the transverse direction.
Taking into account distortions related to the expansion of the Universe,
one may compare the two sizes and constrain the cosmological parameters,
first of all, deviations from flatness.  Clearly, it is not an easy task
to find an object whose longuitudinal and transverse dimensions are
certainly equal; however, one may measure characteristic dimensions of
some astrophysical structures which, averaged over large samples, should
be isotropical. The most precise measurement of this kind
\cite{DoubleGal}
uses double galaxies whose orbits are randomly oriented in space while the
orbital motion is described by the Newton dynamics.

From the general-relativity point of view, the flat Universe represents a
rather specific solution which is characterized by a particular total
energy density (the so-called critical density,
$\rho_{c}\sim 5\times 10^{-6}~\frac{\mbox{GeV}}{\mbox{cm}^{3}}$).
At the same time, the estimates of the energy density related to matter
contribute
$\sim 0.25 \rho_{c}$,
that is the remaining three fourths of the energy density of the Universe
are due to something else. This contribution, whose primary difference
form the matter contribution is in the absence of clustering (that is
of concentration in stars, galaxies, clusters etc.),
carries a not fully successful name of ``dark energy''.

The question about the nature of the dark energy is presently open. The
technically most simple explanation is that the accelerated expansion of
the Universe results from a nonzero vacuum energy (in general relativity,
the reference point on the energy axis is relevant!), that is the
so-called cosmological constant. From the particle-physics point of view,
the dark-energy problem is, in this case, twofold. In the absence of
special cancellations, the vacuum energy density should be of order of the
characteristic scale of relevant interactions $\Lambda$, that is
\[
\rho\sim \frac{\Lambda^{4}}{c^{3} \hbar^{3}}.
\]
The observed value of $\rho $ corresponds to $\Lambda\sim 10^{-3}$~eV,
while characteristic scales of the strong
($\Lambda_{\rm QCD}\sim 10^{8}$~eV) and electroweak ($v\sim 10^{11}$~eV)
interactions are many orders of magnitude higher. One side of the problem
(known for a long time as ``the cosmological-constant problem'') is to
explain how the contributions of all these interactions to the vacuum
energy cancel. In principle, some symmetry may be responsible for this
cancellation: for instance, the energy of a supersymmetric vacuum in field
theory is always zero. Unfortunately, the supersymmetry, even if it has
some relation to the real world, should (as discussed in
Sec.~\ref{sec:hierarchies}), be broken at the scale not smaller than $\sim
v$, and the contributions to the vacuum energy should have then the same
order.  On the other hand, the observed accelerated expansion of the
Universe tells us that the cancellation is not complete and hence there is
a new energetic scale in the Universe $\sim 10^{-3}$~eV. The explanation
of this scale is a task which cannot be completed within the frameworks of
SM, where all parameters of the dimension of energy are orders of
magnitude higher. If this scale is given by a mass of some particle, the
properties of the latter should be very exotic in order both to solve the
problem of the accelerated expansion of the Universe and not to be found
experimentally. For instance, one of suggested explanations
\cite{chameleon} introduces a scalar particle whose effective mass depends
on the density of medium (this particle is called ``chameleon''). By
itself, the dependence of the effective mass on the properties of medium
is well-known (for instance, the dispersion relation of photon in plasma
is modified in such a way that it gets a nonzero effective mass). In our
case, due to interaction with the external gravitational field, the
chameleon has a short-distance potential in relatively dense medium (e.g.\
at the Earth), which prohibits its laboratory detection, but at large
scales of the (almost empty) Universe the effect of this particle becomes
important. One should also note that a solution to the problem of the
accelerated expansion of the Universe might have nothing to do with
particle physics at all and  be entirely based on peculiar properties of
the gravitational interaction (for instance, on deviations from the general
relativity at large distances).

However, the problem of the accelerated expansion of the Universe is not
exhausted by the analysis of the modern state. There are serious
indications that, at an early stage of its evolution, the Universe
experienced a period of intense exponential expansion, called inflation
(see e.g.\
\cite{RuGorby2, inflation}.
Though the inflation theory is currently not a part of the standard
cosmological model (it awaits  more precise experimental tests), it solves
a number of problems of the standard cosmology and, presently, does not
have an elaborated alternative. Let us list briefly some problems which
are solved by the inflationary model.
\begin{enumerate}
 \item
As it has been already pointed out, various parts of the presently
observed Universe were causally disconnected from each other in the past,
if one extrapolates the present expansion of the Universe backwards in
time. Information between the regions which are now observed in different
directions could not be transmitted, for instance, at the moment when the
Universe became transparent for CMB. At the same time, the CMB is
isotropic up to a high level of accuracy (relative variations of its
temperature does not exceed
$10^{-4}$), the fact that indicates to the causal connection between all
currently observed regions.
\item
Zero curvature of the Universe, from the theoretical point of view, is not
singled out by any condition: the Universe should be flat from the very
beginning, nobody knows why.
\item
The modern Universe is not precisely homogeneous -- the matter is
distributed inhomogeneously, being concentrated in galaxies, clusters and
superclusters of galaxies; a weak anisotropy is observed also in CMB. Most
probably, these structures were developed from tiny primordial
inhomogeneities, whose existence should be assumed as the initial
condition.
\end{enumerate}
These and some other arguments point to the fact that the initial
conditions for the theory of a hot expanding Universe had to be very
specific. A simultaneous solution to all these problems is provided by the
inflationary model which is based on the assumption about an exponential
expansion of the Universe which happened before the hot stage. From the
theory point of view, this situation is fully analogous to the present
accelerated expansion, but the energy density, which determines the
acceleration rate, was much higher. It may be related to the presence of a
new, absent in SM, scalar field, the inflaton. If it has a relatively flat
(that is, weakly depending from the field value) potential and the value
itself slowly changes with time, then the energy density of the inflaton
provides for the required exponential expansion. For a particle physicist,
at least two questions arise: first, what is the nature of the inflaton,
and second, what was the reason for the inflation to stop and not to
continue until now.

To summarize, we note that a large number of observations related to the
structure and evolution of the Universe cannot be explained if the
particle physics is described by SM only: one needs to introduce new
particles and interactions. Jointly with the observation of neutrino
oscillations, these facts constitute the experimental basis for the
confidence in incompleteness of SM. At the same time, presently none of
these experimental results point to a specific model of new physics, so
one is guided also by purely theoretical arguments when constructing
hypothetical models.

\section{Aesthetic difficulties: the origin of parameters.}
\label{sec:hierarchies}
\subsection{Electroweak interaction and the Higgs boson.}
\label{sec:EW}
Results of high-precision measurements of electroweak-theory parameters,
in particular at the LEP accelerator, confirm the predictions of SM, based
on the Higgs mechanism. At the same time, the only SM particle which has
not been discovered experimentally, is the Higgs boson. Its mass is a free
parameter of the model and is not directly related to any of measurable
parameters, so the lack of signs of the Higgs boson in data may be simply
explained by its mass: the energies and luminosities of available
accelerators might be insufficient to create this particle with a
significant probability.

At the same time, purely theoretical concerns suggest that the Higgs boson
should not be too heavy. It is related to the fact that, without the
account of the Higgs scalar, the scattering amplitudes of massive $W$
bosons grow  as $E^{2}$ with energy $E$. As a result, at energies somewhat
higher than the $W$ mass, the perturbation theory fails and all model
predictions start to depend on unknown higher-order contributions; the
theory finds itself in a strong-coupling regime and loses predictivity.
The contribution from the Higgs boson, however, cancels the part of the
amplitude which grows with energy, so only the constant term remains,
 $\sim g^{2} M_{H}^{2}/(4 M_{W}^{2})$, where $M_{H}$ and $M_{W}$
are masses of the Higgs and $W$ bosons, respectively, and $g$ is the
$SU(2)_{\rm L}$ gauge coupling constant. Therefore, to keep
calculability,
$M_{H}$
should not be too large; a quite reliable limit is
$M_{H}\lesssim 800$~GeV.
Even more restrictive limits come from the radiative corrections to the
potential of the Higgs boson itself. In the leading order of perturbation
theory, the self-interaction constant of the Higgs boson has a pole at the
energy scale
$$
Q\sim v \exp \left[\frac{4\pi^2 v^2}{3M_H^2}   \right].
$$
This means that at energies $\Lambda \le Q$, the contributions of new
particles or interactions should change the coupling behaviour to avoid
divergence. The requirement
$\Lambda \ge 1$~TeV results in the limit
$M_{H}\lesssim 550$~GeV. Note that this means that the SM Higgs boson
should be discovered at LHC.

\begin{figure}
\centering
\includegraphics[width=0.475\columnwidth]{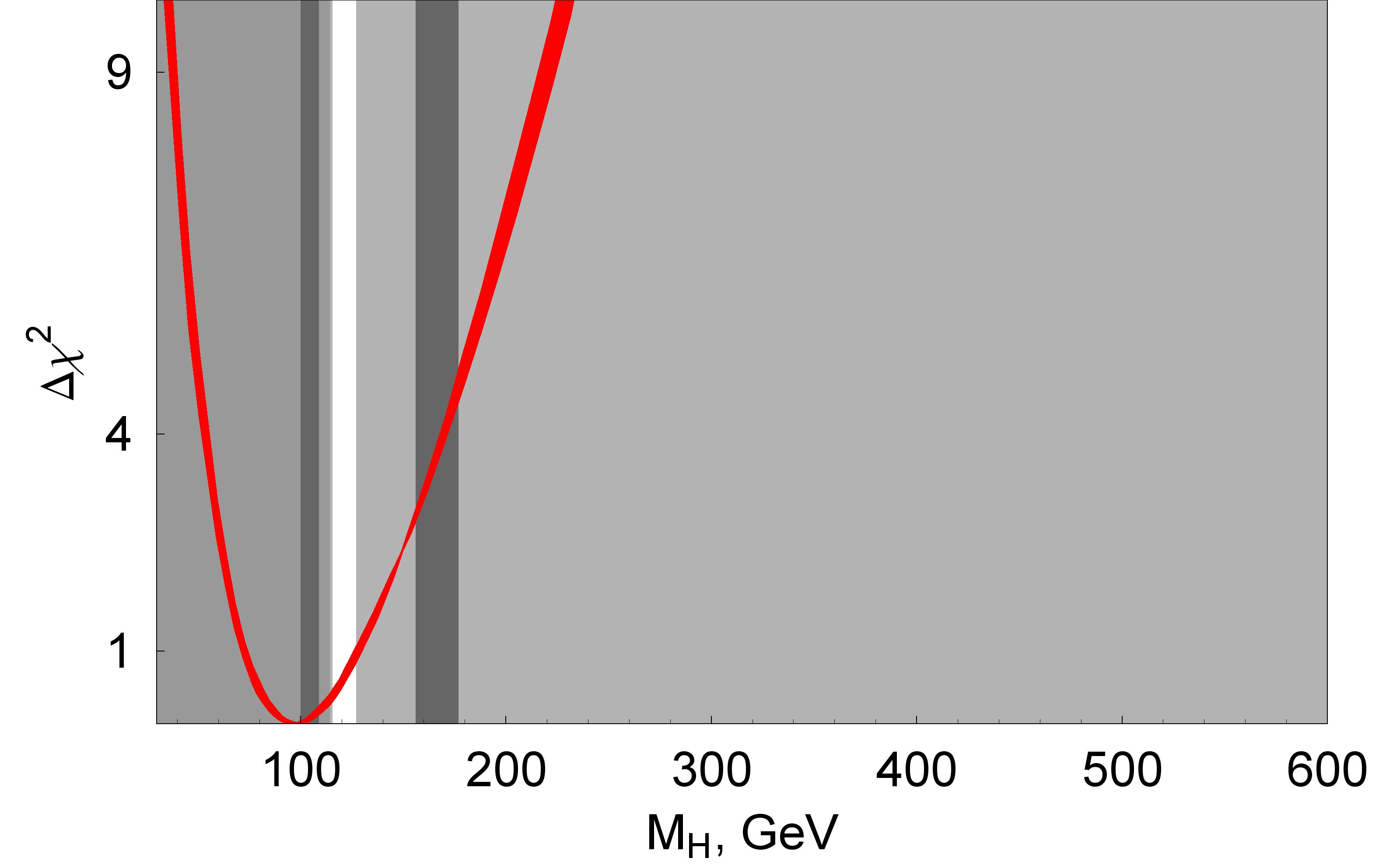}
~~~~
\includegraphics[width=0.475\columnwidth]{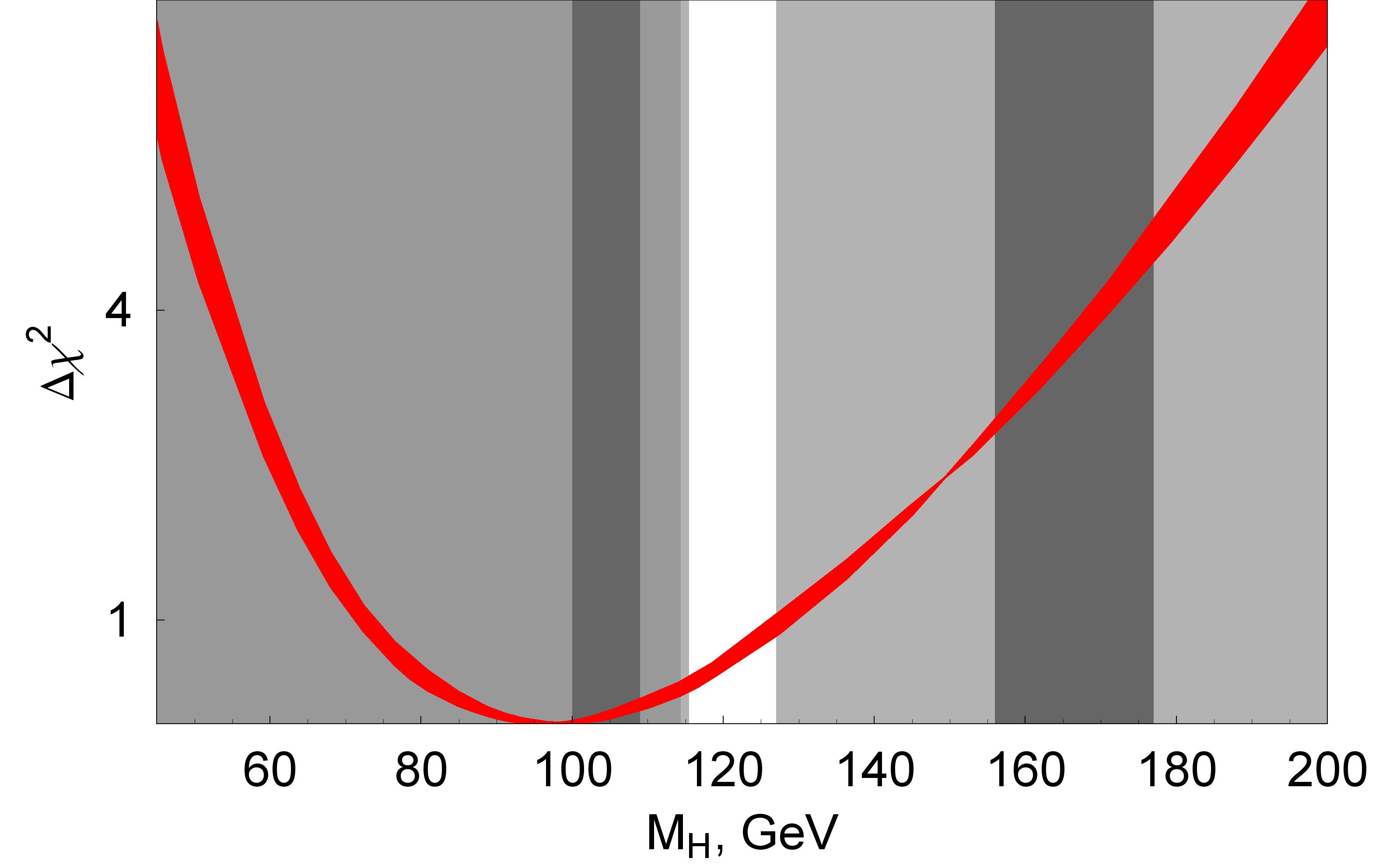}
\caption{
\label{fig:newBlueBand}
The Higgs boson mass expected from indirect data and constrained from
direct searches (see text). The left panel shows all experimental limits;
the right one is a zoom of the most interesting region,
$M_H<200$~GeV.}
\end{figure}

Maybe even more interesting situation is related to the experimental data
on the search of the Higgs particle. The latter may reveal itself not only
directly, being produced at colliders, but also indirectly, through the
influence of the virtual Higgs bosons on numerous observables. Though this
influence is not large, a numebr of electroweak observables are known with
a very high precision, so that their joint analysis may constrain the mass
of yet undiscovered Higgs particle. Let us look at
Fig.~\ref{fig:newBlueBand},
which is based on the analysis of indirect experimental data%
\footnote{See also the regularly updated webpage at
http://gfitter.desy.de/GSM .} as of September 2011. The horizontal axis
gives the possible Higgs-boson mass; the shaded regions of $M_{H}$ are
excluded, as of December 2011, from direct experimental search of the
Higgs boson at colliders at the 95\% confidence level
(the light band $M_H<114$~GeV -- LEP \cite{LEP-Higgs},
the light bands
114~GeV$<M_H<115.5$~GeV and
127~GeV$<M_H<600$~GeV -- LHC
\cite{CMS-Higgs, ATLAS-Higgs}, the dark bands 100~GeV$<M_H<109$~GeV and
156~GeV$<M_H<$177~GeV -- Tevatron \cite{Tevatron-Higgs}).
The curve \cite{Gfitter2011}
demonstrates how well a given value of
$M_{H}$ agrees with a combination of all \textit{other} than the
direct-search experiments
(the lower
$\Delta \chi^{2}$, the better agreement;
the curve width represents the uncertainty in theoretical predictions).
One can see that the most preferable value of $M_{H}$ is already
experimentally excluded! Clearly, this does not mean a catastrophe because
a narrow range of slightly less preferable values are allowed, but it
motivates theoretical physicists to think about possible alternative
explanations of the electroweak symmetry breaking
\cite{Grojean-UFN}.
One should note that it is rather difficult to discover a light,
115~GeV$<M_{H}<127$~GeV, Higgs boson at LHC: unlike for a heavy one,
several years of work might be required.

The lack of the Higgs boson with the expected mass and the prospect
of further restriction of the allowed mass region at LHC are important, but
far not principal arguments in favour of alternative theoretical models of
the electroweak symmetry breaking, whose history goes back for decades.
The point is that the Higgs boson is the only SM scalar particle (all
others are either fermions or vectors). A scalar particle brings to a
theory a number of unfavoured properties some of which we have just
mentioned above, while others will be discussed below.
That is why alternative mechanisms of the
electroweak symmetry breaking use, as a rule, only fermionic and vector
fields.

A class of hypothetical models in which the vacuum expectation value of
the Higgs particle is replaced by the vacuum expectation value of a
two-fermion operator with the same quantum numbers are called technicolor
models (see e.g.\ \cite{technicolor}).  The replacement of the scalar by a
fermion condensate looks quite natural if one recalls that in the
historically first, and surely realized in Nature, example of the Higgs
mechanism, the Ginzburg--Landau superconductor,  the condensate of the
Cooper pairs of electrons plays the role of the Higgs boson.

The base for the construction of technicolor models is provided by the
analogy to QCD. Indeed, an unbroken nonabelian gauge symmetry, similar to
$SU(3)_{\rm C}$, may result in confinement of  fermions and to formation of
bound states (in QCD these are hadrons, bound states of quarks). In fact,
in QCD a nonzero vacuum expectation value of the quark condensate also
appears, but its value, of order $\Lambda_{\rm QCD}\sim$200~MeV, is much
less than the required electroweak symmetry breaking scale ($v\approx
246$~GeV). Therefore one postulates that there exists another gauge
interaction, in a way resembling QCD, but with a characteristic scale of
order $v$. The corresponding gauge group $G_{\rm TC}$ is called a
technicolor group. The bound states, technihadrons, are composed from the
fundamental fermions, techniquarks $T$, which feel this interaction. The
techniquarks carry the same quantum numbers as quarks, except instead of
$SU(3)_{\rm C}$, they transform as a fundamental representation of $G_{\rm
TC}$. Then, the vacuum expectation value $\langle \bar T T \rangle$ breaks
$SU(2)_{\rm L} \times U(1)_{\rm Y} \to U(1)_{\rm EM}$ in such a way that
the correct relation between masses of the $W$ and $Z$ bosons is fulfilled
automatically. A practical implementation of this beautiful idea faces,
however, a number of difficulties which result in a complification of the
model. First, the role of the Higgs boson in SM is not only to break the
electroweak symmetry: its vacuum expectation value also gives masses to
all charged fermions. Attempts to explain the origin of fermion masses in
technicolor models result in significant complication of the model and,
in many cases, in contradiction with experimental constraints on the
flavour-changing processes. Second, many parameters of the electroweak
theory are known with very high precision (and agree with the usual Higgs
breaking), while even a minor deviation from the standard mechanism
destroys this well-tuned picture. To construct an elegant and viable
technicolor model is a task for future which will become relevant if the
Higgs scalar will not be found at LHC.

In another class of models (suggested in
\cite{Higgs5D} and further developed in numerous works which are reviewed,
e.g., in
\cite{Grojean-UFN}),
the Higgs scalar appears as a component of a vector field. Since the
vacuum expectation value of a vector component breaks Lorentz invariance,
this mechanism works exclusively in models with extra space dimensions.
For instance, from the four-dimensional point of view, the fifth component
of a five-dimensional gauge field behaves as a scalar, and giving a vacuum
expectation value to it breaks only five-dimensional Lorentz invariance
while keeping intact the observed four-dimensional one. Symmetries of the
five-dimensional model, projected onto the four-dimensional world, protect
the effective theory from unwanted features related to the existence of a
fundamental scalar particle. These models also have a number of
phenomenological problems which can be solved at a price of significant
complication of a theory.

The so-called higgsless models \cite{Higgsless} (see
also \cite{Grojean-UFN})  are rather close to these multi-dimensional
models, though differ from them in some principal points. The higgsless
models are based on the analogy between the mass and the fifth component
of momentum in extra dimensions: both appear in four-dimensional effective
equations of motion similarly. In the higgsless models, the nonzero
momentum appears due to imposing some particular boundary conditions in a
compact fifth dimension. In the end, these boundary conditions are
responsible for breaking of the electroweak symmetry. Unlike in
five-dimensional models, where the Higgs particle is a component of a
vector field, the physical spectrum of the effective theory in higgsless
models does not contain the corresponding degree of freedom. These models
have some phenomenological difficulties (related e.g.\ to precise
electroweak measurements). Another shortcoming of this class of models is
considerable arbitraryness in the choice of the boundary conditions, which
are not derived from the model but are crucial for the electroweak
breaking.

Finally, we note that a composite Higgs boson may be even more complex
than just a fermion condensate: it may be a bound state which includes
strongly coupled gauge fields. Description of these bound states requires
a quantitative understanding of nonperturbative gauge dynamics. Taking
into account the analogy between strongly coupled four-dimensional
theories and weakly coupled five-dimensional ones (which will be discussed
in Sec.~\ref{sec:QCD:dual}, these models may even happen to be equivalent
to multidimensional models described above.

\subsection{The gauge hierarchy.}
\label{sec:gauge-hierarchy}
Each of the main interactions of particles has its own characteristic
energy scale. For the strong interaction it is
$\Lambda_{\rm QCD}\sim 200$~MeV, the scale at which the QCD running
coupling becomes strong; this scale determines masses of hadrons made of
light quarks. The scale of the electroweak theory is determined by the
vacuum expectation value of the Higgs boson,
$v\approx 246$~GeV, which determines, through the corresponding coupling
constants, the masses of the $W$ and $Z$ bosons and of SM matter fields.
For gravity, the characteristic scale is the Planck scale $M_{\rm Pl}\sim
10^{19}$~GeV, determined by the Newton constant of the classical
gravitational interactions.

These three scales are related to known forces. Extensions of SM give
motivation to some other interactions and, consequently, to other scales.
First of all it is
$M_{\rm GUT} \sim 10^{16}$~GeV, the scale of the suggested Grand
Unification of interactions. In several models explaining neutrino masses
there exists a scale $M_{\nu}$;
sometimes the scale
$M_{\rm PQ}$, related to the $CP$ invariance of the strong interaction, is
also introduced. Values of these two scales are model dependent but roughly
$M_{\rm PQ}\sim M_{\nu }\sim 10^{14}$~GeV.

The gauge hierarchy problem (see also
\cite{SMprimer, Ru-UFN1, Ru-UFN3no-cosm})
consists in the disproportionality of these scales:
$$(\Lambda_{\rm QCD}, v) \ll (M_{\rm Pl},M_{\rm GUT},M_{\rm PQ},M_{\nu })$$
and in a range of related questions which may be divided into three
groups.

\textbf{1. The origin of the hierarchy}:
why the scales of the strong and electroweak interactions are smaller than
others by many orders of magnitude? That is, why, for instance, all SM
particles are practically massless at the gravity scales? It is
possible, in the frameworks of the Grand-Unification hypothesis, to get a
reasonable explanation of the relation
$\Lambda_{\rm QCD} \ll M_{\rm GUT}$. It is based on the logarithmic
renormalization-group dependence of the gauge coupling constant from
energy $E$. In the leading approximation, this dependence for the
strong-interaction coupling
$\alpha _{3}$ reads
\[
\alpha_{3}(E) = \frac{\alpha_{\rm GUT}}{1+\beta_{3}\alpha_{\rm GUT}{\rm
ln}(E/M_{\rm GUT})},
\]
where $\beta_{3}$ is a positive coefficient which depends on the set of
strongly interacting matter fields (in SM,
$\beta_{3}= 11/(12\pi)$), while
$\alpha_{\rm GUT} \sim 1/30$ is the value of the coupling constant of a
unified gauge theory at the energy scale
$\sim M_{\rm GUT}$. The scale $\Lambda_{\rm QCD}$, where $\alpha_{3}$
becomes large, may be determined in this approximation as
\[
\Lambda_{\rm QCD}=M_{\rm GUT} \exp\left(-\frac{1}{\beta_{3} \alpha_{\rm
GUT}} \right)
\]
and the exponent provides for the required hierarchy. However, a similar
analysis is not succesful for the electroweak interaction, whose coupling
constants are small at the scale $v$. The latter is unrelated to any
dynamical scale and is introduced in the theory as a free parameter.

\textbf{2. The stability of the hierarchy.}
In the standard mechanism of the electroweak breaking, the characteristic
scale
$v=M_{H}/\sqrt{2\lambda }$, where $\lambda $
is the self-interaction constant of the Higgs boson. Together with
$M_{H}$, the scale $v$ gets, in SM, quadratically divergent radiative
corrections,
\[
\delta v^{2}\sim \delta M_{H}^{2} = f(g) \Lambda_{\rm UV} ^{2},
\]
where $f(g)$ is a symbolic notation for some known combination of the
coupling constants (in SM, $f(g)\approx 0.1$), and  $\Lambda_{\rm UV}$ is
the ultraviolet cutoff which may be interpreted as an energy scale above
which SM cannot give a good approximation to reality. This scale may be
related to one of the scales
$M_{\rm Pl}, M_{\rm GUT}$ etc.\ discussed above;
in the assumption of the absence of the ``new physics'', that is of
particle interactions  other than those already discovered (SM and
gravity), one should take
$\Lambda_{\rm UV}\sim M_{\rm Pl}$. Therefore, since
$v^{2}=v_{0}^{2} -\delta v^{2}$, where $v_{0}$ is the parameter of the
tree-level lagrangian, the hierarchy $v^{2} \ll M_{\rm Pl}^{2}$
appears as a result of cancellation between two huge contributions,
$v_{0}^{2}$ and $\delta v^{2}$. Each of them is of order
$f(g)M_{\rm Pl}^{2}\sim 10^{33}v^{2}$,
that is the cancellation has to be precise up to
$10^{-33}$ in every order of the perturbation theory. This
\textit{fine tuning} of parameters of the model, though technically
possible, does not look natural. One may revert this logic and say that to
avoid fine tuning in SM one should have
\begin{equation}
f(g) \Lambda_{\rm UV}^{2} \sim v^{2} ~~~ \Rightarrow ~~~ \Lambda_{\rm UV}
\sim \mbox{TeV}.
\label{Eq:LambdaUV=TeV}
\end{equation}
The relation (\ref{Eq:LambdaUV=TeV}) gives a base for the optimism of
researchers who expect the discovery of not only the Higgs
boson but also some new physics beyond SM from the LHC.

\begin{figure}
\centering
\includegraphics[width=0.95\columnwidth]{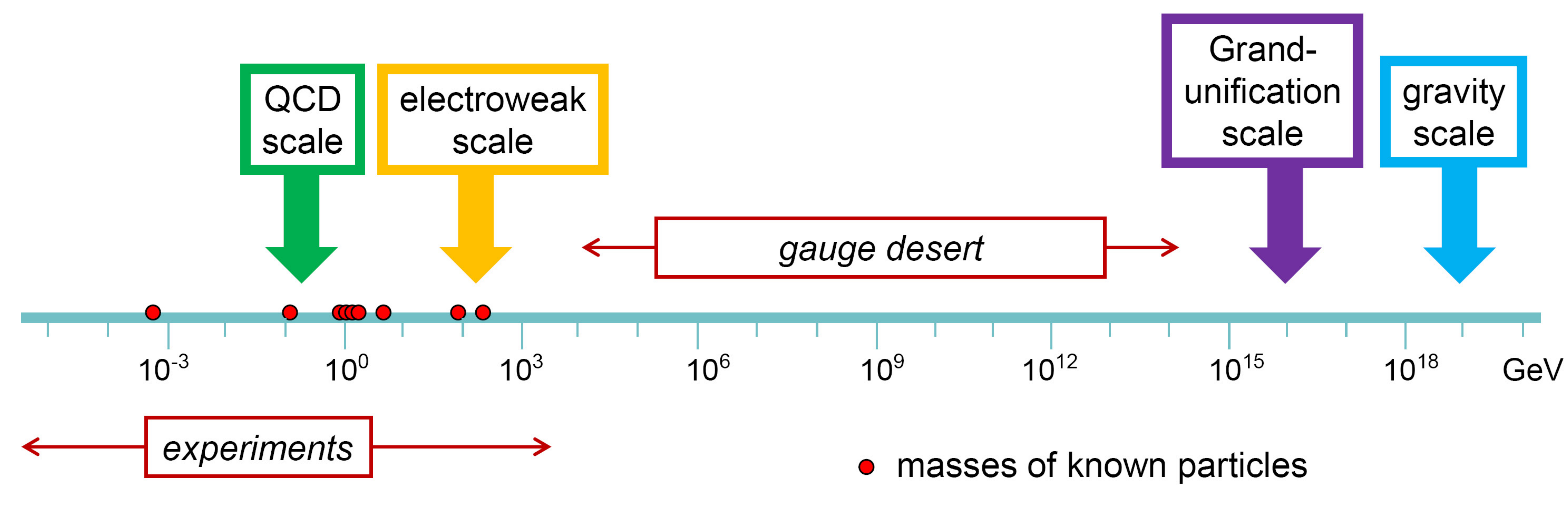}
\caption{
\label{fig:GaugeHierarchy}
Hierarchy of scales of gauge interactions.}
\end{figure}

\textbf{3. The gauge desert.}
The third aspect of the same problem is related to the presumed absence of
particles with masses (and of interactions with scales) between ``small''
($\Lambda_{\rm QCD}, v$) and ``large'' ($M_{\nu }, M_{\rm GUT}, M_{\rm
Pl}$) energetic scales, see Fig.~\ref{fig:GaugeHierarchy}.
All known particles are settled in a relatively narrow region of masses
$\lesssim v$, beyond which, for many orders of magnitude, lays the
so-called \textit{gauge desert}. Clearly, one may suppose that the heavier
particles simply cannot be discoverd due to insufficient energy of
accelerators, but this suggestion is not that easy to accomodate within the
standard approach. Indeed, new relatively light ($\sim v$) particles which
carry the SM quantum numbers are constrained by the electroweak precision
measurements. Also, the latest Tevatron and first LHC results on the
direct search of new quarks strongly constrain the range of their allowed
masses (see
\cite{1104.3874} and references therein). In particular, for the fourth
generation of matter fields similar to the known three, the mass of its
up-type quark should exceed 338~GeV, while that of a down-type quark
should exceed 311~GeV. The mass of the corresponding charged lepton cannot
be lower than 101~GeV
\cite{PDG2010}. The mass of the fourth-generation standard neutrino
should exceed one half of the $Z$-boson mass, as it has been already
discussed above. At the same time, these values of masses of the
fourth-generation charged fermions cannot have the same origin as those
for the first three generation, because to generate masses much larger
than $v$, Yukawa constants much larger than one are required. Since the
methods to calculate nonperturbative corrections to masses are yet
unknown, for this case one cannot be sure that these masses can be
obtained at all in a usual way. Moreover, SM fermion masses exceeding
the electroweak breaking scale are forbidden by the
$SU(2)_{L}\times U(1)_{Y}$ gauge symmetry: a mechanism generating these
masses would also break the electroweak symmetry at a scale $>v$. Addition
of matter fields which do not constitute full generations may be
considered as an essential extension of SM. Finally, addition of new
matter affects the energy dependence of the gauge coupling constants and
spoils their perturbative unification (unless one adds either full
generations or other very special sets of particles of roughly the same
mass which constitute full multiplets of a unified gauge group). We see
that attempts ``to inhabit the gauge desert'' inevitably result in
significant steps beyond SM while the desert itself does not look natural.

Attempts to solve the gauge hierarchy problem may be also divided into
several large groups.

\begin{figure}
\centering
\includegraphics[width=0.75\columnwidth]{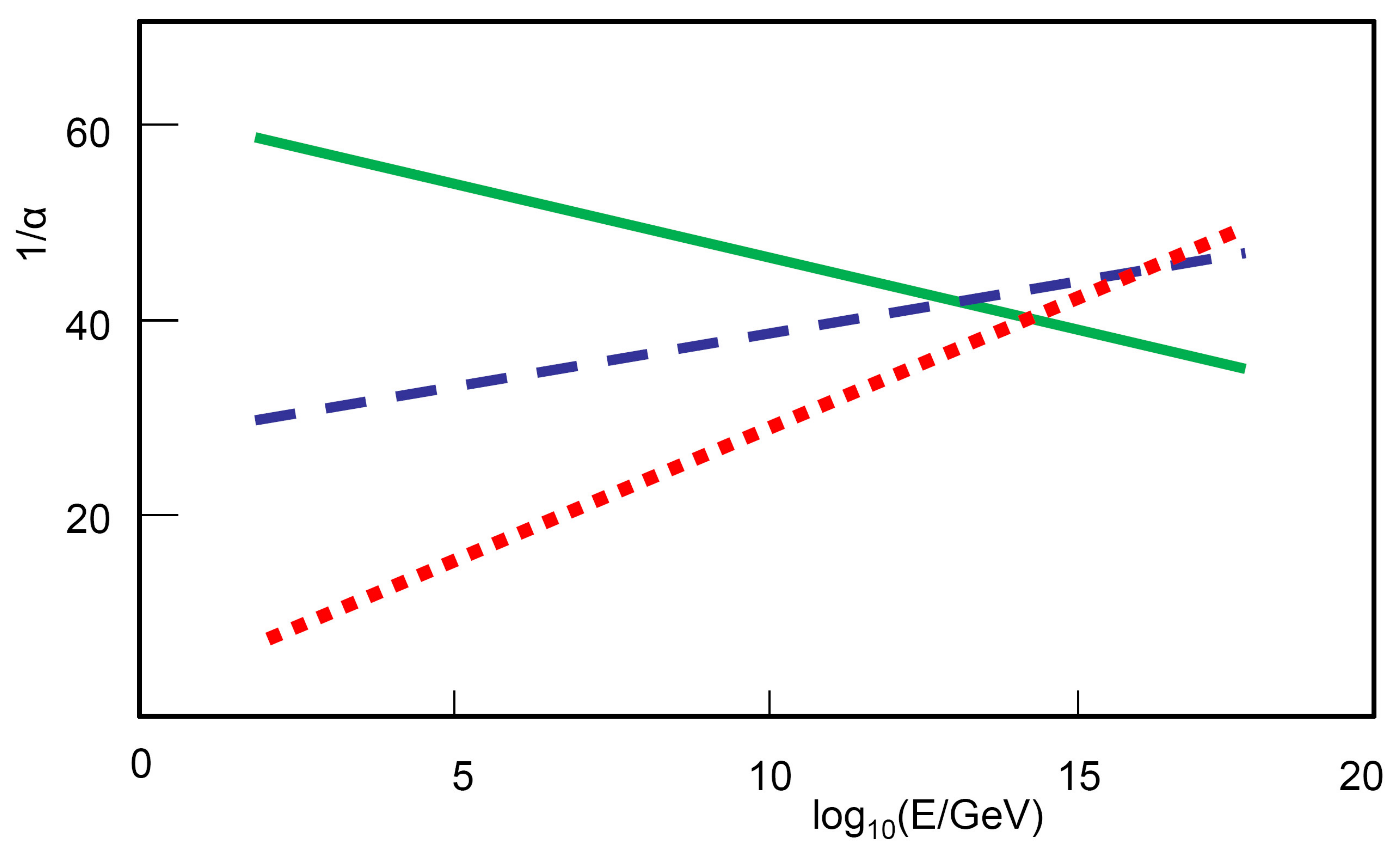}
\caption{
\label{fig:runningCouplings}
The energy-scale dependence of coupling constants of SM gauge interactions
$U(1)_{\rm Y}$
(the full line), $SU(2)_{\rm W}$ (the dashed line) and $SU(3)_{\rm C}$
(the dotted line) in the leading order.}
\end{figure}

{\bf 1.}
The most radical approach, rather popular in recent years, is to assume
that the high-energy scales are simply absent in Nature. For a theoretical
physicist, the most easy scales to refuse are
$M_{\nu }$ and $M_{\rm PQ}$, because they do not appear in all models
explaining neutrino masses and $CP$ conservation in strong interactions,
respectively. $M_{\rm GUT}$ is a
bit more difficult: the Grand Unification of interactions gets support not
only from aesthetic expectations (electricity and magnetism unified to
electrodynamics, electrodynamics and weak interactions unified to
the electroweak theory, etc.) and the arguments related to the electric
charge quantization (see e.g.\ \cite{ChengLi}), but also from the analysis
of the renormalization-group running of the three SM gauge coupling
constants which get approximately the same value at the scale $M_{\rm
GUT}$ (see e.g.\ \cite{SMprimer, ChengLi}). It is worth noting that at the
 plot (see Fig. \ref{fig:runningCouplings})
of
$\alpha_{1,2,3}$ as functions of energy in SM, the three lines do not
intersect at a strictly one point, however, for the case of evolution for
many orders of magnitude in the energy scale, already an approximate
unification is a surprise. To make three lines intersect at one point
precisely, one needs a free parameter, which may be introduced in the
theory with some new particles, e.g.\ with masses $\sim$TeV (this happens
in particular in models with low-energy supersymmetry, see below).
Therefore, the most amusing is not the precise unification of couplings in
an extended theory with additional parameters but the approximate
unification already in SM. It is not that easy to keep this miraculous
property and at the same time to lower the
$M_{\rm GUT}$ scale in order to avoid the hierarchy
$v \ll M_{\rm GUT}$. Indeed, the addition of new particles which affect the
renormalization-group evolution either spoils the unification or, to the
leading order, does not change the
$M_{\rm GUT}$ scale (note that in SM, the unification occurs
in the perturbative regime and higher corrections does not change the
picture significantly). The only possibility is to give up the
perturbativity (the so-called ``strong unification''
\cite{Ross,RuST}).  In this latter approach, the addition of a large number
of new fields in full multiplets of a certain unified gauge group results
in increasing of the coupling constants at high energies; QCD stops to be
asymptotically free at energies higher than the masses of the new
particles. In the leading order, all three coupling constants have, in
this case, poles at high energies; the unification of SM couplings
guarantees that the three poles coincide and are located at $M_{\rm GUT}$.
However, this leading-order approximation has nothing to do with the real
behaviour of constants in the strong-coupling regime, so the theory may
generate a new scale $M_{s}$ at which $\alpha _{1,2,3}$ become strong,
this scale being an ultraviolet analog of $\Lambda_{\rm QCD}$. For a
sufficiently large number of additional matter fields, $M_{s}$ may be
sufficiently close to the electroweak scale $v$: in certain cases, it
might be that $M_{s}\ll M_{\rm GUT}$ (a nonperturbative fixed point). In
this scenario, low-energy observable values of the coupling constants
appear as infrared fixed points and do not depend on unknown details of
the strong dynamics. Note that the Grand-unified theory may have degrees
of freedom very different from SM in this case.

In the recent decade, the models became quite popular in which the
hierarchy problem is solved by giving up the large parameter
$M_{\rm Pl}$. This parameter is related to the gravitational law and any
attempt to change the parameter requires a change in the Newtonian
gravity. This may be achieved, for instance, if the number of space
dimensions exceeds three but, for some reason, the extra dimensions remain
unseen (see e.g.\ a review
\cite{RuUFNextradim}).
Indeed, assume that the extra dimensions are compact and have a
characteristic size $\sim R$, where $R$ is sufficiently small. Then it is
easy to obtain the relation
\begin{equation}
M_{\rm Pl}^2 \sim R^\delta M_{\rm Pl, 4+\delta}^{2+\delta},
\label{Eq:MPl}
\end{equation}
where $\delta$ is the number of extra space dimensions,
$M_{\rm Pl,4+\delta}$ is the fundamental parameter of the
$(4+\delta )$-dimensional theory of gravity, while
$M_{\rm Pl}$ now is the effective four-dimensional Planck mass. Already in
the beginning of the past century, in works by Kalutza
\cite{Kaluza}, subsequently developed by Klein \cite{Klein}, possible
existence of these extra dimensions, unobservable because of small $R$,
has been discussed. This approach assumed that
$R\sim 1/M_{\rm Pl}$ (and therefore $M_{\rm Pl}\sim M_{\rm Pl, 4+\delta}$)
and has become well-known and popular in the second part of the 20th
century in context of various models of string theory, which however did
not result in succesful phenomenological applications by now. We will
discuss, in a little more detail, another approach which allows to make
$R$ larger without problems with phenomenology. It is based on the idea of
localization of observed particles and interactions in a 4-dimensional
manifold of a
$(4+\delta )$-dimensional spacetime \cite{Akama, RuSha1, Visser}.

From the field-theoretical point of view, the localization of a
$(4+\delta)$-dimensional particle means that the field describing this
particle satisfies an equation of motion with variables related to the
observed four dimensions (call them
$x_{\mu},~\mu=0,1,2,3$) separated from those related to $\delta$ extra
dimensions ($z_{A},~A=1,\dots \delta$) and the solution for the
$z$-dependent part is nonzero only in a vicinity (of the size
$\sim \Delta$) of a given point in the $\delta$-dimensional space (without
loss of generality, one may consider the point $z=0$), while the
$x$-dependent part satisfies the usual four-dimensional equations of
motion for this field. As a result, the particles described by the field
move along the four-dimensional hypersurface corresponding to our world
and do not move away from it to the extra dimensions for distances
exceeding $\Delta$. This may happen if the particles are kept on the
four-dimensional hypersurface by a force from some extended object which
coincides with the hypersurface. This solitonlike object is often called
brane, hence the expression ``braneworld''. The readers of
\textit{Physics Uspekhi}  may find a more detailed description of this
mechanism in \cite{RuUFNextradim}.

Based on the topological properties of the brane, localisation of light
(massless in the first approximation) scalars and fermions in four
dimensions\footnote{Note that recently, a fully analogous mechanism of
localisation in one- or two-dimensional manifolds has been tested
\textit{experimentally} for a number of solid-state systems (the quatum
Hall effect, topological superconductors and topological insulators,
graphene), see e.g. \cite{solid}.} implies that many direct experimental
bounds on the size of extra dimensions in a Kalutza-Klein-like model
restrict now the region $\Delta$ accessible for the observed particles,
instead of the size $R$ of the extra dimension. In \cite{ADD}, it has
been suggested to use this possibility, for $R\gg \Delta$, to remove,
according to Eq.~(\ref{Eq:MPl}), a large fundamental scale $M_{\rm Pl}$
and the corresponding hierarchy. It has been pointed out that in this
class of models, $R$ is bound from above mostly by nonobservation of
deviations from the Newtonian gravity at short distances; experiments now
exclude the deviations at the scales of order 50~$\mu$m only
\cite{PRL98(2007)021101} (it was $\sim 1$~mm at the moment when the model
was suggested). This allows, according to Eq.~(\ref{Eq:MPl}), to have
$M_{\rm Pl,4+\delta}\sim$TeV, that is almost of the same order as $v$.
Models of this class are well studied from the phenomenological point of
view but have two essential theoretical drawbacks. The first one is
related to the apparent absence of a reliable mechanism of localization of
\textit{gauge} fields in four dimensions. The only known field-theoretical
mechanism for that \cite{DvaliShifman} is based on some assumptions about
the behaviour of a multidimensional gauge theory in the strong-coupling
regime. Though these assumptions look realistic, they currently cannot be
considered as well-justified. The second difficulty is aesthetic and is
related to the appearence of a new dimensionful parameter $R$: the
hierarchy $v \ll M_{\rm Pl}$ happens to be simply reformulated in terms of
a new unexplained hierarchy $1/R \ll M_{\rm Pl, 4+\delta}$.

To a large extent, these difficulties are overcome in somewhat more
complicated models, in which the spacetime cannot be presented as a direct
product of our four-dimensional Minkowski space and compactified extra
dimensions
\cite{RuSha2, Gogber, Lisa1}. The principal difference of this
approach from the one discussed above is that the gravitational field of
hte brane in extra dimensions is not neglected. For $\delta=1$ and in the
limit of a thin brane, one obtains the usual five-dimensional
general-relativity equations. These equations have, in particular,
solutions with four-dimensional Poincare invariance. The metrics in these
solutions is exponential in the extra-dimensional coordinate (the
so-called anti-de-Sitter space),
\begin{equation}
ds^{2}=\exp(-2k|z|) dx^{2}-dz^{2}
\label{Eq:AdS}
\end{equation}
where $ds^{2}$ and $dx^{2}$
are the squares of the five-dimensional and usual four-dimensional
 (Min\-ko\-w\-ski) intervals, respectively. For a finite size $z_{c}$ of
 the fifth dimension, the relation between the fundamental scales is now
$$
M_{\rm Pl}\sim \exp(k z_c) M_{\rm Pl,5}.
$$
If fundamental dimensionful parameters of the five-dimensional gravity
satisfy
$M_{\rm Pl, 5}\sim k \sim v$, one may \cite{Lisa1}
explain the hierarchy $v/M_{\rm Pl}$ for $z_c\approx 37/k$,
that is instead of the fine tuning with the precision of
$10^{-16}$, one now needs to tune the parameters up to
$\sim 0.1$.
It is interesting that in models of this kind with two or more extra
dimensions, it is possible
\cite{Lisa6D-loc} to localize gauge fields on the brane in the
weak-coupling regime, contrary to the case of the factorizable
geometry.

{\bf 2.}
A completely different approach to the problem of stabilization of the
gauge hierarchy is to add new fields which cancel quadratic divergencies
in expressions for the running SM parameters. The best-known realization
of this approach is based on supersymmetry (see e.g.\ reviews
\cite{SUSYufnNevzorov, SUSYufnWe, SUSYKazakov}),
which provides for the cancellation of divergencies due to opposite signs
of ferminonic and bosonic loops in Feynman diagrams.

The requirement of supersymmetry is very restrictive for the mass spectrum
of particles described by the theory. Namely, together with the observed
particles, their superpartners, that is particles with the same masses and
different spins, should be present. The absence of scalar particles with
masses of leptons and quarks and of fermions with masses of gauge bosons
means that unbroken supersymmetry does not exist in Nature. It has been
shown, however, that it is possible to break supersymmetry while keeping
the cancellation of quadratic divergencies. This breaking is called
``soft'' and naturally results in massive superpartners.

In the minimal supersymmetric extension of SM (MSSM; see e.g.\
\cite{SUSYKazakov}), each of the SM fields has a superpartner with a
different spin: the Higgs boson corresponds to a fermion, higgsino;
matter-field fermions correspond to scalar squarks and sleptons; gauge
bosons correspond to fermions which transform in the adjoint
representation of the gauge group and are called gauginos (in particular,
gluino for $SU(3)_{C}$, wino and zino for the $W$ and $Z$ bosons, bino for
the hypercharge $U(1)_{Y}$ and photino for the electromagnetic gauge group
$U(1)_{\rm EM}$). For the theory to be selfconsistent (absence of
anomalies related to the higgsino loops), and also to generate fermion
masses in a supersymmetric way, the second Higgs doublet is introduced,
which is absent in SM. The cancellation of quadratic divergencies may be
easily seen in Feynman diagrams: in the leading order, closed fermion
loops have the overall minus sign and cancel the contributions from loops
of their superpartner bosons. This cancellation is precise as long as the
masses of particles and their superpartners are equal; otherwise the
contributions differ by an amount proportional to the difference between
squared masses of superpartners,
$\Delta m^2$.
The condition of stability of the gauge hierarchy then requires that
$\frac{g^2}{16 \pi^2}\Delta m^2 \lesssim v^2$, where $g$
is the coupling constant in the vertex of the corresponding loop (the
maximal, $g\sim 1$, coupling constant is that of the top quark). We arrive
to an important conclusion which motivates in part the current interest to
phenomenological supersymmetry: if the problem of stabilization of the
gauge hierarchy is solved by supersymmetry, then the superpartner masses
cannot exceed a few TeV, which means that they might be experimentally
found in the nearest future.

The MSSM lagrangian, in the limit of unbroken supersymmetry, satisfies all
symmetry requirements of SM, including the conservation of the lepton and
baryon numbers. At the same time, the SM gauge symmetries do not forbid,
for this set of fields, certain interaction terms which violate the lepton
and baryon numbers. The coefficients at these terms should be very small
in order to satisfy experimental constraints, for instance, those related
to the proton lifetime. It is usually assumed that these terms are
forbidden by an additional global symmetry $U(1)_{R}$. When supersymmetry
is broken, this
$U(1)_R$ breaks down to a discreet $Z_2$
symmetry called $R$ parity. With respect to the $R$ parity, all SM
particles carry charges $+1$ while all their superpartners carry charges
$-1$. The $R$-parity conservation leads to the stability of the lightest
superpartner (see Sec.~\ref{sec:dark-matter}).

The soft supersymmetry-breaking terms are introduced in the MSSM
lagrangian explicitly. They include usual mass terms for gaugino and
scalars as well as trilinear interactions of the scalar fields. In
addition to the SM parameters, about 100 independent
real parameters are therefore introduced. In general, these new couplings
with arbitrary parameters may result in  nontrivial flavour physics. The
absence of flavour-changing neutral currents and of processes with
nonconservation of leptonic quantum numbers, as well as limits from the
$CP$ violation, narrow the allowed region of the parameter space
significantly.

One may note the following characteristic features of the phenomenological
supersymmetry.

(1). The coupling-constant unification at a high energy scale becomes more
precise as compared to SM, if superpartners have masses $\sim v$ as
required for the stability of the gauge hierarchy.

(2). In the same regime, the gauge desert between $\sim 10^3$~GeV and
$\sim 10^{16}$~GeV is still present.

(3).
In MSSM, there is a rather restrictive bound on the mass of the lightest
Higgs boson. In the leading approximation of perturbation theory, it is
$M_H<M_Z$. The account of loop corrections allow to relax it slightly, but
in most realistic models
$M_H<150$~GeV is predicted. The absence of a light Higgs boson discussed
in Sec.~\ref{sec:EW} is a much more serious problem for supersymmetric
theories than for SM.

(4).
The phenomenological model described above explains the stability of the
gauge hierarchy but not its origin. The small parameter $v/M$, where
$M=M_{\rm GUT}$ or $M=M_{\rm Pl}$,
does not require tuning in every order of perturbtion theory but should be
introduced in the model by hand, that is cannot be derived, nor expressed
through a combination of numbers of order one. At the same time, if the
supersymmetry breaking is moderate, as required to solve the
quadratic-divergency problem, it may be explained dynamically and related
to nonperturbative effects which become important at a characteristic
scale of
$$
\Lambda \sim \exp \left(- O\left(1/g^2 \right) \right)M,
$$
where $g$ is some coupling constant. If $g$ is small, then the
supersymmetry breaking scale is also small,
$\Lambda \ll M$. In a number of realistic models it is possible to get, up
to powers of the coupling constants, $v\sim \Lambda$ dynamically (by means
of radiative corrections) and to explain therefore the origin of the gauge
hierarchy. However, in the MSSM frameworks, there is no place for
nonperturbative effects of the required scale: these effects are relevant
only for QCD and with $\Lambda \sim \Lambda_{\rm QCD} \ll v$.
The dynamical supersymmetry breaking should take place in a new sector,
introduced expressly for this purpose and containing a new strongly
coupled gauge theory with its own set of matter fields. No sign of this
sector is seen in experiments and one consequently supposes that the
interaction between the SM (or MSSM) fields and this sector is rather weak
and becomes significant only at high energies, unreachable in the present
experiments. This interaction is responsible for the soft terms, that is
for mediation of the supersymmetry breaking from the invisible sector to
the MSSM sector. One distinguishes the gravity mediation (at Planck
energies) and the gauge mediation of supersymmetry breaking.
Gravity-mediated and gauge-mediated models have quite different
phenomenology.

\begin{figure}
\centering
\includegraphics[width=0.85\columnwidth]{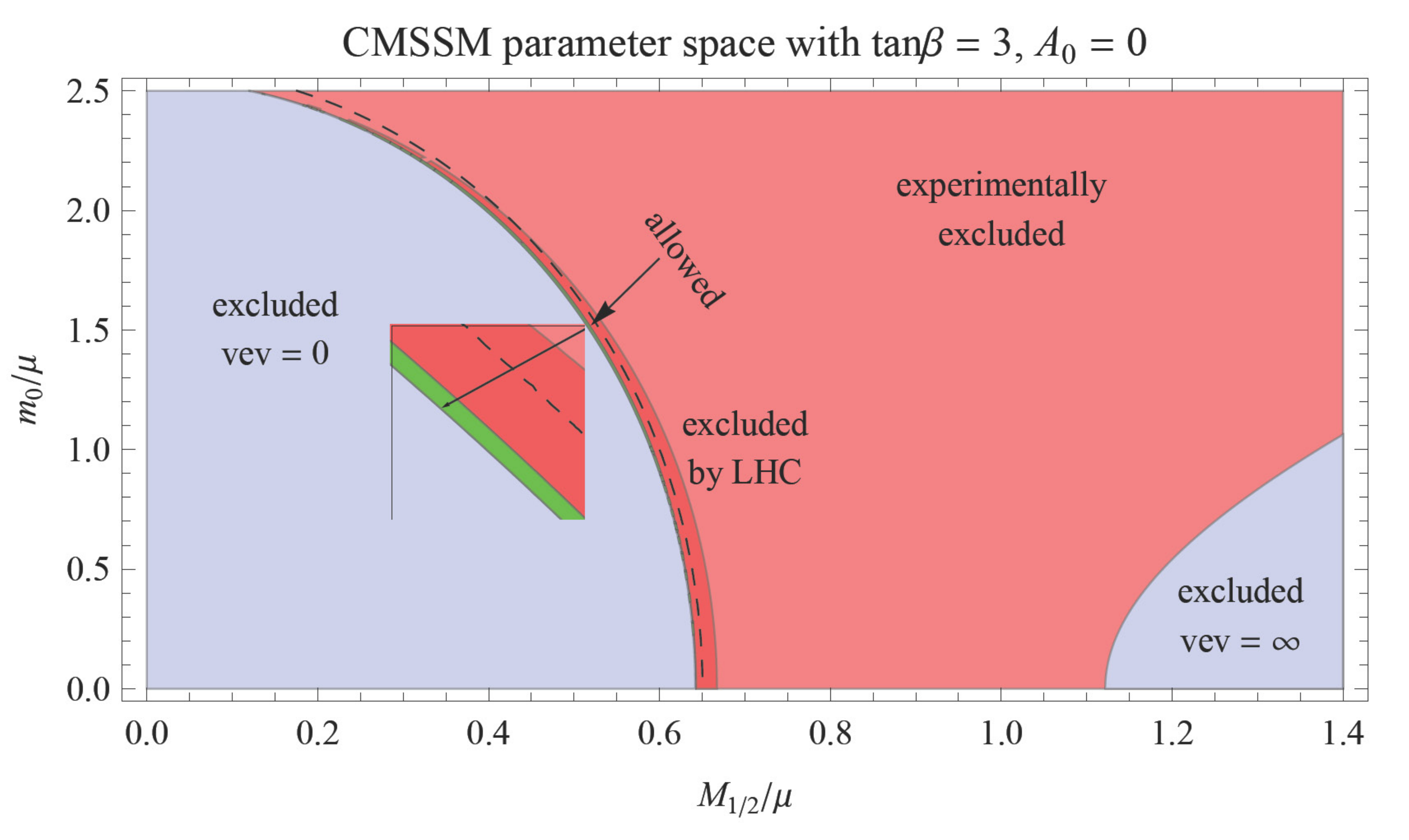}
\caption{
\label{fig:Strumia}
Constraints on the MSSM parameters \cite{Strumia1101.2195}
in one popular scenario (see text). The allowed region is the narrow strip
which can be seen in the zoomed panel.
}
\end{figure}

We see that MSSM, with addition of a sector which breaks supersymmetry
dynamically and of a certain interaction between this hidden sector and
the observable fields, may explain the origin and stability of the gauge
hierarchy, if the masses of superpartners are not very high
($\lesssim$~TeV). Note that the searches for supersymmetry in accelerator
experiments put serious constraints on the low-energy supersymmetry.
Already the fact that superpartners have not been seen at LEP implied that
a significant part of the theoretically allowed MSSM parameter space was
excluded experimentally. Subsequent results of Tevatron and especially the
first LHC data squeeze the allowed region of parameters significantly, so
that for ``canonical'' supersymmetry, only a very narrow and not fully
natural region of possible superpartner mass remains allowed. In
Fig.~\ref{fig:Strumia},
theoretical and experimental (as of summer 2011) constraints on the MSSM
parameters are plotted for one rather natural and popular scenario of
gravity-mediated supersymmetry breaking. The masses of all scalar
superpartners at the
$M_{\rm GUT}$ energy scale are equal to $m_{0}$ in this scenario, while
masses of all fermionic superpartners are
$M_{1/2}$. Their ratios to the supersymmetric mixing matrix of the Higgs
scalars, $\mu$, are given in the plot. In a scenario which explains the
gauge hierarchy, the MSSM parameters and the $Z$-boson mass should be of
the same order; for instance, in the model which corresponds to the
illustration, the following relation holds,
$$
M_Z^2 \simeq 0.2 m_0^2 +1.8 M_{1/2}^2 - 2 \mu^2.
$$
The LHC bound, $M_{1/2} \gtrsim 420$~GeV, results in the requirement of
not fully natural cancellations since
$1.8 M_{1/2}^2 \gtrsim 40 M_Z^2$. Together with the absence of a light
Higgs boson discussed in Sec.~\ref{sec:EW}, this ``little hierarchy''
problem makes the approach based on supersymmetry less motivated than it
looked some time ago, though there exist variations of supersymmetric
models where this dificulty is overcome.

{\bf 3.}
The Higgs field may be a pseudo-Goldstone boson. The Goldstone theorem
guarantees a massless (even with the account of radiative corrections!)
scalar particle for each generator of a broken global symmetry. A weak
explicit violation of this symmetry allows to give a small mass to this
scalar to get the so-called pseudo-Goldstone boson. The same mechanism
results in a low but nonzero mass of some composite particles in a
strongly-interacting theory (for instance, of the $\pi$ meson). A direct
application of this approach to the Higgs boson is not possible because
the interaction of a pseudo-Goldstone particle with other fields contains
derivatives and is very different from the SM interactions. Realistic
models of this kind with large coupling constants and with interactions
without derivatives, at the same time free from quadratic divergencies,
are called the ``Little Higgs models'' (see e.g.\
\cite{LittleHiggs} and references therein). Diagram by diagram, the
absence of quadratic divergencies occurs due to complicated cancellations
of contributions of a number of particles with masses of order TeV, in
particular of additional massive scalars. Note that to reconcile a large
number of new particles with experimental constraints, in particular with
those from the precision electroweak measurements, the model requires
significant complications.

{\bf 4.} Composite models: besides the Little Higgs models, a composite
Higgs scalar is considered in a number of other constructions, see e.g.\
\cite{TonyReview}. In some rather popular models with composite quarks and
leptons, the SM matter fields, together with the Higgs boson (or
even without it) represent low-energy degrees of freedom of a strongly
coupled theory, like hadrons may be considered as low-energy degrees of
freedom of QCD. The mass scales of the theory, $v$ in particular, are
determined by the scale $\Lambda$ at which the running coupling constant
of the strongly-coupled theory becomes large, analogously to
$\Lambda_{\rm QCD}$. The hierarchy
$\Lambda \ll M_{\rm Pl}$ is now determined by the evolution of couplings
in the fundamental theory. These models generalize, to some extent, the
technicolour models, having more freedom  in its construction at the price
of even more complications in the quantitative analysis. Note that (at
least) in some supersymmetric gauge theories, low-energy degrees of
freedom may include also gauge fields, so in principle, one may consider
models in which all SM particles are composite (see e.g.\ \cite{RuST,
TonyReview}). On the other hand, the correspondence between strongly
coupled four-dimensional models and weakly-coupled five-dimensional
theories (see Sec.~\ref{sec:QCD:dual}) may open prospects for a
quantitative study of composite models. It might even happen that the
approaches to the gauge-hierarchy problem, based on assumptions of the
extra space dimensions, are equivalent to the approaches which invoke
strongly coupled composite models. As in other approaches, to explain the
hierarchy, the scale $\Lambda$ should not exceed significantly the
electroweak scale $v$, so that the LHC constraints on compositeness of
quark and leptons (roughly $\Lambda \gtrsim (4 \dots 5)$~TeV) may again be
problematic.

{\bf Conclusion.} All known scenarios which explain the origin and
stability of the gauge hierarchy without extreme fine tuning, predict new
particles and/or interactions at the energy scale not far above the
electroweak scale. Absence of experimental signs of these particles,
especially with the account of the first LHC data, questions the ability
of these scenarios to solve the hierarchy problem. If the LHC finds the
Higgs scalar but will not confirm predictions of any of the models
discussed above, nor will find signs of some other, yet not invented,
mechanism, then one would have to reconcider the question of the
naturalness of the fine tuning. A principally different position, based on
the anthropic principle, is seriously discussed but lays beyond the scope
of our consideration.

\subsection{The fermion mass hierarchy.}
\label{sec:mass-hierarchy}
\begin{figure}
\centering
\includegraphics[width=0.85\columnwidth]{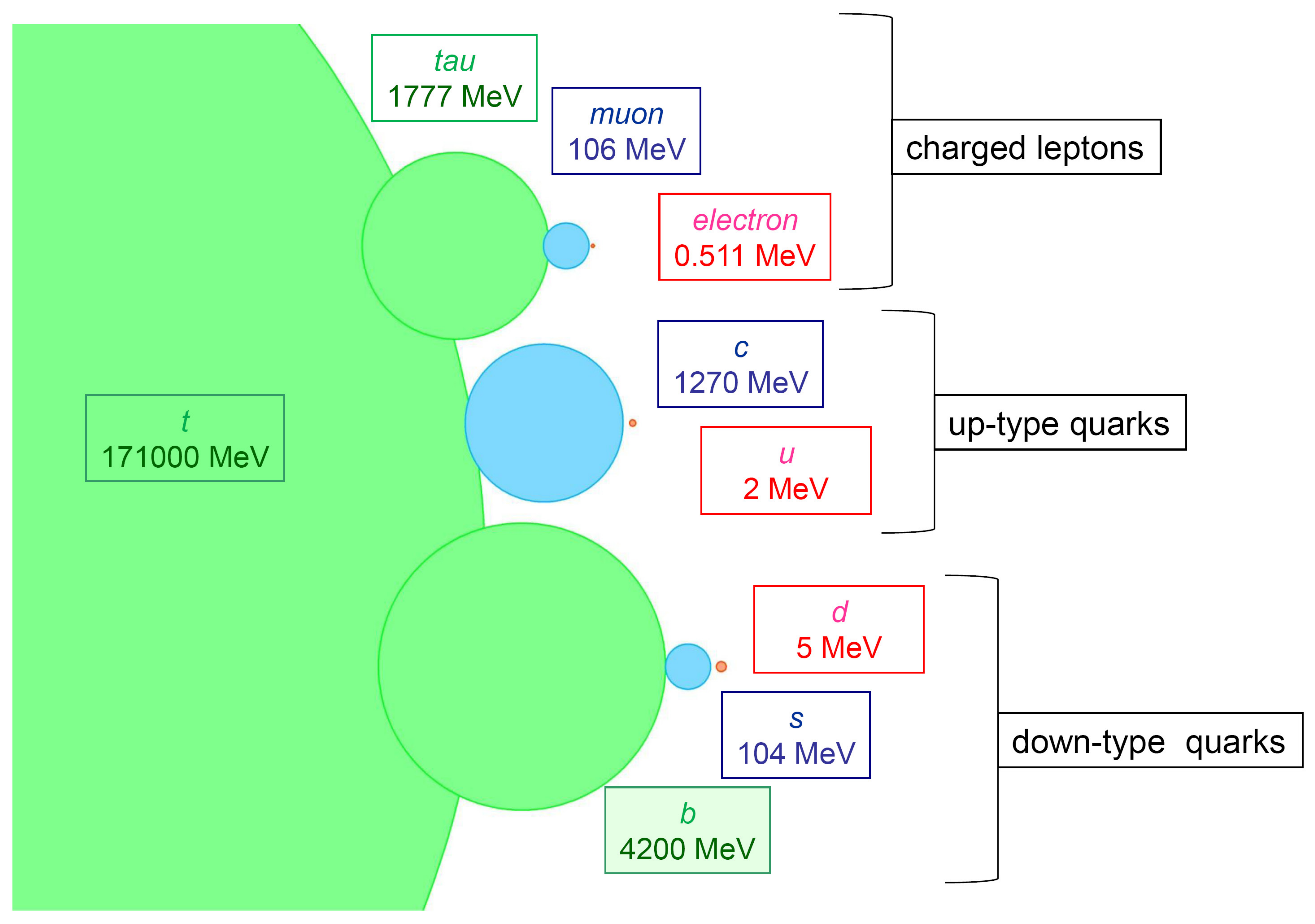}
\caption{
\label{fig:chFermionMasses}
Masses of the charged SM fermions. The area of each circle is proportional
to the mass of the corrsponding particle.}
\end{figure}

As it has already been pointed out, the SM fermionic fields, quarks and
leptons, comprise three generations, that is three sets of particles with
identical interactions but with very different masses (see
Fig.~\ref{fig:chFermionMasses} for a pictorial illustration).
The hierarchy of these masses is one of the biggest puzzles of particle
physics. Indeed, for instance, the electron
($m_e=0.511$~MeV), the muon ($m_\mu=105.7$~MeV) and the tau lepton
($m_\tau=1777$~MeV)  carry identical gauge quantum numbers. For quarks, it
is convenient to determine the mass matrix whose diagonal elements
determine the masses of the quarks of three generations with identical
interactions while combinations of non-diagonal elements provide for the
possibility of mixing between generations. The hierarchical structure
appears both in the diagonal elements (which differ by orders of
magnitude) and in the off-diagonal ones (the mixing is suppressed). In the
SM frameworks, neutrino are strictly massless and the mixing of charged
leptons is absent, but the same hierarchical structure is seen in the set
of masses of charged leptons.

As we have discussed in Sec.~\ref{sec:neutrino},
the experiments of the past decade not only established confidently the
fact of the neutrino oscillations (pointing therefore to nonzero neutrino
masses and giving the first laboratory indication to the incompleteness of
SM) but also opened the possibility of a quantitative study of neutrino
masses and of the mixing in the leptonic sector. It is interesting that the
neutrino masses and the leptonic mixings also have the hierarchical
structure, but it is very different from the corresponding hierarchy in
the quark sector: contrary to the suppressed quark mixings, the leptonic
mixing is maximal; the hierarchy of neutrino masses is at the same time
moderate. A modern theory which succesfully explains the fermion masses
should motivate both hierarchical structures and explain why they are
different.

Meanwhile, even without the neutrino sector, the intergeneration mass
hierarchy is very difficult to explain. A natural idea is to suppose that
there is an extra global symmetry which relates the fermionic generations
to each other and which is spontaneously broken; however, this approach is
not succesful because it implies the existence of a massless Goldstone
boson, the so-called familon, whose parameters are strictly constrained by
experiments
\cite{PDG2010}.

A model of fermion masses should explain only the origin of the hierarchy:
its stability is provided automatically by the fact that all radiative
corrections to the fermion-Higgs Yukawa constants, to which the fermion
masses are proportional, depend on the energy logarithmically, that is
weakly; this does not, however, make the issue significantly less
complicated.

\begin{figure}
\centering
\includegraphics[width=0.65\columnwidth]{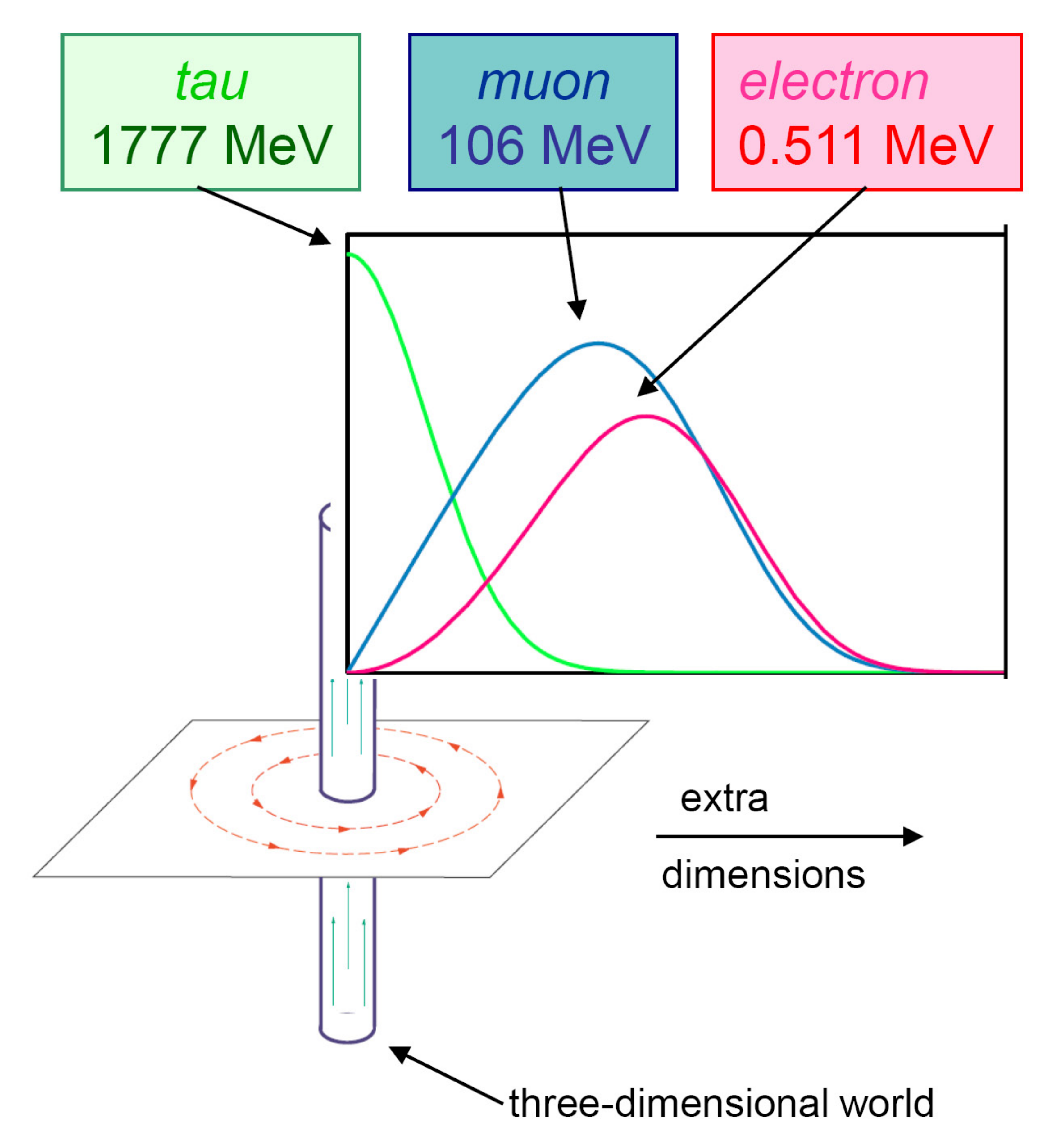}
\caption{
\label{fig:6d-model}
A model with extra space dimensions which explains the mass hierarchy.
}
\end{figure}

An explanation of the hierarchy may be obtained in a model with extra
space dimensions (Fig.~\ref{fig:6d-model}),
in which a single generation of particles in six-dimensional spacetime
effectively describes three generations in four dimensions \cite{LT, FLT}.
Each multidimensional fermionic field has three linearly independent
solutions which are localized on the four-dimensional hypersurface  and
have different behaviour close to the brane. Denoting as $r,\theta$ the
polar coordinates in two extra dimensions and considering the brane at
$r=0$, one gets for the three solutions at $r\to 0$,
$$
u_0\sim \mbox{const}=r^0 \mbox{e}^{i 0 \theta}, ~~~
u_1\sim r^1 \mbox{e}^{i 1 \theta}, ~~~
u_2\sim r^2 \mbox{e}^{i 2 \theta}.
$$
The Higgs scalar has a vacuum expectation value $v(r)$ which depends on
$r$ and is nonzero only in the immediate vicinity of the brane. The
effective observable fermion masses are proportional to the overlap
integrals
$$
m_i \propto \int \! dr\, d \theta \, v(r) |u_i|^2 (r,\theta)
$$
of the coordinate-dependent vacuum expectation value $v$ and
extra-dimensional parts of the fermionic wave functions which correspond
to the three localized solutions ($i=0,1,2$ enumerates three generation of
fermions). One can see from Fig.~\ref{fig:6d-model} that the resulting
$m_{i}$ are hierarchically different. Therefore, in this model the mass
hierarchy follows from the linear independence of eigenfunctions of the
Dirac operator in a particular external field. The same model
automatically describes the required structure of neutrino masses and
mixings
\cite{FuSin}.
Presently, this model is the only one known in which the hierarchy of
families of both charged fermions and neutrinos are obtained on the common
grounds. Note that, contrary to other multidimensional models (e.g.\
\cite{Shmalz}), in this model the number of free parameters is smaller
than the number of parameters it describes.

Compared to the hierarchy of masses of particles with identical
interactions from different generations, the question of the difference of
masses of particles within a generation is much easier. For instance, the
difference between masses of the $\tau$ lepton and the $b$ and $t$ quarks
may be explained by different (because of different quantum numbers)
renormalization-group evolution of the Yukawa couplings, so that at the
Grand-unification scale these constants are equal while at low energies
they are different.

\section{Theoretical challenges in the description of hadrons.}
\label{sec:QCD}
\subsection{Problems of the perturbative QCD.}
\label{sec:QCD:problems}
In this section, we discuss the question about the practical applicability
of the quantum field theory to the description of interactions with large
coupling constants, and in particular to the low-energy limit of QCD. It
would not be  an exaggeration to say that most of the theoretical
achievements in the quantum field theory in the past two decades were
related to this problem. Before proceeding to the discussion of these
achievements, let us note that despite a significant progress, the problem
of description of strong interactions at low energies in terms of QCD is
not solved, so the development of the corresponding methods remains one of
the basic tasks of the quantum field theory.

Recall that QCD, which describes the strong interaction at high energies,
is a gauge theory with the gauge group $SU(3)_{C}$ and $N_{f}=6$ fermions,
quarks, which transform under its fundamental representation, and the
same number of antiquarks transforming under the conjugated
representation. A peculiarity of the model is that the asymptotic states,
in terms of which the quantum theory is constructed, do not coincide with
the fundamental fields in terms of which the Lagrangian is written, that
is with fermions (quarks) and gauge bosons (gluons). Contrary, the
observable particles do not carry the $SU(3)_{C}$ quantum numbers (this
phenomenon is called confinement). The observable strongly interacting
particles are hadrons, whose classification and interactions allow to
interpret them as bound states of quarks. At the same time, the theory
which describes interaction of quarks, QCD, is unable to calculate
properties of these bound states. Intuitively, it seems possible to relate
confinement and formation of hadrons with the energy dependence of the QCD
gauge coupling constant which grows up with the decrease in energy (that
is with the increase in distance; the so-called asymptotic freedom) and
becomes large,
$\alpha_{s}\sim 1$, at the scale
$\Lambda_{\rm QCD}\sim 150$~MeV: when the distance between quarks is
increased, the force between them increases as well, and maybe this force
binds them to hadrons. This picture is however not fully consistent
because at
$\alpha_{s}\gtrsim 1$, the perturbative expansion stops to work and the
true energy dependence of the coupling constant is unknown. Indeed, there
exist examples of theories with asymptotic freedom but without confinement
\cite{AsymptFreedomNoConfinement}.

\begin{figure}
\centering
\includegraphics[width=0.55\columnwidth]{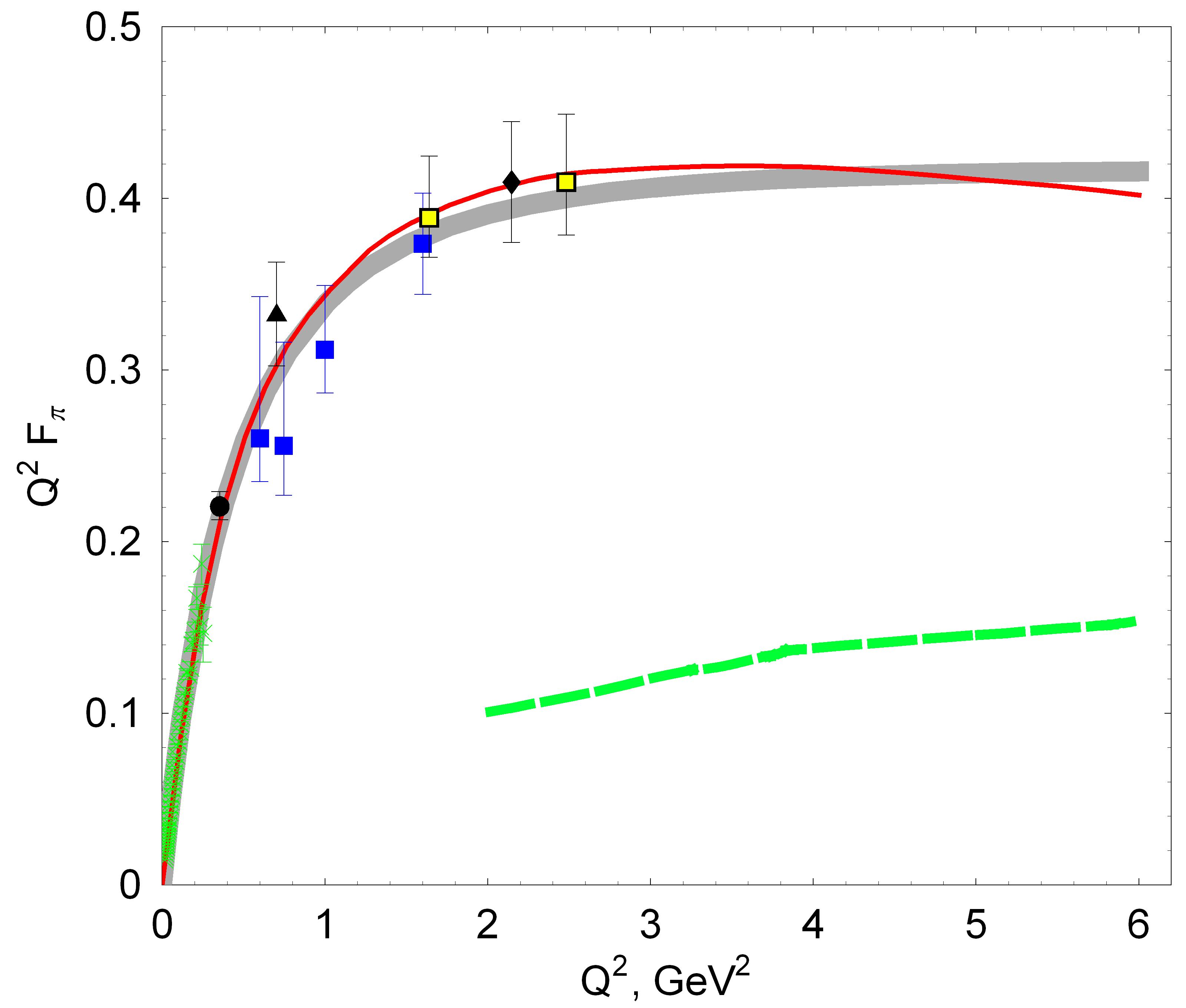}
\caption{
\label{fig:formfactor}
Electromagnetic pion formfactor \cite{KrTr}:
experimental data versus theoretical calculations, perturbative (QCD, the
dashed line) and nonperturbative (full lines representing working models
which are not derived from QCD). Up to the energy scale $\sim 2$~GeV,
there are no signs of approaching the perturbative regime.}
\end{figure}

\begin{figure}
\centering
\includegraphics[width=0.55\columnwidth]{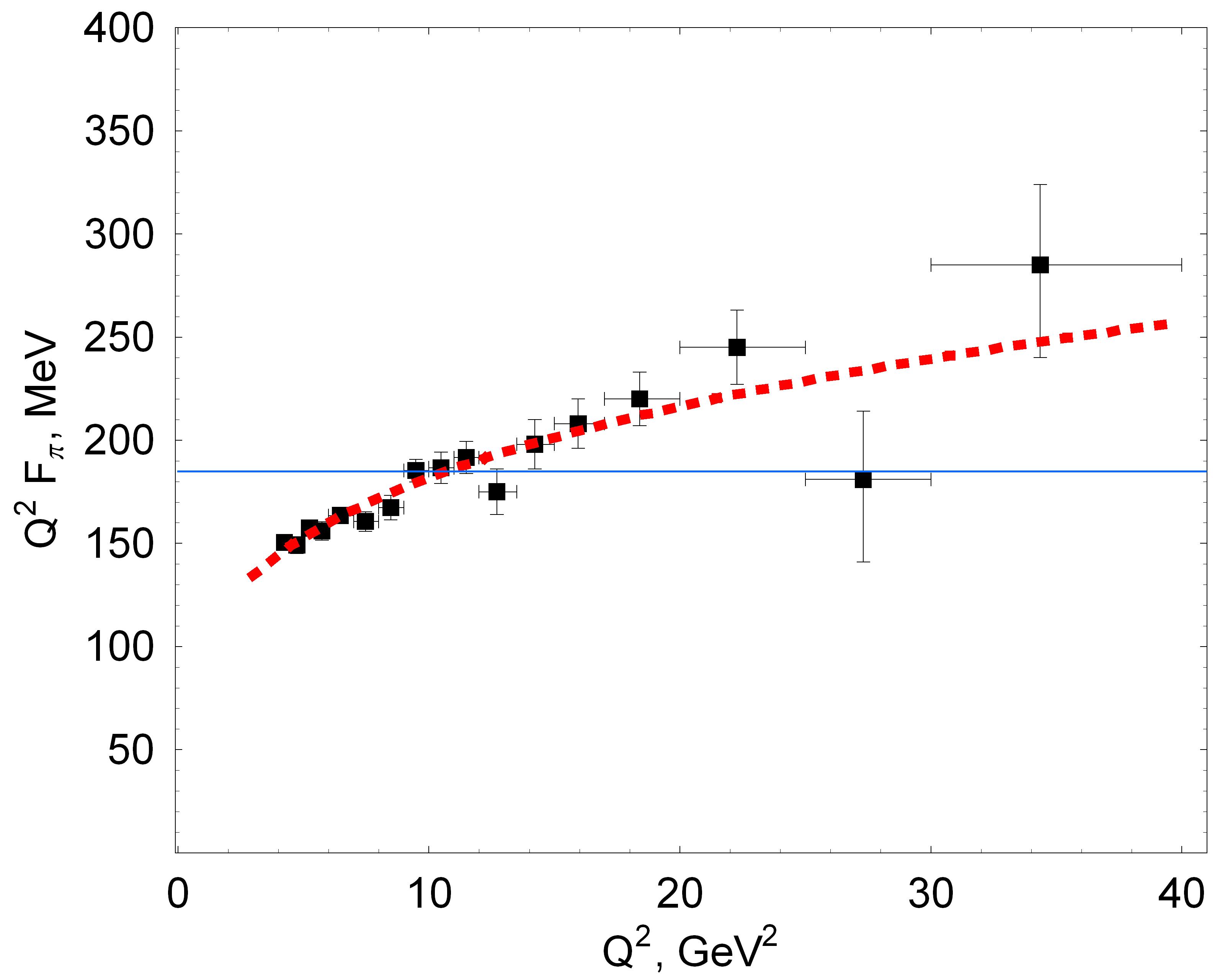}
\caption{
\label{fig:babar}
The transitional form factor of the $\pi$ meson which describes the
process $\pi^{0}\to \gamma \gamma$: experimental data
\cite{BABAR0905.4778} versus calculations of perturbative QCD.
The QCD predicts the behaviour  $Q^{2}F(Q^{2})\sim \mbox{const}$
(the horizonthal full line); at least up to $\sqrt{Q^2}\sim 4$~GeV, the
experiment points to $Q^{2}F(Q^{2})\sim
(Q^{2})^{0.5}$ (the dotted line). }
\end{figure}

To understand the nature of confinement and to describe properties of
hadrons from the first principles (and, in the end, to answer whether QCD
is applicable to the description of hadrons), one require the methods of
the field theory which do not make use of the expansion in powers of the
coupling constants (non-perturbative methods). It is natural to assume
(and it was assumed for a long time) that the perturbative QCD has to
describe well the physics of strong interactions at characteristic
energies above few hundred MeV, because the coupling constant becomes
large at $\sim 150$~MeV. A number of recent experimental results related
to the measurement of the form factors of $\pi$ mesons question the
applicability of perturbative methods at considerably higher momentum
transfer (a few GeV). In general, formfactors are the coefficients by
which the true amplitude of a process with composite or extended particles
involved differs from the same amplitude calculated for point-like
particles with the same interaction. These coefficients are determined by
the internal structure of particles (for instance, by the distribution of
the electric charge); their particular form depends on the process
considered and on the value of the square of the momentum transfer,
$Q^{2}$. A full theory describing the interaction which keeps the
particles in the bound state should allow for derivation of form factors
from the first principles. The results of the experimental determination
of formfactors of $\pi$ mesons related to various processes are given in
Figs.~\ref{fig:formfactor}, \ref{fig:babar}.
One may see that the perturbative QCD experience some difficulties in
explaining the experiment at the momentum transfer
$\lesssim 4$~GeV.

Approaches to nonperturbative description of QCD may be divided into two
classes: (1)~calculations in QCD beyond the perturbation theory (the only
available method here is the numerical calculation of the functional
integral on the lattice) and (2)~construction of an effective theory in
terms of degrees of freedom which correspond to obsevrable particles. In
the latter case the main unsolved question is, as a rule, to justify the
connection of the effective theory to QCD. To some extent, a progress in
this direction became possible within the concept of dual theories
discussed below.

\subsection{The lattice results.}
\label{sec:QCD:lattice}
The Feynman functional integral is a formally strict approach to the
quantization of fields, equivalent to other approaches in the domain of
applicability of the perturbation theory. It is natural to suppose that in
the nonperturbative domain, this method also reproduces the results which
would be obtained within the standard frameworks if the means to get them
existed. Numerical calculation of the functional integral is possible in
lattice calculations in which the continuous and infinite spacetime is
replaced by a finite discrete lattice (see e.g.\ \cite{Lattice:encycl}).
In modern calculations, the lattices
$32^{3}\times 64$, that is 32 points in each of the space coordinates and
64 points in time, are used. For physics applications, it is very
important that the gauge invariance may be defined in the lattice theory
in a strict way.

One of the first serious achievements of the lattice field theory was a
discovery that the lattice model with symmetries and field content of QCD
exhibits confinement
\cite{Lattice:confinement}.
Subsequent works allowed to refine which particular field configurations
are responsible for confinement; the work on this question continues.

\begin{figure}
\centering
\includegraphics[width=0.65\columnwidth]{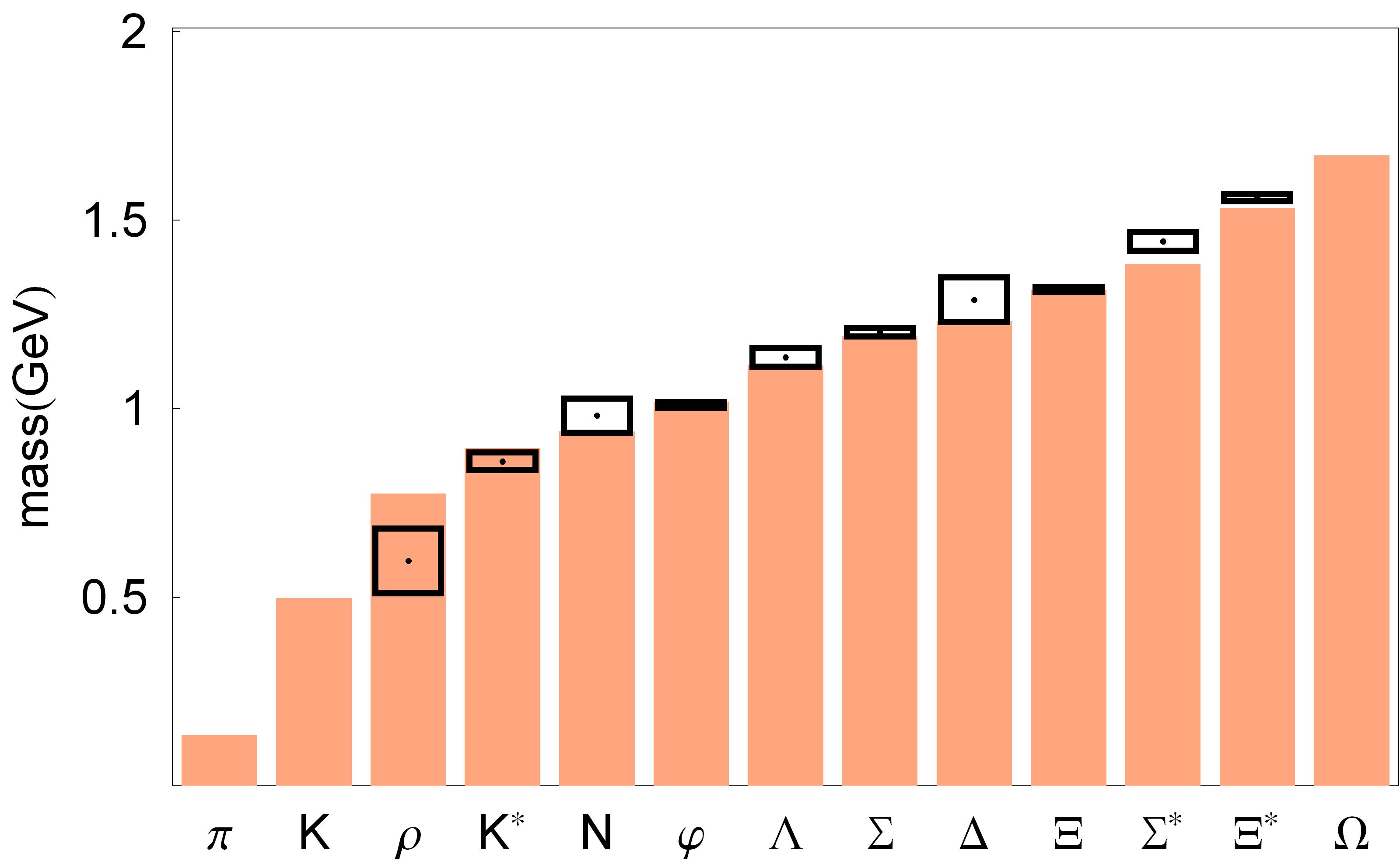}
\caption{
\label{fig:lattice}
Results of the lattice calculations of the hadron masses. Masses of
$\pi$, $K$ and $\Omega$
mesons are taken as input parameters. The calculations have been performed
in the three-quark approximation,
$m_{u}=m_{d}\ne m_{s}$. The histogram gives the experimentally measured
values of masses
\cite{PDG2010},
the points (with the error-bar rectangles) represent the results of
calculations~\cite{0911.2561}.}
\end{figure}

The lattice approach allows to calculate the values of masses and decay
constants of hadrons, and in recent years, a significant progress in this
direction has been achieved (see Fig.~\ref{fig:lattice}).
The most precise for today results
\cite{0911.2561} are obtained for the so-called ``2+1'' parametrization in
which masses of $u$ and $s$ quarks are free parameters, the $d$-quark mass
is assumed to be equal to that of the $u$ quark and the contributions of
heavy $c$, $b$ and $t$ quarks are neglected. Besides  these two
parameters
($m_{u}=m_{d}$ and $m_{s}$),
there is one more, the physical length which corresponds to a unit step of
the lattice. To determine the masses of hadrons, these three parameters
should be specified, so in real calculations one assumes that the masses
of, say,
$\pi$, $K$ and $\Omega$
mesons are known while all other masses and decay constants are expressed
through them. One might try to fix masses of heavier particles and to
calculate those of the lightest ones, but for a confident calculation of
masses of light hadrons a large lattice is required. Currently, the mass
of the $\pi$ meson may be calculated only up to an order of magnitude in
this way.

At high temperature, one expects a transition to the state in which quarks
cannot be confined in hadrons, that is a phase transition. In reality,
these conditions appear in nuclei collisions at high-energy colliders;
probably, they also took place in a very early Universe. By means of the
lattice methods, the existence of this phase transition has been
demonstrated, its temperature has been defined and the dependence of the
order of the phase transition from the quark masses has been studied
\cite{Lattice:PhaseTrans, Fodor}.

It is an open theoretical question to prove that the continuum limit of a
lattice field theory exists (that is the physical results do not depend on
the way in which the lattice size tends to infinity and the lattice step
tends to zero) and coincides with QCD. It may happen that this proof is
impossible in principle unless one finds an alternative way to work with
QCD at strong coupling. However, there exist a series of arguments
suggesting that the lattice theory indeed describes QCD (first of all, it
is the fact that the lattice calculations reproduce experimental results).
At the same time, theoretically, the difference between the lattice and
continuum theories is large; for instance, topologically stable in the
continuum theory configurations, instantons, which determine the structure
of vacuum in nonabelian gauge theories, are not always stable on the
lattice; the lattice description of chiral fermions (automatic in a
continuum theory) requires complicated constructions etc.

\subsection{Dual theories: supersymmetric duality and holography.}
\label{sec:QCD:dual}

In the past two decades, in attempts to relate low-energy models of strong
interactions to QCD, theorists created a number of succesful
descriptions of dynamics of theories with large coupling constants in
terms of other theories, in which the perturbation theory works. These
theories, called dual to each other, have coupling constants
$g_{1}$ and $g_{2}$,
for which $g_{1} \sim 1/g_{2}$;
the knowledge of the Green functions of one theory allows to calculate,
following some known rules, the Green functions of another. Note that the
theory dual to QCD has not been constructed up to now.

The simplest example of duality (see e.g.\
\cite{Duality:encycl})
is a theory of the electromagnetic field with magnetic charges. The
Maxwell equations in vacuum are invariant with respect to the exchange of
the electric field
\textbf{E} and the magnetic field
\textbf{B}:
\begin{equation}
{\bf E} \mapsto {\bf B}, ~~~~ {\bf B} \mapsto -{\bf E}.
\label{Eq:E-B-duality}
\end{equation}
This duality breaks down in the presence of electic charges and currents.
It may be restored, however, if one assumes that sources of the other kind
exist in Nature, namely magnetic charges and currents which correspond to
their motion. The self-consistency of the theory requires the Dirac
quantization condition: the unit electic charge $e$ and the unit magnetic
charge $\tilde e$ have to satisfy the relation
$e\tilde e=2\pi$. The charge $e$ is the coupling constant of the usual
electrodynamics while the magnetic charge $\tilde e$ is the coupling
constant of the theory of interaction of magnetic charges which is
obtained from electrodynamics by the duality transformation
(\ref{Eq:E-B-duality}). Therefore, the weak coupling of electric charges,
$e\ll 1$, corresponds to the strong coupling of magnetic ones,
$\tilde e=2\pi/e \gg 1$.

The electromagnetic duality is based on the geometrical properties of
abelian gauge fields which cannot be directly transferred to the
nonabelian case, which is the most interesting phenomenologically. In a
way similar but much more complicated dualities appear in supersymmetric
nonabelian gauge theories. The best known one is the ``electromagnetic''
duality in $SU(2)$ supersymmetric theory with two supercharges ($N=2$)
which is related to the names of Seiberg and Witten
\cite{SeibergWitten}. From the particle-physics point of view, this model
is a $SU(2)$ gauge theory with scalar and fermionic fields
transforming under the adjoint representation of the gauge group, whose
interaction is invariant under special symmetry. For this model, the
effective theory has been calculated which describes the interaction of
light composite particles at low energies and the correspondence has been
given between the effective low-energy and fundamental degrees of freedom.
Like QCD, the fundamental theory is asymptotically free and is in the
strong-coupling regime at low energies; the effective theory describes
weakly interacting composite particles.

The success of the Seiberg-Witten model gave rise to a hope that the
low-energy effective theory for a nonsupersymmetric gauge model with
strong coupling, for instance for QCD, may be obtained from the problem
already solved by means of addition of supersymmetry-breaking terms to the
lagrangians of both the fundamental and the dual theories. The first step
in this direction was to consider $N=1$ supersymmetric gauge theories.
Earlier, starting from mid-1980s, a number of exact results have been
obtained in these theories by making use of (gouverned by supersymmetry)
analitical properties of the effective action \cite{Affleck}. In contrast
with the case of $N=2$ supersymmetry, this is insufficient for the
reconstuction of the full effective theory, but the models dual to
supersymmetric gauge theories with different gauge groups and matter
content have been suggested \cite{Seiberg}. Contrary to the $N=2$ case,
it is impossible to prove the duality here, but the conjecture withstood
all checks carried out. Moreover, it has been shown that the addition of
small soft breaking terms in the Lagrangians of $N=1$ theories corresponds
to a controllable soft supersymmetry breaking in dual models
\cite{softSeiberg}. Unfortunately, one may prove that with the increase of
the supersymmetry-breaking parameters (for instance, when superpartner
masses tend to infinity, so the $N=1$ theory becomes QCD), a phase
transition happens and the dual description stops to work, so the
straightforward application of this approach to QCD is not possible
\cite{PeskinDua}. Also, it is worth noting that the approach does not
allow for a quantitative description of dynamics at intermediate energies,
when the coupling constants of dual theories are both large. Nevertheless,
these methods themselves, as well as the physics intuition based on their
application, have played an important role in the development of other
modern approaches to the study of dynamics of strongly-coupled theories.

One of the theoretically most beautiful and practically most prospective
approaches to the analysis of dynamics of strong interactions at low and
intermediate energies is the so-called holographic approach. Its idea is
that the dual theories may be formulated in spacetime of different
dimensions, in such a way that, for instance, the four-dimensional
dynamics of a theory with large coupling constant is equivalent to
the five-dimensional dynamics of another theory which is weakly coupled (in
a way similar to the two-dimensional description of a three-dimensional
object with a hologram). The best-known realization of this approach is
based on the AdS/CFT correspondence \cite{Maldacena, WittenAdSCFT}, a
practical realization of the duality between a strongly coupled gauge
theory with a four-dimensional conformal invariance (CFT $=$ conformal
field theory) and a multidimensional supergravity with weak coupling
constant. The four-dimensional conformal symmetry includes the Poincare
invariance supplemented by dilatations and inversions. An example of a
nontrivial four-dimensional conformal theory with large coupling constant
$g$ is the $N=4$ supersymmetric Yang-Mills theory with the gauge group
$SU(N_{c})$ which, in the limit $N_{c}\to\infty$, $g^{2}N_{c}\gg 1$,
appears to be dynamically equivalent to a certain supergravity theory
living on the ten-dimensional AdS$_{5}\times S_{5}$ manifold, where
AdS$_{5}$ is the (4+1)-dimensional space with the anti-de-Sitter metrics
(\ref{Eq:AdS}) and $S_{5}$ is the five-dimensional sphere (the $S_{5}$
factor is almost irrelevant in applications, hence the name, AdS/CFT
correspondense). In the limit considered, these two models are equivalent.
To proceed with phenomenological applications, one has to break the
conformal invariance. As a result, the theory has less symmetries, so the
results proven by making use (direct or indirect) of these symmetries are
downgraded to conjectures. Nevertheless, this not fully strict approach
(called sometimes AdS/QCD) brings interesting phenomenological results.

An example is provided by a five-dimensional gauge theory determined at a
finite interval in the $z$ coordinate of the
AdS$_{5}$ space (other geometries of the extra dimensions are also
considered). For the $SU(2)\times SU(2)$ gauge group and a special matter
set one gets the effective theory with QCD symmetries. The series of the
Kalutza-Klein states corresponds to the sequence of mesons whose masses
and decay constants may therefore be calculated directly in the
five-dimensional theory. This approach was succesful; it allows to
calculate various physical observables (in particular, the $\pi$-meson
formfactor discussed above) which agree reasonably with data. A
disadvantage of the method is that the duality between QCD and the
five-dimensional effective theory is not proven. As a result, the choice
of the latter is somewhat arbitrary. An undisputable advantage of this
approach is its phenomenological success achieved without a large number
of tuning parameters, as well as the possibility to calculate observables
for intermediate energies and not only in the zero-energy limit. One may
hope that in the future, a low-energy effective theory for QCD might be
\textit{derived} in the frameworks of this approach.

\section{Conclusions.}
\label{sec:concl}
The Standard model of particle physics gives an excellent description of
almost all data obtained at accelerators already for several decades.  At
the same time, results of both a number of non-accelerator experiments
(neutrino oscillations) and astrophysical observations cannot be explained
in the frameworks of SM and undoubtedly point to its incompleteness. A
more complete theory, yet to be constructed, should allow for a derivation
of the SM parameters and for explanation of their, theoretically not fully
natural, values. The main unsolved problem of SM itself is to describe the
dynamics of gauge theories at strong coupling which would allow to apply
QCD to the description of hadrons at low and intermediate energies.

One may hope that in the next few years, the particle theory will get
additional experimental information both from the Large Hadron Collider, a
powerful accelerator which is bound to explore the entire range of
energies related to the electroweak symmetry breaking, and from numerous
experiments of smaller scales (in particular, those studying neutrino
oscillations, rare processes etc.) and astrophysical observations.
Possibly, this information will allow to construct a succesful extension
of the Standard Model already in the coming decade.

This work was born (and grew up) from a review lecture given by the author
at the Physics department of the Moscow State University. I am indebted to
V.~Belokurov who suggested to convert this lecture into a printed text,
read the manuscript carefully and discussed many points. I thank
V.~Rubakov and V.~Troitsky for attentive reading of the manuscript and
numerous discussions, to M.~Vysotsky and M.~Chernodub for useful
discussions related to particular topics and to A.~Strumia for his kind
permission to use Fig.~\ref{fig:Strumia}. The work was supported in part
by the RFBR grants
10-02-01406 and 11-02-01528, by the FASI state contract
02.740.11.0244, by the grant of the President of the Russian Federation
NS-5525.2010.2  and by the ``Dynasty'' foundation.


\begin{thebibliography}{157}

\bibitem{LHC}
Krasnikov N V, Matveev V A
\textit{New physics at the Large Hadron Collider}
Moscow, Krasand, 2011 (in Russian);
Krasnikov N V, Matveev V A
\textit{Phys.\ Usp.}\  \textbf{47} 643 (2004)

\bibitem{SMprimer}
Burgess C P, Moore G D
{\it The standard model: A primer,}
Cambridge University Press, 2006

\bibitem{ChengLi}
Cheng T P, Li L F
{\it Gauge Theory Of Elementary Particle Physics,}
Oxford, Clarendon, 1984

\bibitem{RuGorby1}
Gorbunov D S, Rubakov V A
\textit{
Introduction to the theory of the early universe: hot big bang theory,}
World Scientific, 2011

\bibitem{KobayashiNobel}
Kobayashi M
{\it Phys.\ Usp.}\ {\bf 52} 12 (2009)

\bibitem{PDG2010}
Nakamura K {\it et al.}  [Particle Data Group]
  {\it J.\ Phys.\ G} {\bf 37} 075021 (2010)

\bibitem{Giunti-book}
 Giunti C, Kim C W
 {\it Fundamentals of Neutrino Physics and Astrophysics,}
Oxford University Press 2007

\bibitem{Bilenky-UFN}
Bilenky S M \textit{Phys.\ Usp.}\ \textbf{46} 1137 (2003)

\bibitem{Akhmedov-UFN}
Akhmedov E K \textit{Phys.\ Usp.}\ \textbf{47} 117 (2004)

\bibitem{Kudenko-UFN}
Kudenko Yu G
\textit{Phys.\ Usp.}\
 \textbf{54} 549 (2011)

\bibitem{1107.3846}
Evans J J
  arXiv:1107.3846 [hep-ex]

\bibitem{g880}
Pontecorvo B
 \textit{Sov.\ Phys.\ JETP}
  {\bf 6} 429 (1957)

\bibitem{g881}
Pontecorvo B
 \textit{Sov.\ Phys.\ JETP}
{\bf 7} 152 (1957)

\bibitem{maki}
Maki Z, Nakagawa M, Sakata S
  {\it Prog.\ Theor.\ Phys.}\  {\bf 28} 870 (1962)

\bibitem{g883}
Pontecorvo B
 \textit{Sov.\ Phys.\ JETP}
{\bf 26}, 984 (1968)

\bibitem{g567}
Gribov V N, Pontecorvo B
  {\it Phys.\ Lett.\  B} {\bf 28} 493 (1969)

\bibitem{g236}
Bilenky S M,  Pontecorvo B
 \textit{Sov.\ J.\ Nucl.\ Phys.}\  {\bf 24} 316 (1976)

\bibitem{g239}
Bilenky S M, Pontecorvo B
  {\it Lett.\ Nuovo Cim.}\  {\bf 17} 569 (1976)

\bibitem{g404}
Eliezer S, Swift A R
  {\it Nucl.\ Phys.\  B} {\bf 105} 45 (1976)

\bibitem{g466}
Fritzsch H, Minkowski P
  {\it Phys.\ Lett.\  B} {\bf 62} 72 (1976)

\bibitem{g801}
Mikheev S P, Smirnov A Y
\textit{Sov.\ J.\ Nucl.\ Phys.}\
{\bf 42} 913 (1985)

\bibitem{g802}
Mikheev S P, Smirnov A Y
  {\it Nuovo Cim.\  C} {\bf 9} 17 (1986).

\bibitem{g1065}
 Wolfenstein L
{\it   Phys.\ Rev.\  D} {\bf 17} 2369 (1978)

\bibitem{Homestake}
Davis R J, Harmer D S, Hoffman K C
  {\it Phys.\ Rev.\ Lett.}\  {\bf 20} 1205 (1968)

\bibitem{Kamiokande}
Hirata K S {\it et al.}  [KAMIOKANDE-II Collaboration]
  {\it Phys.\ Rev.\ Lett.}\  {\bf 63} 16 (1989)

\bibitem{SAGE}
 Abazov A I {\it et al.}  [SAGE Collaboration]
  {\it Phys.\ Rev.\ Lett.}\  {\bf 67} 3332 (1991)

\bibitem{GALLEX}
 Hampel W {\it et al.}  [GALLEX Collaboration]
  {\it Phys.\ Lett.\  B} {\bf 447} 127 (1999)

\bibitem{hep-ex/0404034}
Ashie Y {\it et al.}  [Super-Kamiokande Collaboration]
  {\it Phys.\ Rev.\ Lett.}\  {\bf 93} 101801 (2004)

\bibitem{Kamiokande-atm}
Hirata K S {\it et al.}  [KAMIOKANDE-II Collaboration]
  {\it Phys.\ Lett.\  B} {\bf 205} 416 (1988)

\bibitem{IMB}
 Casper D {\it et al.}
  {\it Phys.\ Rev.\ Lett.}\  {\bf 66} 2561 (1991)

\bibitem{hep-ex/9611007}
Allison W W M {\it et al.}
{\it   Phys.\ Lett.\  B} {\bf 391} 491 (1997)

\bibitem{hep-ex/9807005}
Ambrosio M {\it et al.}  [MACRO Collaboration]
  {\it Phys.\ Lett.\  B} {\bf 434} 451 (1998)

\bibitem{hep-ex/9812014}
Fukuda Y {\it et al.}  [Super-Kamiokande Collaboration]
  {\it Phys.\ Rev.\ Lett.}\  {\bf 82} 2644 (1999)

\bibitem{nucl-ex/0204008}
Ahmad Q R {\it et al.}  [SNO Collaboration]
  {\it Phys.\ Rev.\ Lett.}\  {\bf 89} 011301 (2002)

\bibitem{0801.4589}
Abe S {\it et al.}  [KamLAND Collaboration]
  {\it Phys.\ Rev.\ Lett.}\  {\bf 100} 221803 (2008)

\bibitem{1109.0763}
  Aharmim B {\it et al.}  [SNO Collaboration]
  arXiv:1109.0763

\bibitem{1104.1816}
Bellini G \textit{ et al.} [Borexino Collaboration]
  arXiv:1104.1816 [hep-ex].

\bibitem{hep-ex/0501064}
Ashie Y {\it et al.}  [Super-Kamiokande Collaboration]
  {\it Phys.\ Rev.\  D} {\bf 71} 112005 (2005)

\bibitem{SuperKnu2010}
Takeuchi Y \textit{et al.} [Super-Kamiokande Collaboration]
talk at Neutrino-2010, Athens, 14-19 June 2010.

\bibitem{hep-ex/0606032}
Ahn M H {\it et al.}  [K2K Collaboration]
  {\it Phys.\ Rev.\  D} {\bf 74} 072003 (2006)

\bibitem{hep-ex/0607088}
Michael D G {\it et al.}  [MINOS Collaboration]
  {\it Phys.\ Rev.\ Lett.}\  {\bf 97} 191801 (2006)

\bibitem{1103.0340}
Adamson P {\it et al.}  [The MINOS Collaboration]
 {\it  Phys.\ Rev.\ Lett.}\  {\bf 106} 181801 (2011)

\bibitem{1006.1623}
Agafonova N {\it et al.}  [OPERA Collaboration]
 {\it  Phys.\ Lett.\  B} {\bf 691} 138 (2010)

\bibitem{PRL101-141801}
Fogli G L
\textit{ et al.,}
 {\it  Phys.\ Rev.\ Lett.}\  {\bf 101} 141801 (2008)

\bibitem{T2K}
Abe K {\it et al.}  [T2K Collaboration]
 {\it  Phys.\ Rev.\ Lett.}\  {\bf 107} 041801 (2011)

\bibitem{MINOStheta13}
Adamson P {\it et al.}  [MINOS Collaboration]
  arXiv:1108.0015 [hep-ex]

\bibitem{1106.6028}
 Fogli G L
\textit{ et al.,}
  arXiv:1106.6028 [hep-ph]

\bibitem{LSND}
Aguilar A {\it et al.}  [LSND Collaboration]
  {\it Phys.\ Rev.\  D} {\bf 64} 112007 (2001)

\bibitem{KARMEN}
Church     E D
\textit{ et al.,}
 {\it  Phys.\ Rev.\  D} {\bf 66} 013001 (2002)

\bibitem{MiniBooNEanti}
Aguilar-Arevalo A A {\it et al.}  [The MiniBooNE Collaboration]
  {\it Phys.\ Rev.\ Lett.}\  {\bf 105}  181801 (2010)

\bibitem{MiniBooNEantiNew}
Djurcic Z,
talk at {\it 13th International Workshop on Neutrino Factories, Super
Beams and Beta Beams}, Geneva, 1--6 August 2011

\bibitem{ReactorAnomaly}
Mention G
\textit{ et al.}
  {\it Phys.\ Rev.\  D} {\bf 83} 073006 (2011)

\bibitem{MiniBooNEnu}
Aguilar-Arevalo A A {\it et al.}  [MiniBooNE Collaboration]
 {\it  Phys.\ Rev.\ Lett.}\  {\bf 103} 111801 (2009)

\bibitem{MINOSantiNu}
Thomas J,
talk at {\it Lepton-Photon 2011}, Mumbai, 22--27 August 2011

\bibitem{SuperKantiNu}
Abe K {\it et al.}  [Kamiokande Collaboration]
  arXiv:1109.1621

\bibitem{GALLEXcalibration}
Anselmann P {\it et al.}  [GALLEX Collaboration.]
  {\it Phys.\ Lett.\  B} {\bf 342} 440 (1995);
Kaether F {\it et al.}
 {\it  Phys.\ Lett.\  B} {\bf 685} 47 (2010)

\bibitem{SAGEcalibration}
Abdurashitov D N {\it et al.} [SAGE Collaboration]
  {\it Phys.\ Rev.\ Lett.}\  {\bf 77} 4708 (1996);
Abdurashitov D N {\it et al.} [SAGE Collaboration]
 {\it  Phys.\ Rev.\  C} {\bf 73} 045805 (2006)

\bibitem{GiuntiGallium}
Giunti C, Laveder M
{\it  Phys.\ Rev.\  C} {\bf 83} 065504 (2011)

\bibitem{1008.4750}
Giunti C, Laveder M
  {\it Phys.\ Rev.\  D} {\bf 82} 113009 (2010)

\bibitem{MiniBooNEexcess}
Aguilar-Arevalo A A {\it et al.}  [MiniBooNE Collaboration]
  {\it Phys.\ Rev.\ Lett.}\  {\bf 102} 101802 (2009)

\bibitem{Lobashev-season}
Lobashev V M {\it et al.}
  {\it Phys.\ Lett.}\  B {\bf 460} 227 (1999)

\bibitem{MiniBooNEseason}
Aguilar-Arevalo A A   {\it et al.} [MiniBooNE Collaboration]
  [arXiv:1109.3480 [hep-ex]]

\bibitem{OPERAsuperluminal}
Adam T {\it et al.}  [OPERA Collaboration]
  arXiv:1109.4897 [hep-ex]

\bibitem{hep-ph/0201134}
Strumia A
  {\it Phys.\ Lett.\  B} {\bf 539} 91 (2002)

\bibitem{hep-ph/0207157}
Maltoni M \textit{ et al.}
 {\it Nucl.\ Phys.\  B} {\bf 643} 321 (2002)

\bibitem{1007.4171}
Akhmedov E, Schwetz T
  {\it JHEP} {\bf 1010} 115 (2010)

\bibitem{hep-ph/0010178}
Murayama H, Yanagida T
  {\it Phys.\ Lett.\  B} {\bf 520} 263 (2001)

\bibitem{BSh}
Bogolyubov N N, Shirkov D V
\textit{ Introduction to the theory of quantized fields,}
Intersci.\ Monogr.\ Phys.\ Astron.\  {\bf 3} 1 (1959)

\bibitem{Tsukerman}
Tsukerman I S
 \textit{Phys.\ Usp.}\ {\bf 48} 825 (2005);
Tsukerman I S
arXiv:1006.4989 [hep-ph]

\bibitem{1108.1799}
Diaz J S, Kostelecky A
  arXiv:1108.1799 [hep-ph]

\bibitem{1002.4452}
Engelhardt N, Nelson A E, Walsh J R
  {\it Phys.\ Rev.\  D} {\bf 81} 113001 (2010)

\bibitem{1009.0014}
Kopp J, Machado P A N, Parke S J
 {\it Phys.\ Rev.\  D} {\bf 82} 113002 (2010)

\bibitem{0805.2234}
 Schwetz T
  arXiv:0805.2234 [hep-ph]

\bibitem{1012.3478}
 Yasuda O
  arXiv:1012.3478 [hep-ph]

\bibitem{TroitskNuMass}
 Aseev V N {\it et al.},
  arXiv:1108.5034 [hep-ex]

\bibitem{MainzNuMass}
Kraus C {\it et al.}
  \textit{Eur.\ Phys.\ J.\  C} {\bf 40} 447 (2005)

\bibitem{NuMassCMB}
Hannestad S
\textit{ et al.}
 \textit{ JCAP} {\bf 1008} 001 (2010)

\bibitem{RuGorby2}
Gorbunov D S, Rubakov V A
  {\it Introduction to the theory of the early universe, Cosmological
  perturbations and inflationalry theory},
World Scientific, 2011

\bibitem{Ru-UFN1}
Rubakov V A
\textit{Phys.\ Usp.}\ \textbf{42} 1193 (1999)

\bibitem{Ru-UFN2}
Rubakov V A
\textit{Phys.\ Usp.}\ \textbf{54} 633 (2011)

\bibitem{Sakharov}
Sajharov A D
 \textit{JETP Lett.},  {\bf 5} 24 (1967)

\bibitem{RuShaUFN}
Rubakov V A, Shaposhnikov M E
\textit{Phys.\ Usp.}\
\textbf{39} 461 (1996)

\bibitem{Rubin}
Rubin V C, Thonnard N, Ford W K
\textit{ Astrophys.\ J.}\
{\bf 238} 471 (1980)

\bibitem{Begeman}
Begeman K G
\textit{ Astron.\ Astrophys.}\
\textbf{ 223}
47 (1989)

\bibitem{DSS}
The digitized sky survey (DSS), in \cite{Hubble-archive}.

\bibitem{Zwicky}
Zwicky F
\textit{ Astrophys.\ J.}\
{\bf 86} 217 (1937)

\bibitem{Lensing}
Limousin M \textit{ et al.}
\textit{ Astrophys.\ J.}\
\textbf{ 668} 643 (2007);
http://www.dark-cosmology.dk

\bibitem{Hubble-archive}
The Multimission Archive at the Space Telescope Science Institute (MAST),
http://archive.stsci.edu/ .
STScI is operated by the Association of Universities for Research in
Astronomy, Inc., under NASA contract NAS5-26555.

\bibitem{Chandra-archive}
The Chandra Data Archive (CDA), http://asc.harvard.edu/cda/ .

\bibitem{Bullett}
Clowe D \textit{ et al.}
\textit{ Astrophys.\ J.}\
 {\bf 648} L109 (2006)

\bibitem{Bullett1}
Bradac M {\it et al.}
\textit{ Astrophys.\ J.}\
  {\bf 652} 937 (2006)

\bibitem{MACHOlensing}
Alcock C {\it et al.}  [MACHO Collaboration]
\textit{ Astrophys.\ J.}\
  {\bf 542} 281 (2000)
;
Tisserand P {\it et al.}  [EROS-2 Collaboration]
\textit{ Astron.\ Astrophys.}\
{\bf 469} 387 (2007)

\bibitem{SN1}
Riess A G {\it et al.}  [Supernova Search Team Collaboration]
  {\it Astron.\ J.}\  {\bf 116} (1998) 1009

\bibitem{SN2}
Perlmutter S {\it et al.}  [Supernova Cosmology Project Collaboration]
  {\it Astrophys.\ J.}\  {\bf 517} 565 (1999)

\bibitem{PerlmPhysToday}
Perlmutter S
{\it Physics Today} {\bf 56} (4) 53 (2003)

\bibitem{astro-ph/9604143}
Riess A G, Press W H, Kirshner R P
  {\it Astrophys.\ J.}\  {\bf 473} 88 (1996)

\bibitem{astro-ph/9608192}
Perlmutter S {\it et al.}  [Supernova Cosmology Project Collaboration]
  Astrophys.\ J.\  {\bf 483} (1997) 565

\bibitem{UNION2}
Amanullah R {\it et al.}
 {\it Astrophys.\ J.}\  {\bf 716} 712 (2010);
data for Fig.~\ref{fig:SN} are taken from http://supernova.lbl.gov/Union/

\bibitem{lensingNew}
Jullo E {\it et al.}
{\it Science} {\bf 329} 924 (2010)

\bibitem{CMBflatness}
Komatsu E {\it et al.}  [WMAP Collaboration]
  {\it Astrophys.\ J.\ Suppl.}\  {\bf 192} 18 (2011)

\bibitem{DoubleGal}
Marinoni C, Buzzi A
{\it Nature} {\bf 468} 539 (2010)

\bibitem{chameleon}
Khoury J, Weltman A
  {\it Phys.\ Rev.\  D} {\bf 69} 044026 (2004)

\bibitem{inflation}
Linde A D
{\it Particle physics and inflationary cosmology,}
Harwood Academic Publishers, 1990

\bibitem{LEP-Higgs}
Barate R {\it et al.}  [LEP Working Group for Higgs boson searches, ALEPH,
                  DELPHI, L3 and OPAL Collaborations]
{\it  Phys.\ Lett.\  B} {\bf 565} 61 (2003)

\bibitem{CMS-Higgs}
Sharma V {\it et al.} [CMS Collaboration],
talk at {\it Lepton-Photon 2011}, Mumbai, 22--27 August 2011

\bibitem{ATLAS-Higgs}
Nisati A {\it et al.} [ATLAS Collaboration],
talk at {\it Lepton-Photon 2011}, Mumbai, 22--27 August 2011

\bibitem{Tevatron-Higgs}
Verzocchi M {\it et al.}\
[CDF and D0 Collaborations],
talk at {\it Lepton-Photon 2011}, Mumbai, 22--27 August 2011

\bibitem{Gfitter2011}
Baak M {\it et al.}
  arXiv:1107.0975

\bibitem{Grojean-UFN}
Grojean C
{\it Phys.\ Usp.}\ {\bf 50} 1 (2007)

\bibitem{technicolor}
Lane K
  arXiv:hep-ph/0202255

\bibitem{Higgs5D}
Manton N S {\it Nucl. Phys. B} {\bf 158} 141 (1979)

\bibitem{Higgsless}
Csaki С
{\it et al.}
  {\it Phys.\ Rev.\  D} {\bf 69} 055006 (2004)

\bibitem{Ru-UFN3no-cosm}
Rubakov V A
{\it Phys.\ Usp.}\ {\bf 50} 390 (2007)

\bibitem{1104.3874}
Atwood D, Gupta S K, Soni A
  arXiv:1104.3874 [hep-ph]

\bibitem{Ross}
Ghilencea D, Lanzagorta M, Ross G G
  {\it Phys.\ Lett.\  B} {\bf 415} 253 (1997)

\bibitem{RuST}
Rubakov V A, Troitsky S V
  arXiv:hep-ph/0001213

\bibitem{RuUFNextradim}
Rubakov V A
{\it Phys.\ Usp.}\ {\bf 46} 211 (2003)

\bibitem{Kaluza}
Kaluza T
{\it Sitzungsber. Preuss. Akad. Wiss. Berlin}, Math.-Phys. Kl. (1) 966
(1921)

\bibitem{Klein}
Klein O {\it Z. Phys.}\ {\bf 37} 895 (1926)

\bibitem{Akama}
Akama K {\it Lecture Notes Phys.}\ {\bf 176} 267 (1983)

\bibitem{RuSha1}
Rubakov V A, Shaposhnikov M E {\it Phys. Lett. B} {\bf 125} 136 (1983)

\bibitem{Visser}
Visser M {\it Phys. Lett. B} {\bf 159} 22 (1985)

\bibitem{solid}
von Klitzing K, Nobel lecture (1985);
Fu L, Kane C L
  {\it Phys.\ Rev.\ Lett.}\  {\bf 100} 096407 (2008);
Ghaemi P, Wilczek F arXiv:0709.2626;
Bergman D L, Le Hur K
{\it Phys.\ Rev.\ B} {\bf 79} 184520 (2009);
Volovik G E
  {\it The Universe in a helium droplet},
  {\it Int.\ Ser.\ Monogr.\ Phys.}\  {\bf 117} (2006)

\bibitem{ADD}
Arkani-Hamed N, Dimopoulos S, Dvali G {\it Phys. Lett. B} {\bf 429} 263
(1998)

\bibitem{PRL98(2007)021101}
Kapner D J {\it et al.}
{\it   Phys.\ Rev.\ Lett.}\  {\bf 98} 021101 (2007)

\bibitem{DvaliShifman}
Dvali G, Shifman M
{\it Phys. Lett. B} {\bf 396} 64 (1997)

\bibitem{RuSha2}
Rubakov V A, Shaposhnikov M E
  {\it Phys.\ Lett.\  B} {\bf 125} 139 (1983)

\bibitem{Gogber}
Gogberashvili M {\it Mod. Phys. Lett. A} {\bf 14} 2025 (1999)

\bibitem{Lisa1}
Randall L, Sundrum R {\it Phys. Rev. Lett.}\ {\bf 83} 3370 (1999)

\bibitem{Lisa6D-loc}
Oda I {\it Phys. Lett. B} {\bf 496} 113 (2000)

\bibitem{SUSYufnNevzorov}
Vysotsky M I, Nevzorov R B
{\it Phys.\ Usp.}\ {\bf 44} 919 (2001)

\bibitem{SUSYufnWe}
Gorbunov D S, Dubovsky S L, Troitsky S V
{\it Phys.\ Usp.}\  {\bf 42} 623 (1999)

\bibitem{SUSYKazakov}
Kazakov D I
  arXiv:hep-ph/0012288

\bibitem{Strumia1101.2195}
Strumia A
  {\it JHEP} {\bf 1104} 073 (2011)

\bibitem{LittleHiggs}
 Schmaltz M, Tucker-Smith D
  {\it Ann.\ Rev.\ Nucl.\ Part.\ Sci.}\  {\bf 55} 229 (2005)

\bibitem{TonyReview}
Gherghetta T
  arXiv:1008.2570 [hep-ph].

\bibitem{LT}
Libanov M, Troitsky S
{\it  Nucl.\ Phys.\  B} {\bf 599} 319 (2001)

\bibitem{FLT}
Frere J-M, Libanov M, Troitsky S
  {\it Phys.\ Lett.\  B} {\bf 512} 169 (2001)

\bibitem{FuSin}
Frere J-M, Libanov M, Ling F S
  {\it JHEP} {\bf 1009} 081 (2010)

\bibitem{Shmalz}
Dvali G R, Shifman M A,
  {\it Phys.\ Lett.}\  {\bf B475 } 295 (2000)

\bibitem{AsymptFreedomNoConfinement}
Iwasaki Y {\it et al.}
  {\it Phys.\ Rev.\ Lett.}\  {\bf 69} 21 (1992)

\bibitem{KrTr}
Krutov A F, Troitsky V E, Tsirova N A
  {\it Phys.\ Rev.\  C} {\bf 80} 055210 (2009)

\bibitem{BABAR0905.4778}
Aubert B {\it et al.}  [The BABAR Collaboration]
 {\it  Phys.\ Rev.\  D} {\bf 80} 052002 (2009)

\bibitem{Lattice:encycl}
Di Giacomo A
{\it Lattice gauge theory},
in:
{\it Encyclopedia
of Mathematical Physics}, Academic Press, Oxford (2006)

\bibitem{Lattice:confinement}
Creutz M
  {\it Phys.\ Rev.\  D} {\bf 21} 2308 (1980)

\bibitem{0911.2561}
Aoki S {\it et al.}  [PACS-CS Collaboration]
  {\it Phys.\ Rev.\  D} {\bf 81} 074503 (2010)

\bibitem{Lattice:PhaseTrans}
Kajantie K, Montonen C, Pietarinen E
  {\it Z.\ Phys.\  C} {\bf 9} 253 (1981)

\bibitem{Fodor}
Aoki Y {\it et al.}
  Nature {\bf 443} 675 (2006)

\bibitem{Duality:encycl}
Tsun T S
{\it Electric--magnetic duality},
in:
{\it Encyclopedia
of Mathematical Physics}, Academic Press, Oxford (2006)

\bibitem{SeibergWitten}
Seiberg N, Witten E
  {\it Nucl.\ Phys.\  B} {\bf 426} 19 (1994)
  [Erratum-ibid.\   {\bf 430} 486 (1994)]

\bibitem{Affleck}
Affleck I, Dine M, Seiberg N
  {\it Nucl.\ Phys.\  B} {\bf 241} 493 (1984)

\bibitem{Seiberg}
Seiberg N
  {\it Nucl.\ Phys.\  B} {\bf 435} 129 (1995)

\bibitem{softSeiberg}
Evans N J, Hsu S D H, Schwetz M
  {\it Phys.\ Lett.\  B} {\bf 355} 475 (1995)

\bibitem{PeskinDua}
Aharony O {\it et al.}\
  {\it Phys.\ Rev.\  D} {\bf 52} 6157 (1995)

\bibitem{Maldacena}
Maldacena J M
  {\it Adv.\ Theor.\ Math.\ Phys.}\  {\bf 2} 231 (1998)
  [\textit{Int.\ J.\ Theor.\ Phys.}\  {\bf 38} 1113 (1999)]

\bibitem{WittenAdSCFT}
Witten E
  {\it Adv.\ Theor.\ Math.\ Phys.}\  {\bf 2} 253 (1998)


\end{thebibliography}
\end{document}